\documentclass[twocolumn]{aastex631}

\usepackage{natbib}

\newcommand{\RNum}[1]{\uppercase\expandafter{\romannumeral #1\relax}}

\newcommand{\HII}{\textup{H~\textsc{II}}}
\newcommand{\Hb}{\textup{H}\ensuremath{\beta}}
\newcommand{\Ha}{\textup{H}\ensuremath{\alpha}}

\newcommand{\SIIISII}{[\ion{S}{3}]/[\ion{S}{2}]}
\newcommand{\OIIIOII}{[\ion{O}{3}]/[\ion{O}{2}]}
\newcommand{\temp}[2]{$T_{\rm{e},\text{[{#1}{ \RNum{#2}}]}}$}
\newcommand{\tempDiff}[4]{$\Delta$(\temp{#1}{#2}, \temp{#3}{#4})}
\received{August 29, 2023}
\accepted{March 1st, 2024}

\shorttitle{AASTeX v6.3.1 Sample article}
\shortauthors{Rickards Vaught et al.}
\graphicspath{{./}{figures/}}
\defcitealias{Zurita2021MNRAS.500.2359Z}{Z21}
\defcitealias{Delgado_Desire2023arXiv230513136M}{MD23}
\defcitealias{Berg2020ApJ...893...96B}{CHAOS-IV}
\defcitealias{Rogers:chaos:2021}{CHAOS-VI}
\defcitealias{Garnett1992AJ}{G92}
\defcitealias{BOND}{BOND}
\begin{document}

\title{Investigating the Drivers of Electron Temperature Variations in HII Regions with Keck-KCWI and VLT-MUSE}

\author[0000-0001-9719-4080]{Ryan J. Rickards Vaught}
\affiliation{Center for Astrophysics and Space Sciences, Department of Physics, University of California, San Diego, 9500 Gilman Dr., La Jolla, CA 92093, USA}
\correspondingauthor{Ryan J. Rickards Vaught}
\email{rjrickar@ucsd.edu}

\author[0000-0002-4378-8534]{Karin M. Sandstrom}
\affiliation{Department of Astronomy \& Astrophysics, University of California, San Diego, 9500 Gilman Dr., La Jolla, CA 92093, USA}

\author[0000-0002-2545-5752]{Francesco Belfiore}
\affiliation{INAF -- Osservatorio Astrofisico di Arcetri, Largo E. Fermi 5, I-50157 Firenze, Italy}

\author[0000-0001-6551-3091]{Kathryn Kreckel}
\affiliation{Astronomisches Rechen-Institut, Zentrum f\"{u}r Astronomie der Universit\"{a}t Heidelberg, M\"{o}nchhofstr. 12-14, D-69120 Heidelberg, Germany}

\author[0000-0002-6972-6411]{J. Eduardo Méndez-Delgado}
\affiliation{Astronomisches Rechen-Institut, Zentrum f\"{u}r Astronomie der Universit\"{a}t Heidelberg, M\"{o}nchhofstr. 12-14, D-69120 Heidelberg, Germany}

\author[0000-0002-6155-7166]{Eric Emsellem}
\affiliation{European Southern Observatory, Karl-Schwarzschild Stra{\ss}e 2, D-85748 Garching bei M\"{u}nchen, Germany }
\affiliation{Univ Lyon, Univ Lyon1, ENS de Lyon, CNRS, Centre de Recherche Astrophysique de Lyon UMR5574, F-69230 Saint-Genis-Laval France}

\author[0000-0002-9768-0246]{Brent Groves}
\affiliation{International Centre for Radio Astronomy Research, University of Western Australia, 7 Fairway, Crawley, 6009 WA, Australia}

\author[0000-0003-4218-3944]{Guillermo A. Blanc}
\affiliation{Observatories of the Carnegie Institution for Science, 813 Santa Barbara Street, Pasadena, CA 91101, USA}
\affiliation{Departamento de Astronom\'{i}a, Universidad de Chile, Camino del Observatorio 1515, Las Condes, Santiago, Chile}

\author[0000-0002-5782-9093]{Daniel A. Dale}
\affiliation{Department of Physics \& Astronomy, University of Wyoming, Laramie, WY 82071, USA}

\author[0000-0002-4755-118X]{Oleg~V.~Egorov}
\affiliation{Astronomisches Rechen-Institut, Zentrum f\"{u}r Astronomie der Universit\"{a}t Heidelberg, M\"{o}nchhofstr. 12-14, D-69120 Heidelberg, Germany}

\author[0000-0001-6708-1317]{Simon C.~O.\ Glover}
\affiliation{Universit\"{a}t Heidelberg, Zentrum f\"{u}r Astronomie, Institut f\"{u}r Theoretische Astrophysik, Albert-Ueberle-Str. 2, 69120, Heidelberg, Germany.}

\author[0000-0002-3247-5321]{Kathryn~Grasha}
\altaffiliation{ARC DECRA Fellow}
\affiliation{Research School of Astronomy and Astrophysics, Australian National University, Canberra, ACT 2611, Australia}   
\affiliation{ARC Centre of Excellence for All Sky Astrophysics in 3 Dimensions (ASTRO 3D), Australia}

\author[0000-0002-0560-3172]{Ralf S.\ Klessen}
\affiliation{Universit\"{a}t Heidelberg, Zentrum f\"{u}r Astronomie, Institut f\"{u}r Theoretische Astrophysik, Albert-Ueberle-Str. 2, 69120, Heidelberg, Germany.}
\affiliation{Universit\"{a}t Heidelberg, Interdisziplin\"{a}res Zentrum f\"{u}r Wissenschaftliches Rechnen, Im Neuenheimer Feld 205, D-69120 Heidelberg, Germany.}

\author[0000-0002-3289-8914]{Justus Neumann}
\affiliation{Max-Planck-Institut für Astronomie, Königstuhl 17, D-69117 Heidelberg, Germany}

\author[0000-0002-0012-2142]{Thomas G. Williams}
\affiliation{Sub-department of Astrophysics, Department of Physics, University of Oxford, Keble Road, Oxford OX1 3RH, UK}



\begin{abstract}
\HII\ region electron temperatures are a critical ingredient in metallicity determinations and recent observations reveal systematic variations in the temperatures measured using different ions. We present electron temperatures ($T_e$) measured using the optical auroral lines ([\ion{N}{2}]$\lambda5756$, [\ion{O}{2}]$\lambda\lambda7320,7330$, [\ion{S}{2}]$\lambda\lambda4069,4076$, [\ion{O}{3}]$\lambda4363$, and [\ion{S}{3}]$\lambda6312$) for a sample of \HII\ regions in seven nearby galaxies. We use observations from the Physics at High Angular resolution in Nearby Galaxies survey (PHANGS) obtained with integral field spectrographs on Keck (Keck Cosmic Web Imager; KCWI) and the Very Large Telescope (Multi-Unit Spectroscopic Explorer; MUSE). We compare the different $T_e$ measurements with \HII\ region and interstellar medium environmental properties such as electron density, ionization parameter, molecular gas velocity dispersion, and stellar association/cluster mass and age obtained from PHANGS. We find that the temperatures from [\ion{O}{2}] and [\ion{S}{2}] are likely over-estimated due to the presence of electron density inhomogeneities in \HII\ regions. We measure high [\ion{O}{3}] temperatures in a subset of regions with high molecular gas velocity dispersion and low ionization parameter, which may be explained by the presence of low-velocity shocks. In agreement with previous studies, the $T_{\rm{e}}$--$T_{\rm{e}}$ between [\ion{N}{2}] and [\ion{S}{3}] temperatures have the lowest observed scatter and follow predictions from photoionization modeling, which suggests that these tracers reflect \HII\ region temperatures across the various ionization zones better than [\ion{O}{2}], [\ion{S}{2}], and [\ion{O}{3}].
\end{abstract}


\section{Introduction} \label{sec:intro}
The characterization of abundance variations within galaxies provides insight into the physical processes that drive galaxy and chemical evolution. A galaxy's gas-phase metal abundance (i.e. metallicity) reflects the history of chemical enrichment from stars and the net balance of gas flows (mixing, outflows, inflows of pristine material, etc.). In addition, the metallicity of ISM gas directly controls its cooling and other important ISM physics \citep{Draine2011piim.book.....D, Piembert2017PASP..129h2001P}.

The distribution of gas-phase metals in a galaxy is commonly traced by the abundance of oxygen, nitrogen, sulfur, and other metals using the emission from ionized gas located inside \HII\ regions \citep[e.g.][]{Kennicutt1996,Bresolin2012ApJ...750..122B,Hernandez2013ApJ...777...19H,Ho2017ApJ...846...39H,Kreckel2019ApJ...887...80K, Kreckel2020MNRAS.499..193K, vanloon2021MNRAS.tmp.1227V, Grasha:2022}. There are several indirect methods calibrated using strong optical emission lines to derive an estimate of the \HII\ region metallicity \citep[e.g.][]{2008ApJ...681.1183K, Blanc:2015}. A ``direct'' measure of an \HII\ region metallicity requires knowledge of the electron temperature ($T_{\rm{e}}$) of the gas. Due to their exponential dependence on $T_{\rm{e}}$, one of the ways to infer electron temperature is through the auroral-to-nebular line ratios of collisionally excited lines \citep[CEL;][]{Peimbert1967ApJ...150..825P,Osterbrock:2006,Piembert2017PASP..129h2001P}. Nebular and auroral lines originate from different excited states of ions. Auroral lines are from higher energy levels, but are still accessible for collisional excitation in a $T \sim 10^4$ K gas. If the density of the gas is below the auroral and nebular line critical densities (i.e.\ when collisional de-excitation is negligible), then the auroral-to-nebular line ratio is sensitive to the electron temperature \citep[e.g.][]{Osterbrock:2006}. Given that the excitations to the auroral level are only accessible to electrons of higher energy, auroral line emission can be $> 100$ times weaker than nebular lines \citep{Kennicutt2003ApJ...591..801K,Esteban2004MNRAS.355..229E,Berg2020ApJ...893...96B}. One alternative way to measure $T_{\rm{e}}$ include the ratio between recombination line (RL) emission from H and other species \citep{Peimbert1967ApJ...150..825P,Osterbrock:2006,Piembert2017PASP..129h2001P}. But, because RLs of ions exhibit a much weaker dependence to temperature \citep[$T_{\rm{e}}^{-\kappa}$ where $-0.2 < \kappa <0.2$, ][]{Piembert2017PASP..129h2001P}, the optical RLs useful for use as temperature diagnostics are typically reserved for deep high S/N spectra as RL emission is typically much fainter than the emission from auroral lines.

For ions with optical auroral lines studied in this work, we can measure the temperatures for each ion using the following line ratios:
\begin{center}
\temp{O}{3} $\rightarrow$ [\ion{O}{3}]$\lambda4363$/$\lambda\lambda4959 ,5007$, 
\end{center}
\begin{center}
\temp{O}{2} $\rightarrow$ [\ion{O}{2}]$\lambda\lambda7320,7330$/$\lambda\lambda3726,3729$\footnote{[\ion{O}{2}]$\lambda\lambda7320,7330$ is an unresolved quadruplet with transitions at $\lambda7319$, $\lambda7320$, $\lambda7330$ and $\lambda7331$ \AA.},
\end{center}
\begin{center}
\temp{S}{3} $\rightarrow$ [\ion{S}{3}]$\lambda6312$/$\lambda\lambda9069,9532$, 
\end{center}
\begin{center}
\temp{S}{2} $\rightarrow$ [\ion{S}{2}]$\lambda\lambda4069,4076$/$\lambda\lambda6716,6731$,
\end{center}
\begin{center}
\temp{N}{2} $\rightarrow$ [\ion{N}{2}]$\lambda5756$/$\lambda\lambda6548,6584$. 
\end{center}
The O$^{+}$, N$^+$, S$^+$ ions require energies of $13.6$ eV, $14.5$ eV, and $10.3$ eV to be produced while S$^{++}$ and  O$^{++}$ require energies 23 eV and 35 eV, respectively.

Several effects play competing roles in determining the ionization and temperature structure of \HII\ regions. These include a radially decreasing intensity and hardening of the radiation field (photons closest to 13.6 eV are absorbed first) as well as a change in the ions which dominate gas cooling, and therefore the cooling efficiency \citep{stasinska:1980,Garnett1992AJ}.
Because of the varying degree of ionization within an \HII\ region, a model of three ionization zones---low-, intermediate- and high---is often used to describe them. Because each ionization zone could theoretically have different temperatures, this further stresses the importance of observing multiple auroral lines and developing temperature priorities for use in accurately determining abundances \citep[e.g.][]{Berg2015ApJ...806...16B,Berg2020ApJ...893...96B, Rogers:chaos:2021}. 

Observing the full set of optical auroral lines in an \HII\ region can be challenging. In addition to the large wavelength range needed, $\sim$ 3700--10000 \AA, some auroral lines are weaker than others depending on the metallicity and temperature of the gas. Because of these challenges, it is very important to establish temperature relationships that allow us to infer the conditions of a certain ionization zone from the others. Photoionization modeling \citep[e.g.][]{Garnett1992AJ, BOND} has been used to derive temperature relationships, but standard models consider only simple geometries and homogeneous physical and ionization conditions, that might not be suitable for more complex regions potentially affected by shocks, stellar feedback, or other mechanism that produce density or temperature inhomogeneities \citep{Peimbert1967ApJ...150..825P,Peimbert1991shocks,Binette2012_OIII_shocks,Berg2015ApJ...806...16B,Berg2020ApJ...893...96B,Arellano2020, Nicholls2020PASP..132c3001N, Delgado_Desire2023arXiv230513136M, Delgado2023arXiv230511578M}.

In the presence of temperature fluctuations, the exponential dependence of CEL strengths on temperature will bias auroral-to-nebular temperatures towards higher values than the true average \citep{Peimbert1967ApJ...150..825P,Peimbert:1969}. Such inhomogeneities may be related to the presence of turbulence, density structure, and shocks associated with either stellar winds or radiation--pressure driven expansion. If the sources of temperature inhomogeneities are confined to the central part of the nebula, the effects that these phenomena have on temperature may primarily affect only the high ionization zone. This has been suggested by \citet{Delgado2023arXiv230511578M} who presented evidence for temperature inhomogeneities affecting only the highly ionized gas traced by [\ion{O}{3}]. In a sample of Galactic and extra-galactic \HII\ regions, they observed that differences between [\ion{O}{3}] and [\ion{N}{2}] temperatures correlated with the degree of deviation from the average temperature measured using faint \ion{O}{2} recombination line emission. Furthermore, a strong correlation between the \ion{O}{2} recombination and [\ion{N}{2}] temperatures observed by \citet{Delgado2023arXiv230511578M} implies that temperatures inferred from the [\ion{N}{2}] auroral line accurately measures the average $T_\mathrm{e}$ of the low-ionization zone \citep{Delgado2023arXiv230511578M,Delgado_Desire2023arXiv230513136M}.

Due to the importance of obtaining accurate temperatures for precise abundances, significant effort has been devoted to advancing our understanding of the temperatures of different \HII\ region ionization zones. For example, previous works have found that the scatter between temperatures of different ionization zones may be correlated with other properties of the gas such as the ionization parameter and metallicity \citep{Berg2015ApJ...806...16B, Arellano2020, Berg2020ApJ...893...96B, Yates2020A&A...634A.107Y}. 

To explore these questions, we use deep 3600--9500 \AA\ IFU mapping to measure the set of optical auroral lines and nebular lines for a sample of \HII\ regions. We use observations obtained from the Keck Cosmic Web Imager \citep[KCWI,][]{2018ApJ...864...93M} and Multi-Unit Spectroscopic Explorer \citep[MUSE,][]{MUSE_INST:2010} to measure the electron temperature from all 3 ionization zones in \HII\ regions in nearby galaxies. In Section \ref{sec:Observations} we present our sample galaxies as well as primary and supplemental observations. In Section \ref{sec:reduction} we discuss the reduction of the KCWI data. We assess the quality of the KCWI mosaics in Section \ref{sec:data_quality_assesment}. We construct our \HII\ region sample in Section \ref{sec:hii_cat_intro}. We present the auroral line measurements in Section \ref{sec:auroral_fit}. We derive \HII\ region properties from nebular diagnostics and from ALMA and HST data in Section \ref{sec:hii_region_property_derivations}. The results and discussion are presented in Section \ref{sec:RESULTS} and \ref{sec:DISCUSSION}.

\section{Observations}
\label{sec:Observations}
The analysis presented here makes joint use of multi-wavelength observations of seven galaxies obtained with Keck-KCWI, VLT-MUSE, Atacama Large Millimeter/submillimeter Array (ALMA), and the Hubble Space Telescope (HST).

\subsection{Sample Selection}
The seven galaxies in this analysis are drawn from the PHANGS-MUSE sample \citep{EmsellemMuse}. To date, 90 galaxies make up the full PHANGS sample\footnote{\href{https://sites.google.com/view/phangs/home}{http://phangs.org/} } \citep{2021Leroy_sample}, and 19 have been observed by MUSE. In order to be observed with KCWI in the northern hemisphere, we selected the seven target galaxies from a subset of PHANGS-MUSE galaxies with declination, $\delta> -30^{\circ}$. Table \ref{tab:sample_properties} presents general properties of these galaxies, including distances, masses, sizes, and the angular resolution of the MUSE data.

\subsection{Keck Cosmic Web Imager}
\label{sec:obs}
We observed each galaxy using KCWI on the Keck~II telescope with multiple pointings taken over several nights between the years 2017 and 2021. Clear conditions were present for the majority of observations, except for the nights of October 16 and 17, 2018, which suffered from variable cloud coverage. These poor conditions primarily affect the observations of NGC\,628. The instrument was configured with the ``Large'' slicer and BL grating centered at 4600 \AA. The usable spectral range afforded by this configuration is 3650--5550 \AA\ with a spectral resolution $R\sim 900$, corresponding to a full width at half maximum (FWHM) $\sim 5.1$ \AA\ (or $\sim$ 300 km s$^{-1}$) at the central wavelength. The Large slicer has an angular slice width of 1.35\arcsec. The field of view (FoV) using the Large slicer and BL grating is 33\arcsec\ perpendicular and 20.4\arcsec\ parallel to the slicer.

Because the FoV is small compared with the large angular size of each galaxy, we observed each galaxy over multiple fields. Most fields were observed two times using 1200~s (i.e. 20 min) integration times. The only exceptions were: all fields in NGC\,3627, which were observed five times each with 120~s (2 min) integration times; field 17 in NGC\,628 which was observed 3 times using 1200~s; and field~2 in NGC\,5068 and field~5 in NGC\,1385, both having only a single observation of 1200~s. A half slice width, or 0.675\arcsec, dither was applied between each exposure. %Immediately following the second exposure, 
We observed an off-galaxy region, selected to be free of extended emission and/or bright sources, in order to measure a sky spectrum close in time to the observations. These sky frames, observed using an integration time of $600$~s (i.e.\ 10 min), were used for sky subtraction during data reduction. We summarize the number of fields, exposure times, and dates in Table \ref{tab:obs} of Appendix \ref{appn:observations}. The full data reduction is outlined in Section \ref{sec:reduction}.

\subsection{Multi-Unit Spectroscopic Explorer} \label{sec:MuseInfo}
MUSE observations of these galaxies come from the PHANGS-MUSE survey \citep{EmsellemMuse}. MUSE covers wavelengths between 4800--9500 \AA. Taken in combination, KCWI and MUSE span the full optical spectrum.  The full details of the MUSE data reduction and data products are presented in \cite{EmsellemMuse}, and we provide a brief overview here. The PHANGS-MUSE program observed 19 galaxies using 168 individual 1$'\times1'$ pointings. The median spatial resolution across all pointings is $\sim 50$ pc (or $\sim$ 0.80\arcsec) with a typical spectral FWHM of $\sim 2.5$ \AA\ (but varying with wavelength). The data were reduced using the \texttt{pymusepipe}\footnote{https://pypi.org/project/pymusepipe/} and spectral fitting and analysis was performed using the Data Analysis Pipeline \footnote{https://gitlab.com/francbelf/ifu-pipeline} packages described in \citet{EmsellemMuse}. The individual MUSE pointings were homogenized to a uniform Gaussian point-spread-function (PSF) with FWHM set to the largest FWHM measured for each target, resulting in ``convolved and optimized'' (COPT) mosaics. The PSFs of the COPT mosaics are listed in Table \ref{tab:sample_properties}. We use the COPT mosaics in the following work.
 
 \subsection{ALMA} \label{sec:ALMAInfo}
 Our analysis makes use of Atacama Large Millimeter/submillimeter Array (ALMA) data obtained as part of PHANGS–ALMA \citep{2021Leroy_sample}. PHANGS-ALMA observed the $J=2-1$ rotational transition of $^{12}$CO, hereafter CO, for $90$ galaxies. The details of the data reduction are described in \citet{2021Leroy_data}. We make use of the ALMA datacubes with combined CO measurements from the 12m and 7m arrays plus Total Power (12m+7m+TP). The nominal angular resolution of 12m+7m+TP observations is $\sim 1.3\arcsec$, similar to the resolution of both KCWI and MUSE. The velocity resolution is $2.5$ km s$^{-1}$.
 
 \subsection{HST} \label{sec:HSTInfo}
The PHANGS-HST survey \citep{2022PHANGS_HST} observed\footnote{\citet{2022PHANGS_HST} use of previous NGC\,628 HST imaging obtained as part of the Legacy ExtraGalactic Ultraviolet Survey \citep[Legus;][]{LEGUS:2015}.} our target galaxies using 5 HST filters: F275W (NUV), F336W (U), F438W (B), F555W (V), F814W (I). Of the high-level data products produced from this dataset, we make use of compact star cluster catalogs \citep[][Maschmann \& Lee et al. submitted]{ThilkerCompactClusters} and multi-scale stellar association catalogs \citep{Larson2023}. The association catalog identifies sources using both the V and NUV filters and has been convolved to several physical resolutions (8, 16, 32 and 64~pc, respectively). Following \citet{Scheuermann2023MNRAS.522.2369S} we used the NUV selected, 32~pc catalogs.

\begin{deluxetable*}{cDDDDcc}
\tablecaption{Properties of the PHANGS-MUSE galaxies observed with KCWI.}
\label{tab:sample_properties}
\tablehead{\colhead{Name} & \multicolumn2c{Distance\tablenotemark{a}} & \multicolumn2c{log$_{10}(M_{*})$ \tablenotemark{b}} & \multicolumn2c{$R_{25}$\tablenotemark{c}}  & \multicolumn2c{PSF$_{\mathrm{MUSE}}$\tablenotemark{d}} & \colhead{PSF$_{\mathrm{KCWI}}$\tablenotemark{e}} \\
\colhead{} & \multicolumn2c{[Mpc]} & \multicolumn2c{[$M_{\odot}$]} & \multicolumn2c{[Arcmin]} & \multicolumn2c{[Arcsec]} & \colhead{[Arcsec]}} 
\decimals
\startdata
NGC~628 & 9.8$\pm$0.6  & 10.34$\pm$0.1 & 4.9 &  0.92  & 2.0 $\pm$ 0.4 \\
NGC\,1087 & 15.9$\pm$2.2 & 9.93$\pm$0.1  & 1.5 &  0.92  & 1.2 $\pm$ 0.1 \\
NGC\,1300 & 19.0$\pm$2.3 & 10.62$\pm$0.1 & 3.0 &  0.89* & 1.3 $\pm$ 0.1 \\
NGC\,1385 & 17.2$\pm$2.6 & 9.98$\pm$0.1  & 1.7 &  0.77* & 1.3 $\pm$ 0.1 \\
NGC\,2835 & 12.2$\pm$0.9 & 10.00$\pm$0.1 & 3.2 &  1.15  & 1.4 $\pm$ 0.1   \\
NGC\,3627 & 11.3$\pm$0.5 & 10.83$\pm$0.1 & 5.1 &  1.05  & 1.1 $\pm$ 0.1  \\
NGC\,5068 & 5.2$\pm$0.2  & 9.40$\pm$0.1  & 3.7 &  1.04  & 1.5 $\pm$ 0.4
\enddata
\tablenotetext{a}{From the compilation of \cite{Anand2021MNRAS.501.3621A}}
\tablenotetext{b}{Derived by \cite{2021Leroy_sample}, using GALEX UV and WISE IR photometry.}
\tablenotetext{c}{From LEDA \cite{Makarov2014}.}
\tablenotetext{d}{The FWHM of the Gaussian PSF for the homogenized COPT mosaic from PHANGS-MUSE \citep{EmsellemMuse}.} 
\tablenotetext{e}{The average FWHM of the KCWI PSF for the set of a galaxy's observed pointings.}
\tablenotetext{*}{Denotes galaxies observed with MUSE using ground based adaptive optics.}
\end{deluxetable*}

\section{KCWI Data Reduction}
\label{sec:reduction}
The KCWI observations were reduced using the Version 1.0.1 Python implementation of the KCWI Data Extraction and Reduction Pipeline (KDRP)\footnote{\href{https://github.com/Keck-DataReductionPipelines/KCWI\_DRP}{KCWI DRP-Python}}. It was built using the Keck Data Reduction Pipeline Framework package\footnote{\href{https://github.com/Keck-DataReductionPipelines/KeckDRPFramework}{KeckDRPFramework}} and is a port of the initial IDL reduction pipeline\footnote{\href{https://github.com/Keck-DataReductionPipelines/KCWIDRP}{KCWI DRP-IDL}} \citep{2018ApJ...864...93M}.
The pipeline performs basic CCD reduction including bias and over-scan subtraction, gain correction, cosmic ray removal, dark and scattered light subtraction as well as a flat field correction. 

Following these basic reductions, the KDRP used the continuum bar and Thorium/Argon arc lamp images to generate geometric and wavelength solutions to convert each 2D science image into a spectral datacube. The accuracy of the wavelength solutions were similar across all the observation nights. The average RMS for the derived wavelength solutions was 0.1 \AA.

We derived an inverse sensitivity curve to flux calibrate each datacube from standard star observations. The measured standard deviation between all of the derived inverse sensitivity curves was $\sim9\%$ at $\lambda=4600$\AA. The maximum standard deviation within the wavelength range containing the lines used in this analysis is $\sim11\%$. Details of each standard star observation can be found in Table~\ref{tab:std_obs} of Appendix \ref{appn:observations}. After flux calibration, each datacube was corrected for differential atmospheric refraction.

Because the instrument FoV is much smaller than each galaxy, our images contained no sky pixels. To perform sky subtraction, we observed dedicated sky positions interspersed between science observations.  We assigned the  sky frame closest in time to each science observation to be used for sky subtraction according to the instructions in the KDRP documentation\footnote{https://kcwi-drp.readthedocs.io/en/latest/sky\_subtraction.html}. The KDRP then performed sky subtraction using our preferred frames. The sky in all pixels was averaged together and scaled by the ratio of science-to-sky exposure time to estimate the sky observed in the ``on'' position. The final data products output by the pipeline include flux calibrated science and $1\sigma$ uncertainty cubes, as well as a bad-pixel mask cube.

\subsection{Image Re-Projection}
\label{sec:mosaic}
Next we constructed mosaics from the individual KCWI pointing datacubes. The steps involved included image registration, matching the flux calibration to MUSE, and image co-addition. We also compared the absolute flux calibration of the final KCWI mosaics to MUSE and SDSS

The datacubes output by the KDRP have rectangular pixels with pixel-scale 1.35\arcsec\ $\times$ 0.29\arcsec. We reprojected the cubes onto a square pixel grid using the astronomical mosaic software \texttt{Montage}\footnote{See http://montage.ipac.caltech.edu/}. Prior to running \texttt{Montage}, we converted the KCWI data to surface brightness units (erg s$^{-1}$ cm$^{-2}$ \AA$^{-1}$ sr$^{-1}$) by dividing the flux per pixel by the pixel solid angle in steradians. The reprojected images have a final square pixel grid with a uniform pixel-scale of 0.29\arcsec $\times$ 0.29\arcsec. We validated the flux-conservation in our data before and after reprojection.

\subsection{Image Registration}
\label{sec:registration}
To co-add each galaxy's set of cubes into a spectral mosaic, we placed each cube onto a common world coordinate system (WCS). It is typical to perform image registration by matching the location of foreground stars or background galaxies to known positions found in catalogs. However, in our case most individual fields did not contain a sufficient number of bright point sources. Instead, we performed image registration by maximizing the cross-correlation between individual KCWI fields and overlapping MUSE data in order to match the KCWI pointing astrometry to MUSE. The astrometry of the MUSE galaxy mosaics were validated against wide-field broadband imaging and stellar positions from the Gaia DR1 as described in \cite{EmsellemMuse}. The MUSE astrometry, when compared to broadband imaging, exhibited better than 100 milli-arcseconds RMS. in both R.A. and Dec. 

To cross-correlate KCWI and MUSE, we created synthetic photometry ($P_S$) images from spectral regions where the wavelength coverage of KCWI and MUSE overlap. Because there is some saturation in \Hb\ and [\ion{O}{3}] at the brightest locations in the KCWI data, we masked out those lines in both cubes to avoid any issues with the comparison between the two data cubes.

In order to determine the optimal astrometric shifts to apply to each KCWI science frame, we utilized a two-step grid search operation, first shifting in 1 pixel (or 0.29\arcsec) steps $\pm$ 17 pixels (or $5\arcsec$) from the center of the KCWI pointing in order to find the optimal R.A. and Dec. offsets which maximize the correlation of the KCWI and corresponding MUSE $P_S$ images. After locating 1st-pass shifts, we performed a secondary grid search using finer 1/2 pixel increments over a smaller range ($\pm 1\arcsec$ from the image center). The 0.5 pixel sampling corresponds to 0.145\arcsec\, which is less than the MUSE pixel scale of 0.20\arcsec\ but also corresponds to 1/10 the typical KCWI FWHM which is equal to 1.2\arcsec. Across the galaxy sample, the final offsets correspond to correlation coefficients $>$ 0.9 between KCWI and MUSE.

\subsection{Matching the MUSE Flux Calibration}
\label{sec:match_muse_flux}
In order to correct for any additive and/or multiplicative offsets between the MUSE and KCWI flux calibrations, we compared the surface brightness (SB) calculated in apertures in overlapping position and wavelength. To do this we made use of the $P_S$ images, described in Section \ref{sec:registration}, and measured the surface brightness inside a number of 3\arcsec\ radius apertures located at randomly drawn positions inside the combined KCWI and MUSE coverage. The aperture size was chosen to be large enough to minimize effects arising from the different PSFs of KCWI and MUSE. We determined the best-fit line to the measured $SB_{\mathrm{KCWI}}$ vs. $SB_{\mathrm{MUSE}}$ relationship, where the slope, $m$, and $y$-intercept, $b$, reflect the multiplicative and additive offset between the KCWI and MUSE flux calibration. We applied the correction by multiplying the KCWI datacubes by $m$ and adding $b$ to the full spectrum in each pixel. The average multiplicative and additive offsets were $m=1.03\pm0.02$ and $b=-7.6\pm1.7 \times 10^{-20}\ \mathrm{erg}\ \mathrm{s}^{-1}\ \mathrm{cm}^{-2}\ \text{\AA}^{-1}$. This $\sim 3\%$ deviation from a 1--1 slope and low level of additive offset show that the calibrations were already in good agreement. 

\subsection{PSF of Individual KCWI Fields}
\label{sec:psf+per_pointing}
Because the MUSE mosaics have been homogenized to a uniform Gaussian PSF, we have an image with a known (parameterized) PSF. This is advantageous, as we can directly measure the KCWI PSF per pointing using cross-convolution, following the steps outlined in \cite{EmsellemMuse}. We briefly summarize the procedure here. 1) We produced $P_S$ images of both the MUSE mosaic and the individual KCWI pointings. 2) We re-projected the MUSE cutout onto the KCWI 0.29\arcsec $\times$ 0.29\arcsec\ pixel grid. 3) We convolved the KCWI pointing with a 2D Gaussian Kernel with PSF equal to the reference MUSE PSF. 4) In an iterative manner, the reference MUSE image is then convolved with a 2D Gaussian Kernel, PSF$_k$, where the PSF$_k$ represents the KCWI pointing PSF which is a free-parameter. We then varied the FWHM of this kernel until we maximized the correlation between the KCWI pointing and the reference MUSE image. For the set of images observed for each galaxy, we present the average and standard deviation of the measured PSFs in Table \ref{tab:sample_properties}. The average PSF across the galaxy sample is $1.4'' \pm 0.2''$ which is consistent with the PSFs measured using the Standard Star observations presented in Appendix \ref{appn:observations}. We chose not to homogenize the PSF of the KCWI data. To do so, would mean convolving the KCWI data to the largest observed PSF. In turn, this would increase the mismatch in resolution between KCWI and MUSE as well as lead to larger \HII\ region boundaries. The larger regions have higher susceptibility of contamination from the diffuse ionized gas.

\subsection{Image Co-Addition}
\begin{figure*}
    \centering
    \includegraphics[scale=0.65]{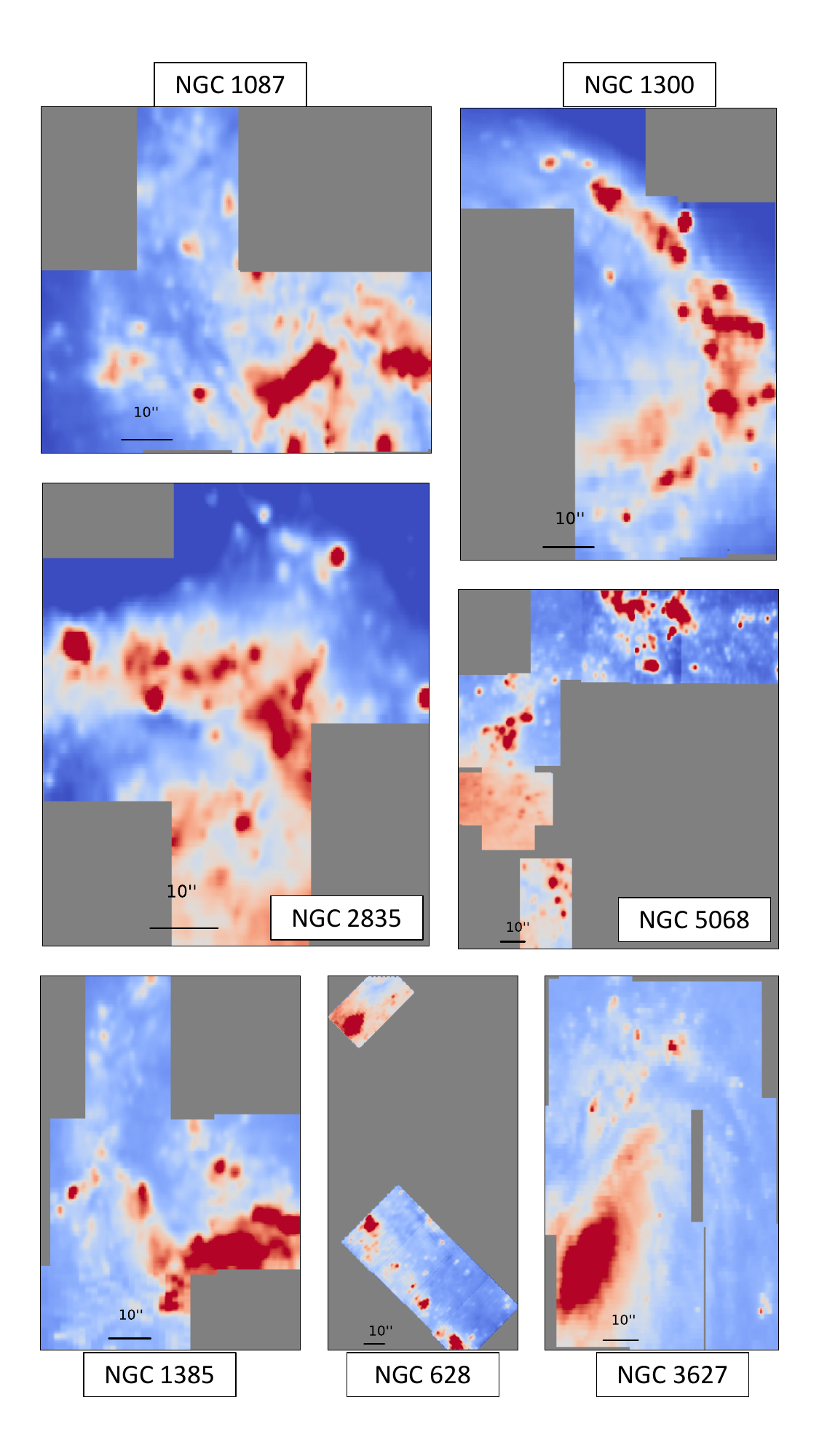}
    \caption{Pseudo $g$-band images of the KCWI mosaics. A 10\arcsec\ scale bar is shown in the bottom of each image. Images are oriented to have North pointing upwards.}
    \label{fig:gband_sample}
\end{figure*}

After the KCWI datacubes had been aligned and flux calibrated to match the MUSE mosaics, the KCWI datacubes were co-added to create KCWI galaxy mosaics. The image co-addition was performed with \texttt{Montage}. The \texttt{mAddCube} call to \texttt{Montage} initiates the co-addition. The co-addition is performed by stacking the aligned images, according to the an output WCS determined by \texttt{Montage}, and then taking the average value between any overlapping pixels. Pseudo $g$-band images for the final mosaics of each galaxy are shown in Figure \ref{fig:gband_sample}.

\label{sec:data_quality_assesment}
\subsection{Absolute Calibration of KCWI Compared to MUSE}
\begin{figure*}[!h]
\centering
\includegraphics[scale=1.1]{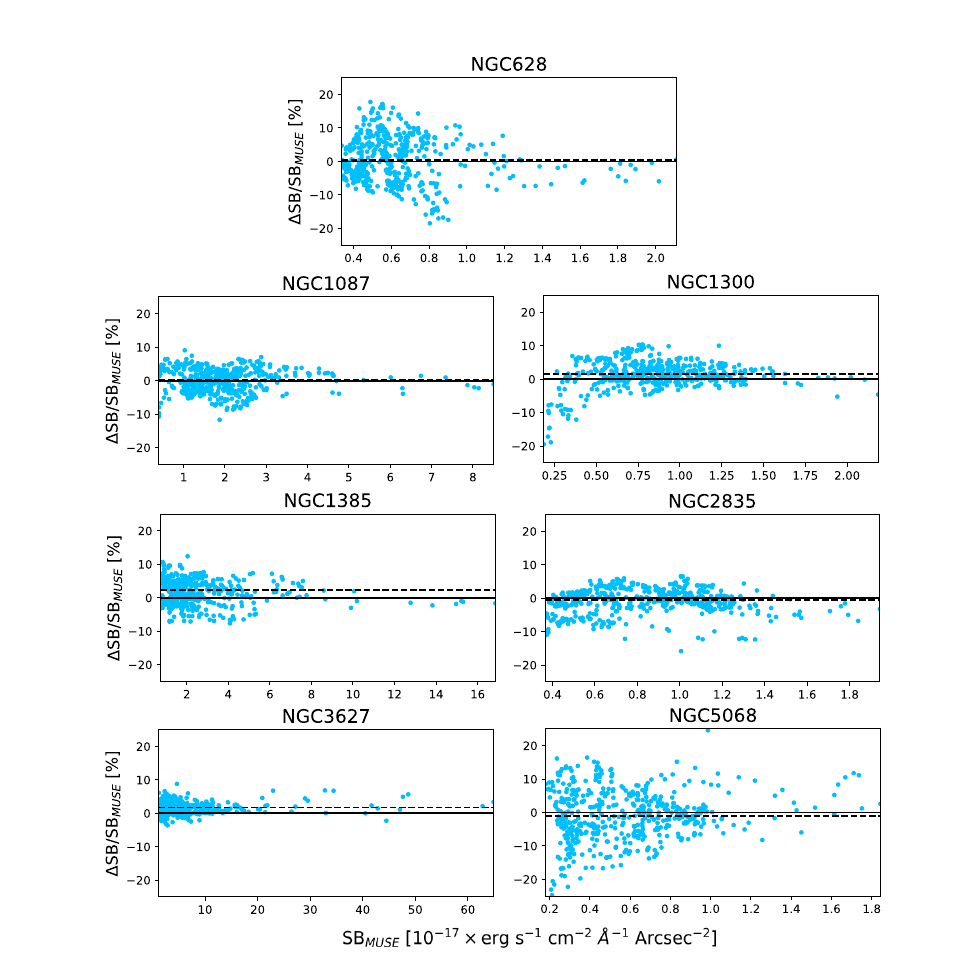}
\caption{Comparison of $P_S$ image surface brightness measured in $r=3$\arcsec\ apertures from synthetic KCWI and MUSE $P_S$ mosaics of the 7 galaxies. In each panel we show the fractional SB differences between KCWI and MUSE versus the SB of MUSE. The median fractional offset (\textit{black-dashed}) is shown relative to the zero line (\textit{black-solid}). Across all galaxies, the median offset is $\sim -0.1\%$.}
\label{fig:muse_comparisons}
\end{figure*}

We assessed the absolute calibration of the co-added KCWI mosaics by comparing the SB's between the KCWI and MUSE mosaics. Comparisons of the MUSE mosaics with SDSS imaging in the $r$-band described in \cite{EmsellemMuse} showed that the MUSE absolute flux calibration is consistent with SDSS calibration within the instrument uncertainty of $0.06$ mag \citep{SDSS_mag_unc}. We calculated the SB inside $r=3\arcsec$ apertures, randomly placed in the KCWI coverage. The SB offsets between the KCWI and MUSE mosaics are shown in Figure \ref{fig:muse_comparisons}. The resulting offsets, summarized in Table \ref{tab:sdss_kcwi}, reveal acceptable agreement between the MUSE and KCWI absolute calibration. The average percent SB offset with respect to the SDSS imaging, $\Delta \mathrm{SB}/\mathrm{SB}_{\mathrm{SDSS}}$, is between $-1.1$\% and $0.7$\% with a median value across all galaxies of $-0.1$\% $\pm$ 4\%.

\begin{figure*}[!h]
\centering
\includegraphics[scale=0.9]{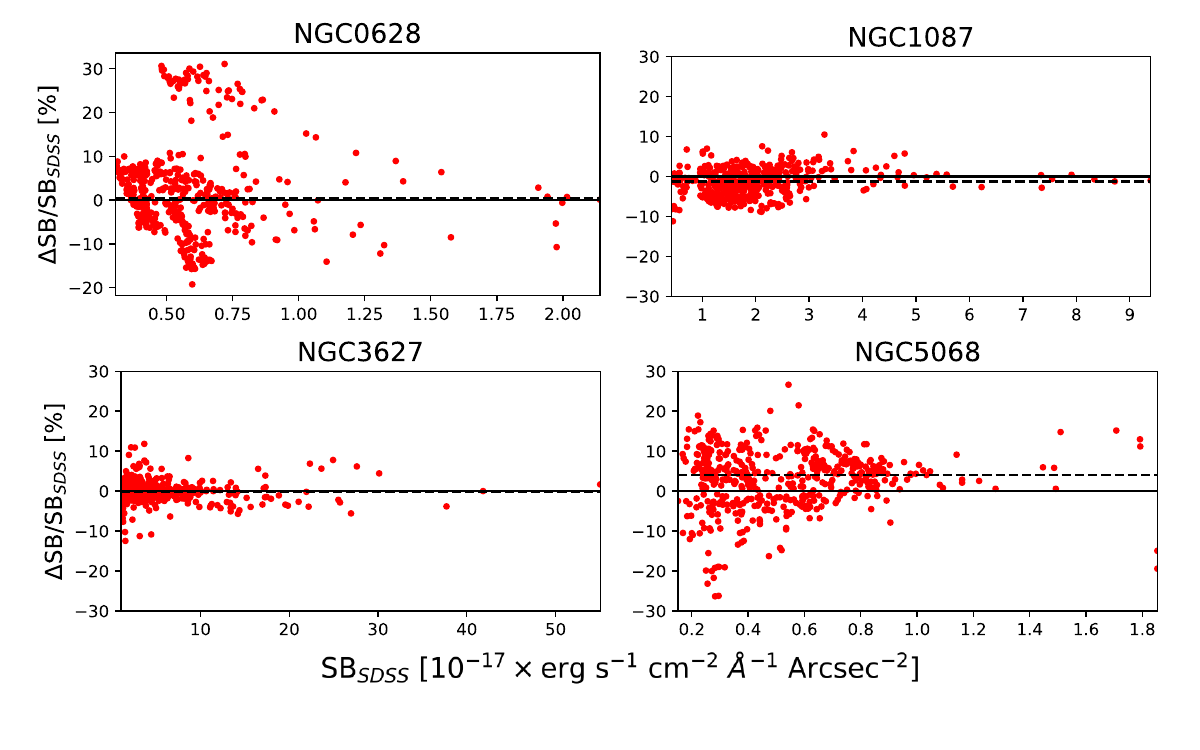}
\caption{Comparison of $g$-band surface brightness measured in $r=3$\arcsec\ apertures from synthetic KCWI $g$-band mosaics and SDSS imaging for the 4 KCWI galaxies with available SDSS data. The surface brightness offset between KCWI and SDSS is shown versus the SDSS $g$-band surface brightness. The median offset (\textit{black-dashed}) is shown relative to the zero line (\textit{solid-black}). Across the sample, the offset between the KCWI and SDSS surface brightness is $\sim 1$\%.}
\label{fig:sdss_comparisons}
\end{figure*}

\subsection{Absolute Calibration of KCWI Compared to SDSS}
We have shown agreement between KCWI and MUSE, but this comparison is only an assessment of the flux calibration in the overlapping wavelength range of KCWI and MUSE. To assess the absolute flux calibration across a wider wavelength range, we compared synthetic \textit{g}-band images of the KCWI mosaics to Sloan Digital Sky Survey \citep[SDSS,][]{SDSS2003AJ....126.2081A} images of the same galaxies. Only four galaxies with KCWI mosaics have  SDSS imaging: NGC\, 628, NGC\,1087, NGC\,3627 and NGC\,5068. We constructed synthetic \textit{g}-band images of these galaxies by convolving the spectrum in each pixel, $F_{\lambda}(x,y)$, with the $g$-band transmission curve, $T_{g}(\lambda)$, according to the following equation:
\begin{equation}
    F_{g}(x,y)=\frac{\int F_{\lambda}(x,y)T_{g}(\lambda)d\lambda}{\int T_{g}(\lambda)d\lambda}.
\end{equation}
The SDSS imaging is presented in units of \textit{nanomaggies} or $f_{\nu} =$ 3.631$\times 10^{-6}$ Jy. To compare with the KCWI data, with native units of flux density, $f_{\lambda}$, we converted to flux density with the following expression \citep{2005PASP..117..421T},
\begin{equation}
f_{\lambda}=\frac{c}{\lambda_{p}^2} f_{\nu},  
\end{equation}
where $\lambda_{p}$ is the pivot wavelength of the band-pass. The pivot wavelength for the \textit{g}-band filter is $\lambda_{p}=4702$ \AA.
The surface brightness for both KCWI and SDSS are measured inside $3\arcsec$ radius apertures. The mean and 1$\sigma$ scatter of the measured surface brightness are shown in Figure \ref{fig:sdss_comparisons}. We found overall agreement of the absolute calibration between KCWI and SDSS for the four galaxies. Summarized in Table \ref{tab:sdss_kcwi}, the median offset across the sample galaxies, $\Delta \mathrm{SB}/\mathrm{SB}_{\mathrm{SDSS}}$ range between $-$1.0\% and 3.0\%  and exhibits scatter between 3\% and 10\%. 
The source of the largest scatter is from the northern
most observation of NGC 628. In spite of the quality of this KCWI field, we find that this field contains a single HII region and, as shown in Figure \ref{fig:hii_masks_0628} of Appendix \ref{appn:comparison_of_muse_and_kcwi_hii_regions}, contains detectable auroral and nebular emission from ions only within the MUSE spectrum. Because of this, including this KCWI field will have no negative impact on the overall analysis. The median SB offset across all galaxies is $1 \pm 7$\% which suggests that the KCWI calibration is in good agreement with SDSS.  

\begin{deluxetable}{cDDDD}[h!]
\tablecaption{Flux Calibration Comparisons between KCWI, SDSS and MUSE.}
\tablehead{\colhead{Name} & \multicolumn2c{$\mu(\Delta$SB)\tablenotemark{a}} & \multicolumn2c{$\sigma(\Delta$SB)\tablenotemark{b}} & \multicolumn2c{$\mu(\Delta$SB$_g$)\tablenotemark{c}} & \multicolumn2c{$\sigma(\Delta$SB$_g$)\tablenotemark{d}} \\
 \colhead{} & \multicolumn2c{[\%]} & \multicolumn2c{[\%]}& \multicolumn2c{[\%]}& \multicolumn2c{[\%]}}
\decimals
\startdata
NGC\,628 & 0.3 & 6.0 &  0.3 & 10.0~\tablenotemark{f} \\
NGC\,1087 & 0.5  & 3.6 & -1.0 & 3.4 \\
NGC\,1300 & 1.5  & 4.0 &  -    & -    \\
NGC\,1385 & -0.6 & 4.3 &  -    & -   \\
NGC\,2835 & -0.4 & 3.6 &  -    &  -  \\
NGC\,3627 & 1.7 & 1.6 &  -0.05 & 2.6 \\
NGC\,5068 & 0.3  & 6.7 &  2.9 & 8.4 
\enddata
\tablenotetext{a}{The median fractional surface brightness offset between KCWI and MUSE in \%.}
\tablenotetext{b}{The standard deviation of the fractional surface brightness offset between KCWI and MUSE in \%.}
\tablenotetext{c}{The median $g$-band surface brightness offset between KCWI and SDSS.}
\tablenotetext{d}{The standard deviation of $g$-band surface brightness offset between KCWI and SDSS.}
\tablenotetext{f}{Removing the problematic field, discussed in Section 3.7, reduces this standard deviation to 7\% which is comparable to the other galaxies.}
\label{tab:sdss_kcwi}
\end{deluxetable}

\section{H~II Region Catalog}
\label{sec:hii_cat_intro}
In order to assess the emission line properties of \HII\ regions, we determine the \HII\ region location and boundaries using H$\beta$ maps constructed from the KCWI spectral datacubes and the image segmentation software \texttt{HIIphot} \citep{Thilker2000AJ....120.3070T}. Although \HII\ region masks have previously been constructed from the MUSE H$\alpha$ maps for these galaxies \citep{Kreckel2019ApJ...887...80K,Santoro2022A&A...658A.188S,Groves2023NebCat, Congiu:2023}, the MUSE angular resolution is higher than that of KCWI. Because of this, the \HII\ region boundaries derived from MUSE may not fully encapsulate the spatial extent of the \HII\ regions observed using KCWI. Furthermore, simply convolving or reprojecting the MUSE \HII\ masks to the KCWI resolution or grid would introduce uncertainty on the boundaries for tightly spaced \HII\ regions. We therefore perform \HII\ region identification directly on the KCWI data.

\subsection{Construction of H$\beta$ Maps}
\label{sec:hii_maps} 
H$\beta$ is the brightest H~I recombination line observed by KCWI, and maps of this emission for the galaxies will be used to define our \HII\ regions. The continuum near and underlying H$\beta$ emission must be removed in order to accurately map its emission. To remove the continuum we used \texttt{LZIFU}, an emission line fitting code designed specifically for use with IFUs \citep{Ho2016}. \texttt{LZIFU} implements and streamlines the penalized pixel-fitting software \citep[\texttt{PPXF},][]{Cappellari2004,Cappellari2017MNRAS.466..798C} for using \texttt{PPXF} on IFU emission line maps. To fit the continuum of the input spectrum \texttt{LZIFU} matches a series of input single-metallicity, $-1.31<$ [Z/H] $< 0.22$, stellar population models \citep[MILES]{LZIFUMiles} that have been redshifted, and convolved to match the input spectrum PSF. \texttt{LZIFU} fits Gaussian models at the location of emission lines. In the output H$\beta$ map, pixels with weak or no H$\beta$ can contain `NaN' values which can be problematic for \texttt{HIIphot}. To avoid these artifacts, we subtract the stellar continuum from each pixel's spectrum and construct the final H$\beta$ maps by integrating the continuum-subtracted spectra between 4856--4876 \AA. The final maps are suitable for \HII\ region identification using \texttt{HIIphot}.

\subsection{H~II Region Identification}
\label{sec:hiiphot}
\texttt{HIIphot} was designed to identify \HII\ regions and complexes (unresolved or blended \HII\ regions) while also minimizing the inclusion of surrounding diffuse ionized gas (or DIG). \texttt{HIIphot} works by first defining ``seeds'' at the location of peak emission in H$\beta$ (or H$\alpha$), then iteratively grows each ``seed'' and terminates only when the gradient of the H$\beta$ (or H$\alpha$) surface brightness distribution matches a termination value, in mandatory units of emission measure (EM), set by the user. The gradient of the surface brightness distribution is a more robust method of stopping uncontrolled growth at lower S/N compared to using only the average local background level. For each galaxy we apply the the same termination gradient, $\Delta=$ 5 EM pc$^{-1}$ or 2.43 $\times\ \rm{erg}\ \rm{s}^{-1}\ \rm{arcsec}^{-2}\ \rm{pc}^{-1}$, as the recent PHANGS-MUSE work by \cite{Santoro2022A&A...658A.188S}. 
 
\texttt{HIIphot} uses the PSF to convolve the input \Hb\ map to different spatial scales to identify seeds. Using a constant PSF for all galaxies can potentially miss valid regions or generate non-physical regions. The PSF of the input H$\beta$ emission map is required by \texttt{HIIphot}. For each galaxy mosaic, we used the average PSF from the its KCWI pointings (see Table \ref{tab:obs} in Appendix \ref{appn:observations}) as the input for \texttt{HIIphot}. The resulting 2D mask returned by \texttt{HIIphot} contains \HII\ regions with smooth and reasonable boundaries, judged by the distinction between clearly separated HII regions, the minimization of spurious small and pixelated regions or runaway growth. In total, \texttt{HIIphot} identifies $\sim$ 688 \HII\ regions or complexes across all of the KCWI mosaics. This number is smaller than the 2169 potential \HII\ regions identified for the Nebular catalog \citep{Kreckel2019ApJ...887...80K,Santoro2022A&A...658A.188S, Groves2023NebCat}, and 2124 potential \HII\ regions from \cite{Congiu:2023}, inside the same KCWI footprints. This is largely due to the differences in angular resolution between KCWI and MUSE, and the three-fold decrease in strength of H$\beta$ emission relative to H$\alpha$. In Figure \ref{fig:lum} of Appendix \ref{appn:comparison_of_muse_and_kcwi_hii_regions} we show histograms of the H$\beta$ luminosity and radii for KCWI and Nebular catalog regions as well a comparisons of the spatial masks in Figures \ref{fig:hii_masks_1087}--\ref{fig:hii_masks_0628} of Appendix \ref{appn:comparison_of_muse_and_kcwi_hii_regions}. Additionally, we also present in Table \ref{tab:total_regions} of Appendix \ref{appn:comparison_of_muse_and_kcwi_hii_regions} the number of regions detected per galaxy.

\begin{figure*}
    \centering
    \includegraphics[scale=0.7]{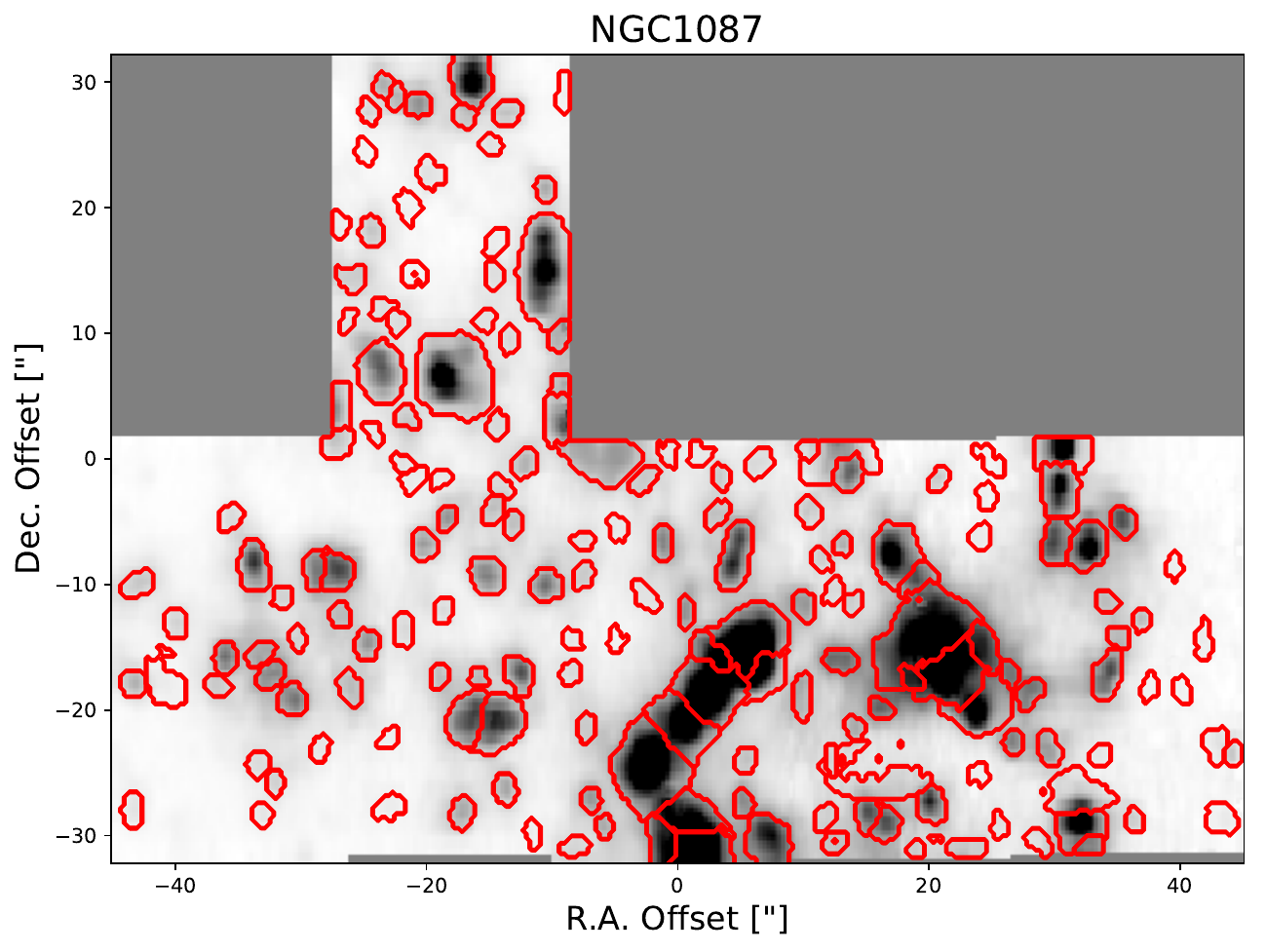}
    \caption{Region boundaries returned by \texttt{HIIPhot}. Pixels within the (\textit{red}) boundaries are identified as corresponding to a potential \HII\ region. The boundaries for the remaining sample galaxies are shown in Appendix \ref{appn:comparison_of_muse_and_kcwi_hii_regions}. The R.A. and Dec. offset are centered on the R.A. and Dec. coordinates 02hr~46.0m~25.53s, -00$^{\circ}$~29\arcmin ~38.8\arcsec.}
    \label{fig:hii_phot_output}
\end{figure*}

\subsection{Generation of Integrated H~II Region Spectra}
\begin{figure*}[!ht]
    \centering
    \includegraphics[width=0.8\textwidth]{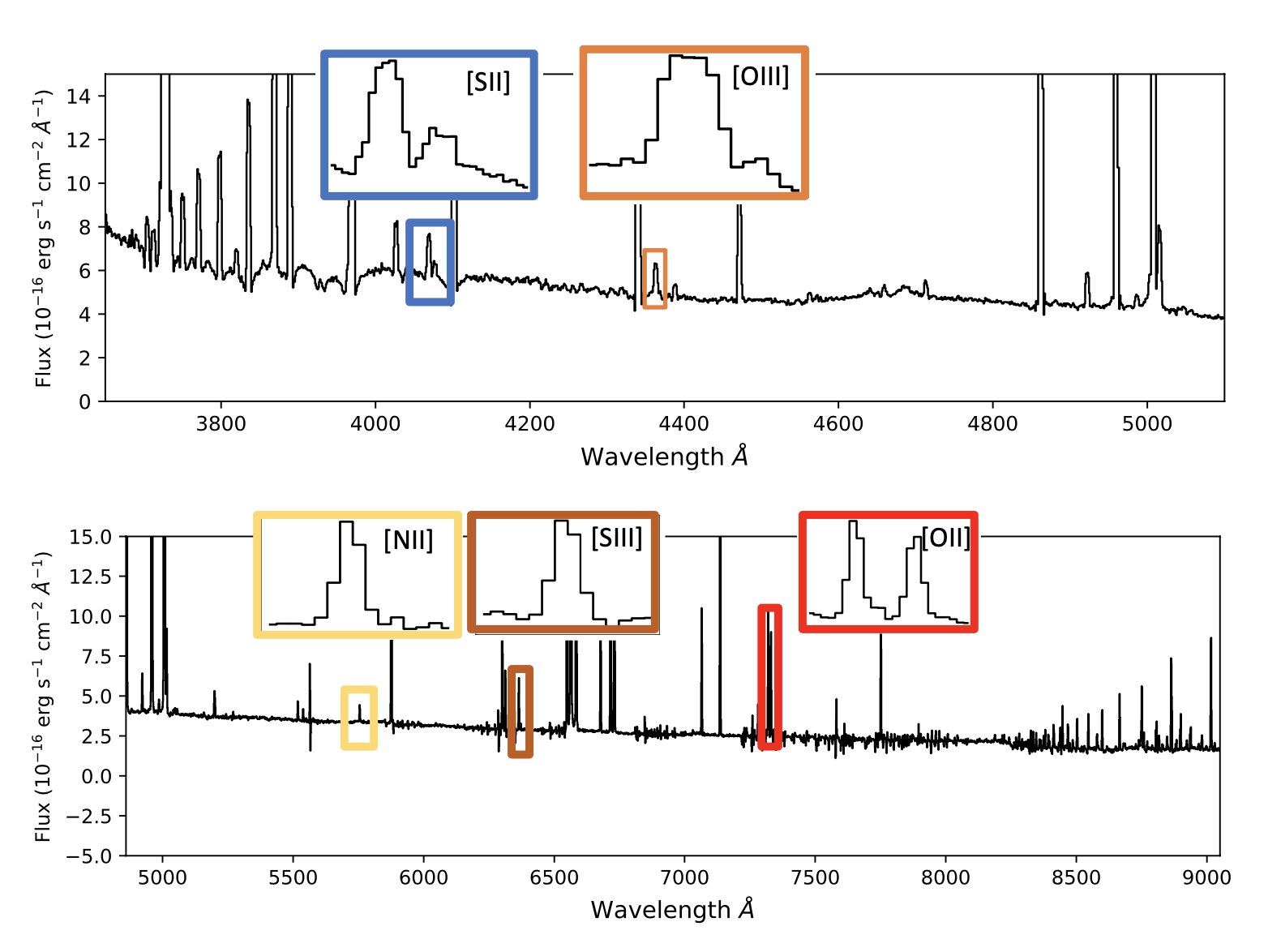}
    \caption{A KCWI, shown in the top panel, and MUSE, shown in the bottom panel, \HII\ region spectrum. Example spectrum for an \HII\ region in NGC\,5068. The full wavelength range afforded by combining both KCWI and MUSE capture the full set of optical auroral lines: [\ion{S}{2}]$\lambda\lambda 4068,4078$, [\ion{O}{3}]$\lambda4363$, [\ion{N}{2}]$\lambda5756$, [\ion{S}{3}]$\lambda6312$ and [\ion{O}{2}]$\lambda\lambda 7320,7330$, which are identified with zoomed insets.}
    \label{fig:spectrum}
\end{figure*}

The KCWI \HII\ region masks, produced by \texttt{HIIphot}, are used to isolate and sum the spectra in pixels belonging to each \HII\ region, resulting in an integrated \HII\ region spectrum. To produce a matching MUSE \HII\ region mask we transformed the \HII\ regions coordinates/boundaries from the KCWI pixel grid onto the MUSE pixel grid. These are then used to construct MUSE integrated spectra for each \HII\ region. The KCWI and MUSE \HII\ region spectra for a single \HII\ region, with the full set of auroral lines highlighted, is shown in Figure \ref{fig:spectrum}.

We also produce integrated variance spectra for each  \HII\ region. The variance spectra for both KCWI and MUSE  \HII\ regions are constructed by propagating the pipeline-produced variance datacubes for pixels contained within each \HII\ region boundary. In the case of MUSE, the datacubes have undergone an additional convolution process in order to generate mosaics with uniform PSF which introduces a correlation between neighboring pixels \citep{EmsellemMuse}. Because of the additional convolution, we generated the MUSE variance spectra, assuming fully correlated conditions, by adding the pixel variance spectra linearly \citep{Taylor1997ieas.book.....T}. 

We verify this choice by comparing the median standard deviation of the propagated MUSE variance spectra, $\sigma_{\mathrm{propagated}}$, to the median standard deviation of the \HII\ region spectrum, $\sigma_{\mathrm{measured}}$, in the emission line free wavelength range $5400-5450$~\AA. We measured an average ratio between the propagated and measured error of $\sigma_{\mathrm{measured}}/\sigma_{\mathrm{propagated}}=$ $1.1 \pm 0.2$, implying that we are appropriately propagating the error. We also perform a similar comparison for KCWI, and find agreement, $\sigma_{\mathrm{measured}}/\sigma_{\mathrm{propagated}}=1.9\pm0.5$, between the measured and propagated error using uncorrelated pixel error propagation. To generate the appropriate variance we propagated the error using uncorrelated, for KCWI, and fully correlated error, for MUSE, propagation methods.
 
 \subsection{H~II Region Stellar and Emission Line Fitting}
\label{sec:fitting}
We modeled the stellar continuum and emission lines of the integrated \HII\ region spectra using the general \texttt{PPXF} toolkit. Although the \texttt{LZIFU} implementation of \texttt{PPXF} allowed for the streamlined, full datacube, fitting of the KCWI H$\beta$ emission line map, the general \texttt{PPXF} toolkit offers more flexibility in the input fitting parameters. For example, we can input a wavelength dependent LSF function as well as fix the kinematics between emission lines of doublets and lines with similar levels of ionization. 
We followed \cite{EmsellemMuse} and fit the emission lines simultaneously with the stellar continuum. This particular fitting recipe was chosen to mirror the philosophy of the Mapping Nearby Galaxies at APO Data Analysis Pipeline \citep[MaNGA DAP,][]{MangaDAP2016AJ....152...83L,EmsellemMuse} and is suggested to mitigate the biases on emission line fluxes introduced from the masking of stellar absorption features around affected lines \citep{Sarzi2006_efit,Oh2011_efit,Belfiore2019_efit}.
\texttt{PPXF} robustly fits the stellar continuum by matching a set of templates to the observed \HII\ region spectrum. The templates originate from the E-MILES library of simple stellar population (SSP) models \citep{Vazdekis2016MNRAS.463.3409V}. The SSP ages were between $0.15-13.5$ Gyr. Each age bin contained SSPs with the following metallicities, $\mathrm{[Z/H]}=[-1.49, -0.96, -0.35, 0.06, 0.26, 0.4]$. Typically, \texttt{PPXF} convolves the SSP templates with a Gaussian model accounting for the spectral resolution of the input spectrum and the stellar velocity dispersion. However, because the KCWI line profile deviates significantly from a Gaussian (see Appendix \ref{sec:lsf}) we convolved the \texttt{PPXF} templates with a 4-moment Gauss-Hermite function while fitting KCWI emission lines. We constrained the fits of $h_4$ and $h_3$ in the Gauss-Hermite functions to values listed in Appendix \ref{sec:lsf}. We performed the \texttt{PPXF} fitting of the KCWI \HII\ region spectra independently from the MUSE \HII\ region spectra. The fractional difference between the \HII\ region H$\beta$ flux for KCWI and MUSE is $-2.3\pm7.5\%$.  

We obtained errors on the emission line fluxes from the output of \texttt{PPXF}. The output errors are considered reliable if the \texttt{PPXF} derived reduced $\chi^2 \approx 1$. Together the resulting fits for both the KCWI and MUSE have an average reduced $\chi_{\mathrm{reduced}}^2 \approx 2.0$, indicative that the input variance spectra are under-estimated. We obtained a better estimate of the errors by re-scaling the returned errors, for each fit to KCWI and MUSE spectra, by a factor of $\sqrt{\chi_{\mathrm{reduced}}^2}$ \citep{Cappellari2004,Cappellari2017MNRAS.466..798C, EmsellemMuse}.

\subsection{Dust Correction}
\begin{figure}
    \centering
    \includegraphics[width=0.44\textwidth]{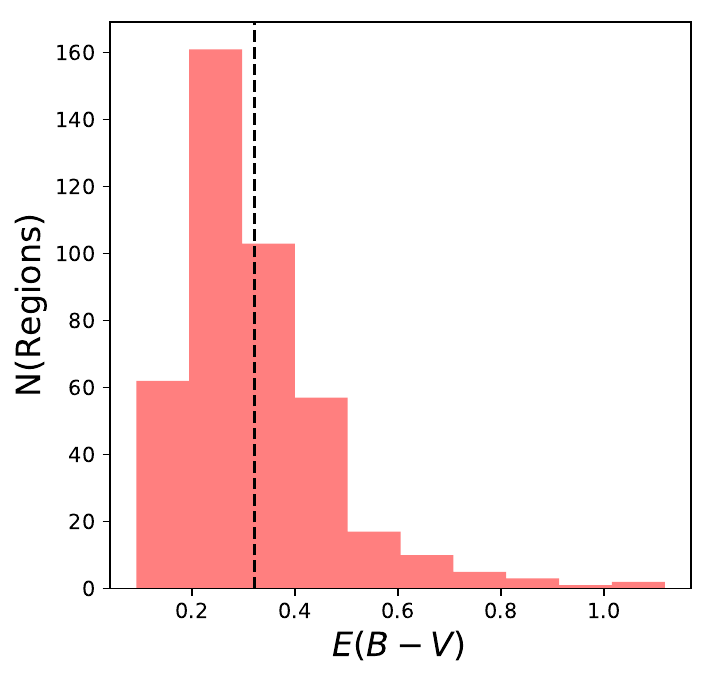}
    \caption{Histogram of derived extinction, E(B-V). We measured E($B-V$) using the \cite{ExtinctionCorr1994ApJ...422..158O} extinction law for the regions identified by \texttt{HIIPhot}. A (\textit{black dashed}) line is located at the mean E($B-V$), 0.30 $ \pm 0.14$ mag.}
    \label{fig:ebv_hist}
\end{figure}

We derive the V-band extinction ($A_V$) for each \HII\ region using the Balmer decrement. To evaluate this decrement while also taking into account the errors on the measured H$\alpha$ and H$\beta$ emission, we construct a distribution of the MUSE H$\alpha$/H$\beta$ ratios by sampling the error for each line. Next, using \texttt{PyNeb}, we calculate the extinction by comparing the average of the H$\alpha$/H$\beta$ distribution to the theoretical value assuming the Case B recombination conditions $n_e=$10$^3$ cm$^{-3}$ and $T_e=10^4$ K \citep{Storey:1995}. We explore how changing the assumed $T_e$ could affect the derived $A_V$ by sampling a range of temperatures between 5000 K and 1.5$\times10^4$ K and find that the standard deviation of $A_V$ for a fixed Balmer decrement is $\sim 0.06$ mag. We apply the wavelength dependent extinction correction assuming a \citet{ExtinctionCorr1994ApJ...422..158O} extinction curve. We present a histogram of the derived E($B$-$V$) in Figure \ref{fig:ebv_hist}. The average E($B$-$V$) for the regions is $0.30$ mag and corresponds to an $A_V\sim0.9$ mag. We find a negligible difference when using KCWI H$\beta$ in place of the MUSE H$\beta$ flux. We add that a recent investigation shows that correcting, or not correcting, the Balmer lines for DIG contamination can impact the measured E($B$-$V$) \citep{Congiu:2023}. For the range of E($B$-$V$) observed, a DIG-corrected E($B$-$V$) may return values of $A_V$~0.05--0.1 mag lower than presented here. We discuss the effects that under/over-estimated extinction may have on the measured electron temperature in Section \ref{res:low_ionization_zone_te_te}. Furthermore, given the good agreement between the integrated KCWI and MUSE \HII\ region fluxes, we replace any saturated integrated \HII\ region KCWI H$\beta$ and [\ion{O}{3}] fluxes with those measured from their integrated MUSE spectrum.

\subsection{Diffuse Ionized Gas}
\label{sec:dig}
Emission from Balmer transitions, [\ion{S}{2}]$\lambda\lambda6716,6731$, 
[\ion{N}{2}]$\lambda6584$, and other lines originating from DIG surrounding the \HII\ regions can contaminate the \HII\ region emission line fluxes of the same transitions. Measuring the DIG contribution to \HII\ region line flux is only beginning to being explored by IFU studies \citep[see][]{Belfiore2022_DIG,Congiu:2023}. In order to remove regions with a large degree of DIG contamination we calculate the contrast between \HII/DIG emission following the scheme outlined in \cite{KreckelChargeExchange}. First, we mask each 
\HII\ region in the MUSE emission line maps for: H$\alpha$, H$\beta$, [\ion{S}{2}], 
[\ion{N}{2}] and [\ion{O}{1}]. For each emission line, we place a $10\arcsec \times 10\arcsec$ aperture around 
each \HII\ region and measure the median DIG flux in pixels with $\rm{S/N} > 3$ and H$\alpha$ 
surface brightness $\rm{Log}_{10}(SB_{\rm{H}\alpha}/[\mathrm{erg}\ \mathrm{s}^{-1}\ \mathrm{kpc}^{-2}]) < 38$ \citep[see][]{Belfiore2022_DIG}. Finally, we calculate the ``integrated" DIG emission by multiplying the median DIG flux by the \HII\ region size.  For each DIG emission line, we calculate the percent contrast between the measured \HII\ region flux and DIG flux.
If the contrast between the \HII\ region flux and DIG flux of any low-ionization emission line is $< 50$\%, we exclude it from the sample.

\subsection{Quality Assessment and Classification of Regions}
\label{sec:hiiconstraints}
\begin{figure*}
    \centering
    \includegraphics[width=0.8\textwidth]{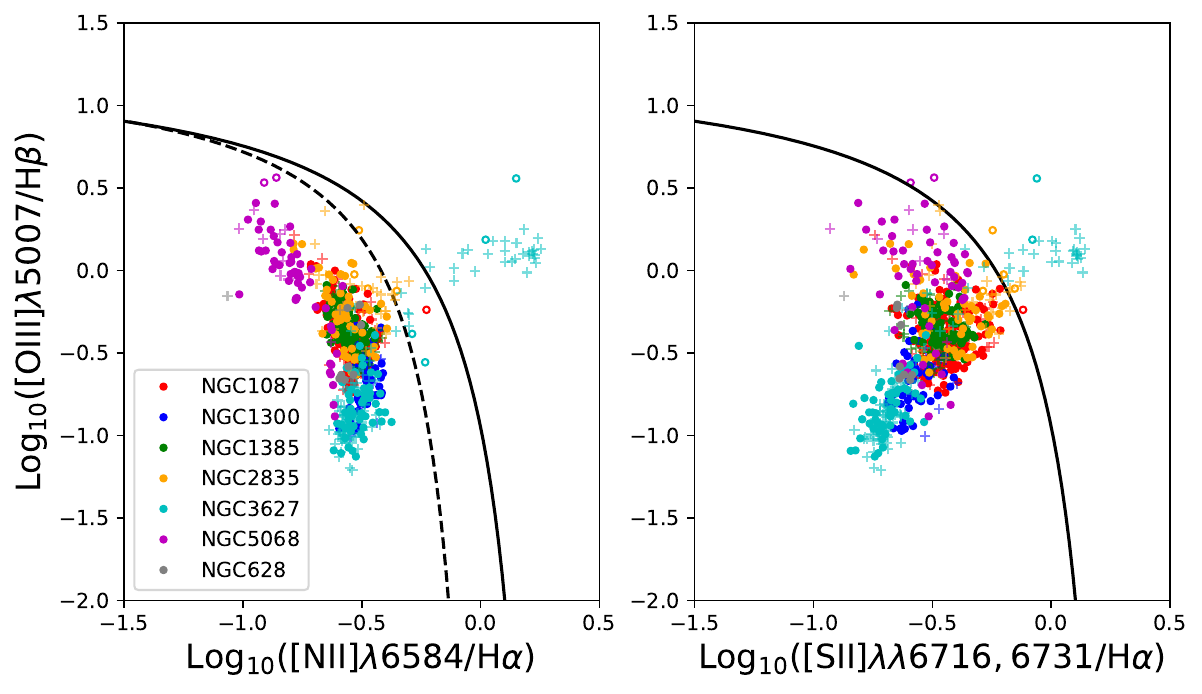}
    \caption{BPT diagrams showing [\ion{O}{3}]/H$\beta$ vs.\ [\ion{N}{2}]/H$\alpha$, in the left panel, and [\ion{O}{3}]/H$\beta$ vs.\ [\ion{S}{2}]/H$\alpha$ ratios, in the right panel, for each region identified by \texttt{HIIphot}. Regions with emission line ratios consistent with photoionization by stars are expected to populate the parameter space below the 
    theoretical \citep[\textit{solid-black,}][]{Kewley2001ApJS..132...37K} and empirical \citep[\textit{dashed-black},][]{KauffmanBPT} classification lines. The \HII\ regions are marked by filled markers. The regions above either of the classification lines are labeled with non-filled markers and are rejected from the catalog. Regions marked by `+' symbols have been rejected by our other considered constraints.}
    \label{fig:my_bpt}
\end{figure*}

The regions identified by \texttt{HIIphot} are potentially a mix of \HII\ regions, planetary nebulae, supernova remnants or low S/N in critical emission lines. We perform a set of cuts to reject non-\HII\ regions and/or low S/N spectra from the catalog of \HII\ regions. 
\begin{itemize}
    
    \item We exclude any \HII\ region whose centroid coordinates are within 2\arcsec\ from the edge of the mosaic. This step removes 86 \HII\ regions
     
    \item  We require the strong lines used for temperature determinations: \Hb, \Ha, [\ion{O}{3}]$\lambda\lambda4959,5007$ 
    [\ion{O}{2}]$\lambda3727$, [\ion{N}{2}]$\lambda\lambda6548,6584$, 
    [\ion{S}{3}]$\lambda9069$ and [\ion{S}{2}]$\lambda\lambda6716,6731$ to be detected above a threshold of S/N $>5$ . This step cuts 72 \HII\ regions from the sample.
    
    \item Using the lines of H$\beta$, H$\alpha$, [\ion{O}{3}] and [\ion{N}{2}] we construct a Baldwin-Phillips-Terlevich \citep[BPT,][]{BPT1981PASP...93....5B} diagram. We require \HII\ regions to be consistent with photoionization by massive stars.  Therefore, we require them to fall below the empirical [\ion{O}{3}]/H$\beta$ vs. [\ion{N}{2}]/H$\alpha$ \citep{KauffmanBPT} and [\ion{O}{3}]/H$\beta$ vs. [\ion{S}{2}]/H$\alpha$ \citep{Kewley2001ApJS..132...37K} lines. The BPT diagram showing the location of each \HII\ region is shown in Figure \ref{fig:my_bpt}. Out of the sample, 11 \HII\ regions are above the empirical and theoretical line cut-offs and are removed.

    \item We exclude \HII\ regions that fail our DIG contrast check in Section \ref{sec:dig}. This step removes 98 \HII\ regions.
    
\end{itemize}

The constraints together remove 267 out of the 688 detected \HII\ regions leaving 421 \HII\ regions remaining for use in future analysis. For these 421 regions, comparison of line ratios in Figure \ref{fig:my_bpt} to model classifications in Figure 3 of \cite{Congiu:2023} suggest that this sample is consistent with their \HII\ region classification. In order to compare the electron temperatures derived from the lines which are critical for $T_{\rm{e}}$--$T_{\rm{e}}$ comparisons we also exclude regions with less than 2 significant (i.e. S/N $>3$, see Section \ref{sec:auroral_fit}) detections in any auroral line. This cut excludes removes 161 regions, leaving a final sample of 260 \HII\ regions. We report in Table \ref{tab:total_regions} of Appendix \ref{appn:comparison_of_muse_and_kcwi_hii_regions} the number of regions with at least 2 auroral lines for each galaxy.

\section{Measurement of Auroral Line Emission}
\label{sec:auroral_fit}
In order to robustly measure the flux and uncertainty for the faint, temperature sensitive auroral lines from [\ion{N}{2}]$\lambda5756$, [\ion{O}{2}]$\lambda\lambda7320,7330$, [\ion{S}{2}]$\lambda\lambda4069,4076$, [\ion{O}{3}]$\lambda4363$, and [\ion{S}{3}]$\lambda6312$ we implement a specific auroral line fitting scheme in place of \texttt{PPXF}. This is necessary because any under/over subtraction of the continuum at the location of the faint auroral lines can bias the measured auroral line flux.

The framework of the auroral line fit is as follows. We first subtract the stellar continuum spectrum fitted by \texttt{PPXF}. This results in an \HII\ region spectrum that contains only emission lines and residuals from the continuum subtraction. Next, we measure the standard deviation of the residuals, $\sigma_{cont}$, in a region near the auroral line but also free of emission. We then perform a large number of fits to the auroral line with random noise added to each wavelength bin, drawn from a normal distribution with standard deviation $\sigma_{cont}$. In each trial, we fit a single Gaussian (or double depending on the auroral line) plus a linear offset to the spectrum. The linear term is needed to correct for any residual continuum present in the continuum-subtracted spectrum. After completing the $N$ trials, we calculate the average integrated line flux, $F_{avg}$, and the standard deviation of the measured fluxes, $\sigma_{avg}$. 
If S/N $>3$ we consider the auroral line detected. In Figure \ref{fig:auroral_line_fits} of Appendix \ref{appn:auroral_fits} we show the model fits and residuals for a region in NGC\,5068. 

Although this general process is performed for all the auroral lines, the auroral line from [\ion{O}{3}] is subject to additional constraints because emission from the [\ion{O}{3}]$\lambda 4363$ can be blended with [\ion{Fe}{2}]$\lambda4360$. This has been observed in both stacked galaxy and individual \HII\ region spectra
\citep{Curti2017MNRAS.465.1384C, Berg2020ApJ...893...96B, Arellano2020}. The strength of the [\ion{Fe}{2}]$\lambda4360$ has been observed to increase with the metallicity of the gas \citep{Curti2017MNRAS.465.1384C}, although continuum pumping fluorescence contributes strongly to its emissivity \citep{rodriguez1999}. [\ion{Fe}{2}]$\lambda4360$ and [\ion{Fe}{2}]$\lambda4288$ arise from the same atomic upper level and their relative intensities are independent of the physical conditions of the gas and depend solely on the atomic transition probabilities. Since $I(\lambda4360)/I(\lambda4288)=0.73$ \citep{Mendoza:2023}, if [\ion{Fe}{2}]$\lambda4360$ is detected, [\ion{Fe}{2}]$\lambda4288$ should be present too.

To remove blending of the [\ion{O}{3}]$\lambda 4363$ line by [\ion{Fe}{2}]$\lambda4360$, we use the fixed ratio, $I(\lambda4360)/I(\lambda4288)=0.73$, to estimate the degree of contamination by measuring the strength of the brighter [\ion{Fe}{2}]$\lambda4288$ emission. For each region we first fit the [\ion{Fe}{2}]$\lambda4288$ line using a single Gaussian plus a linear continuum model. The initial guesses for the line center and width are taken from the parameters of H$\gamma$ returned by the \texttt{PPXF} fits. In the case that the [\ion{Fe}{2}]$\lambda4288$ is not detected above a S/N threshold of 3, we instead use the 3$\sigma$ upper limit. We next fit [\ion{O}{3}]$\lambda 4363$ using a single Gaussian plus constant offset model. The initial guesses for the kinematics of [\ion{O}{3}]$\lambda 4363$ are taken from the \texttt{PPXF} fit of [\ion{O}{3}]$\lambda4958$. After adding random noise, we generate a model for the [\ion{Fe}{2}]$\lambda$4360 using the best fit parameters and errors derived from the fit to [\ion{Fe}{2}]$\lambda4288$, using the fixed $I(\lambda4360)/I(\lambda4288)$ ratio. After generating the [\ion{Fe}{2}]$\lambda$4360 model, we subtract it from the trial spectrum and proceed to then fit for the emission line flux originating from [\ion{O}{3}]$\lambda 4363$. We also note that by performing the above fitting scheme on each region, we may be introducing a bias in the form of a systematic reduction of the [\ion{O}{3}]$\lambda$4363 flux. However, this bias would favor systematically lower \temp{O}{3}, but, shown in Section \ref{sec:RESULTS}, we do not observe any behaviour with \temp{O}{3} that would indicate the presence of such a systematic.

We detect emission from [\ion{Fe}{2}]$\lambda4288$ in 30 \HII\ regions, two of these are in regions with measurable [\ion{O}{3}]$\lambda 4363$. 
The low number of [\ion{Fe}{2}]$\lambda4288$ detections suggest that the combination of high-metallicity, needed for the presence of iron lines, and the exponential dampening of [\ion{O}{3}]$\lambda 4363$ makes the contamination of [\ion{O}{3}]$\lambda 4363$ by [\ion{Fe}{2}]$\lambda4360$ a rare occurrence in spiral galaxies. From the non-detections, we determined that the 3$\sigma$ upper-limit on the [\ion{Fe}{2}]$\lambda4360$ flux is 3.5 $\times 10^{-17} \rm{erg}\ \rm{s}^{-1}\ \rm{cm}^{-2}\ \rm{pc}^{-1}$ which is $\sim 20\%$ of the average [\ion{O}{3}]$\lambda 4363$ flux. We show in Figure \ref{fig:auroral_line_fits_lowsn} Appendix \ref{appn:auroral_fits} an example [\ion{O}{3}]$\lambda 4363$ fit which has had significant [\ion{Fe}{2}]$\lambda4360$ contribution removed.

\section{H~II Region Nebular, Environmental, and Stellar Properties}
\label{sec:hii_region_property_derivations}
We assess the ionized gas physical conditions---$n_{\rm{e}}$, $T_e$, and ionization parameter $U$---of each \HII\ region using a subset of the dust-corrected emission line fluxes. We also measure a number of characteristics of the \HII\ region's environment and local stellar population, as described below.

\subsection{Electron Density}
\label{sec:ne_determination}
Using \texttt{PyNeb} we calculate the electron density $n_{\rm{e}}$ for each \HII\ region using the [\ion{S}{2}]$\lambda\lambda 6716, 6731$ doublet. Based on our constraints discussed in Section \ref{sec:hiiconstraints}, each \HII\ region is guaranteed to have measured emission from this doublet at S/N $>5$. The [\ion{O}{2}]$\lambda\lambda 3726,3729$ doublet is also commonly used to estimate $n_{\rm{e}}$. The atomic levels responsible for [\ion{S}{2}]$\lambda\lambda6716,6731$ and [\ion{O}{2}]$\lambda\lambda3726,3729$ both have critical densities, when collisional and radiative de-excitation are occurring at equal rates, that are of order $10^3$~cm$^{-3}$. The critical densities, as well as other references for the atomic data used, are listed in Table \ref{tab:atomic_references}. Both the [\ion{S}{2}]$\lambda\lambda6716,6731$ and [\ion{O}{2}]$\lambda\lambda3726,3729$ doublets are sensitive to densities $10^{2}\ \mathrm{cm}^{-3} < n_{\rm{e}} < 10^{3.5}\ \mathrm{cm}^{-3}$. However, in the KCWI measurements, the [\ion{O}{2}]$\lambda\lambda3726,3729$ doublet is unresolved, and therefore we do not use it to measure $n_{\rm{e}}$. With the exception of a handful of regions, the measured electron densities are in the low-density limit, $n_{\rm{e}}\ <\ 100$\ $\mathrm{cm}^{-3}$.

\begin{deluxetable*}{cccc}
\tablecaption{Transitions Probabilities, Collision Strengths, and Critical Densities for the relevant emission lines used in the PyNeb Temperature Determinations.}
\label{tab:atomic_references}
\tablehead{\colhead{Ion} & \colhead{Transition Probabilities} & \colhead{Collision Strengths} &  \colhead{$n_{\mathrm{crit,nebular}}$} \\
\colhead{} & \colhead{} & \colhead{} & \colhead{($10^3$ cm$^{-3}$)}}
\startdata
$[$\ion{O}{2}$]$ & \cite{Zeippen1982} & \cite{Kisielius2009} &  2\tablenotemark{a} \\
$[$\ion{O}{3}$]$ & \cite{Froese2004} & \cite{StoreyOIIIColl} &  691 \\
$[$\ion{N}{2}$]$ & \cite{Froese2004} & \cite{Tayal2011} &  88 \\ 
$[$\ion{S}{2}$]$ & \cite{RGJ19} & \cite{TayalSIIIColl} &  3\tablenotemark{b} \\
$[$\ion{S}{3}$]$ & \cite{FFCRecomData} & \cite{Tayal1999ApJ} &  543\tablenotemark{c}
\enddata
\tablenotetext{a}{Average critical density for [\ion{O}{2}]$\lambda\lambda3726,3729$ .}
\tablenotetext{b}{Average critical density for [\ion{S}{2}]$\lambda\lambda6716,6731$ .}
\tablenotetext{c}{Average critical density for [\ion{S}{3}]$\lambda\lambda9069,9532$ .}
\end{deluxetable*}

\subsection{H~II Region Electron Temperatures}
The auroral lines allow the determination of the electron temperatures for the O$^+$, O$^{2+}$, N$^+$, S$^+$ and S$^{2+}$ ions, or \temp{O}{2}, \temp{O}{3}, \temp{N}{2}, \temp{S}{2}, and \temp{S}{3}, respectively. The temperatures are calculated via \texttt{PyNeb} using the collision strengths and transition probabilities, listed in Table \ref{tab:atomic_references}, as well the measured upper-limits of the electron density, $n_{\rm{e}}$, to convert an auroral-to-nebular ratio to temperature. 
The critical densities of the requisite lines, see Table \ref{tab:atomic_references}, used to estimate \temp{O}{3}, \temp{N}{2}, and \temp{S}{3} are high enough such that the auroral-to-nebular lines ratios are insensitive to choice of $n_{\rm{e}} < 10^{4}$~$\mathrm{cm}^{-3}$. The density sensitivity in the lines used for \temp{O}{2} and \temp{S}{2} begins at smaller densities $n_{\rm{e}} \approx 10^{3}$~$\mathrm{cm}^{-3}$ and is further discussed in Section \ref{res:differences}. The uncertainty for each temperature measurement is the standard deviation of the distribution of temperatures constructed by Monte Carlo sampling of the error for each auroral and nebular line included in the temperature determination. We summarize the number of detections and median temperature for each ion in Table \ref{tab:temp_statistics}.

\begin{deluxetable}{ccc}[h]
\label{tab:temp_statistics}
\centering
\caption{Median temperature for each ion.} 
\tablehead{\colhead{T$_{\mathrm{e,Ion}}$} & \colhead{Median\tablenotemark{a}} & \colhead{$N_{\mathrm{regions}}$\tablenotemark{b}} \\
\colhead{} & \colhead{[10$^4$ K]} & \colhead{}}
\startdata
\temp{O}{2} & 0.95$^{+0.22}_{-0.12}$ &  156 \\
\temp{N}{2} & 0.81$^{+0.12}_{-0.09}$ &  245 \\
\temp{S}{2} & 0.94$^{+0.19}_{-0.14}$ &  305  \\
\temp{S}{3} & 0.89$^{+0.30}_{-0.10}$ &  143  \\
\temp{O}{3} & 1.20$^{+0.35}_{-0.16}$ &  26
\enddata
\tablenotetext{a}{The 50th $\pm$ the 16th -- 84th percentile of the measured temperatures.}
\tablenotetext{b}{Number of \HII\ regions with temperature measured from the particular ion.}
\end{deluxetable}

\subsection{H~II Region Ionization Parameters}
\label{sec:ip_determination}
The ionization parameter is an indicator of the strength of the ionizing radiation field. The Str\"omgren sphere descriptions of \HII\ regions define the ionization parameter as $U=Q_0/(4\pi R^2 n_\mathrm{H} c)$, where $Q_0$ is the emission rate of photons capable of ionizing hydrogen (i.e. with energy $> 13.6$ eV), $R$ is the radius of the ionized region, $n_\mathrm{H}$ and $c$ are the hydrogen density and speed of light. However, calculating the ionization parameter using this definition is difficult \citep{Kreckel2019ApJ...887...80K,KreckelChargeExchange} as resolved studies show that \HII\ regions exhibit a range of non-spherical morphologies and non-uniform densities \citep{Wood_hii_region_morph1989ApJS...69..831W}.

Instead, we trace the ionization parameter using both the \SIIISII\ and \OIIIOII\ emission line ratios. Photoionization modeling has shown that both the [\ion{S}{3}]$\lambda\lambda9069,9532$/[\ion{S}{2}]$\lambda\lambda6716,6731$ and [\ion{O}{3}]$\lambda\lambda4959,5007$/[\ion{O}{2}]$\lambda\lambda3626,3729$ correlate with the ionization parameter, $U$ \citep{Dors2011MNRAS.415.3616D}. Although it has a positive correlation with the ionization parameter, the \OIIIOII\ also has a secondary dependence on metallicity, increasing with decreasing metallicity. \SIIISII\ is not as sensitive to metallicity, making the ratio a more reliable tracer of the ionization parameter \citep{Kewley:2002}. Differences between these diagnostics are discussed further in Section \ref{sec:result_ip}. While [\ion{S}{3}]$\lambda9532$ is not observed with MUSE, we measure the [\ion{S}{3}]$\lambda9069$ and assume the fixed theoretical line ratio of [\ion{S}{3}]$\lambda9532/\lambda9069=2.5$ \citep{FFCRecomData} in all calculations. Without observations of [\ion{S}{3}]$\lambda9532$ we can not use this theoretical ratio to assess any impact of atmospheric absorption. Despite this, and discussed in Section \ref{res:te_te_intermediate}, the low scatter between \temp{N}{2} and \temp{S}{3} suggest that the decrease in the [\ion{S}{3}]$\lambda9069$ flux due to atmospheric absorption may be negligible.

\subsection{ALMA-CO: Intensity, Peak Temperature, and Velocity Dispersion}
\label{sec:mom_determination}
Using the ALMA 12m+7m+TP datacubes, we calculate moment 0, and 2 (integrated intensity and velocity dispersion) for molecular gas near each \HII\ region. Because molecular and ionized gas are not entirely co-spatial at our resolution, it is necessary to make a selection to capture gas near the \HII\ region.  One possibility is to match \HII\ regions to molecular clouds via a nearest-neighbor algorithm \citep{Grasha2019MNRAS.483.4707G,Zhang_starless_clumps2021A&A...646A..25Z,Zakardjian_GMC}.
We instead choose to integrate the ALMA spectrum contained in the \HII\ region boundaries in order to measure the properties of the molecular gas closest in projection to the the ionized gas (i.e. in front, behind, or blended due to the resolution of KCWI) and likely affected by the radiative feedback. To measure the CO spectra, we reproject the \HII\ region masks onto the grid of the ALMA datacubes and integrate the ALMA spectra for pixels located inside the footprint of each \HII\ region.

From the ALMA spectrum, we then calculate the integrated intensity, $I_{\rm CO}$, peak temperature, $T_{\rm peak}$ and the velocity dispersion, $\sigma_{v, \mathrm{CO}}$. In order to accurately measure these CO moments in the presence of noise, we construct a signal mask following the basic approach from \citet{2021Leroy_data}. To do this, we locate the velocity channel which contains the peak emission and construct an integration window around this channel by including velocity channels with signal above the $1\sigma$ noise. 

As a check on our analysis of molecular gas, we compare the calculated velocity dispersions to those from a sample of nearest-neighbor matched \HII\ regions-GMC's constructed by \cite{Zakardjian_GMC}. We find that our $\sigma_{v, {\rm CO}}$ span a similar range, up to Log$_{10}$($\sigma_{v, {\rm CO}}/$[km s$^{-1}$])=1.5, with an average and standard deviations of  Log$_{10}(\sigma_{v, {\rm CO}}/$[km s$^{-1}$])=0.88 $\pm$ 0.25; in line with the average from  \cite{Zakardjian_GMC} which suggests that this method of extracting and measuring the CO properties is reasonable.

\subsection{\HII\ Region Compact Clusters and Associations}
\label{sec:hst_matching}
In order to test for correlations between the young stellar populations that power \HII\ regions and their electron temperatures, we compile the stellar mass and age of compact clusters and multi-scale stellar associations within our KCWI \HII\ regions using results from HST observations.

We match the HST clusters to \HII\ regions, with two or more auroral lines, by simply selecting all the clusters whose on-sky coordinates fall inside any \HII\ region's spatial footprint. For the associations, we match the NUV selected, 32 pc scale, stellar associations to the individual \HII\ regions in the same manner as \cite{Scheuermann2023MNRAS.522.2369S}. The associations catalog comes with spatial masks identifying the footprint of all the detected associations. Because the association masks have a finer pixel scale than the KCWI \HII\ region masks, we reproject the KCWI mask onto the association mask pixel grid. We find for cluster matches that only 65 of 260 or (25\%) of our \HII\ regions are matched to a single cluster. 150 of 260 (or 57\%) of our \HII\ regions have zero matches. For association matches, only 85 or 260 (or 33\%) of our \HII\ regions are matched to a single association 45 of 260 (or 17\%) of our \HII\ regions have zero matches.

For the remaining regions with more than one clusters or association match, because we expect that the youngest and most massive clusters or associations contribute most to the overall ionization of the \HII\ region, we assign to each \HII\ region the age (mass) of the youngest (most massive) available cluster/association.

\section{Results}
\label{sec:RESULTS}
We present electron temperatures derived using the auroral-to-nebular line ratios from [\ion{O}{2}], [\ion{N}{2}], [\ion{S}{2}], [\ion{S}{3}], and [\ion{O}{3}]. We construct $T_{\rm{e}}$--$T_{\rm{e}}$ diagrams to compare any multi-ionization zone $T_{\rm{e}}$ relationships to recently measured and/or modeled trends. We then compare the temperatures to properties of the ionized gas such as electron density, $n_{\rm{e}}$, and ionization parameter, $U$. We also relate the temperatures to properties of the molecular gas and stellar populations. We present in Table \ref{tab:auroral_nebular_line_fluxes} a summary of the measured emission lines and derived properties for the \HII\ regions with two or more detected auroral line.

\subsection{Temperature-Temperature Relations}
We show $T_{\rm{e}}$--$T_{\rm{e}}$ relations for our sample \HII\ regions in Figures \ref{fig:temps_low_gals}, \ref{fig:temps_intermediate_gals} and \ref{fig:temps_high_gals}. The left panel displays the individual \HII\ region measurements. To assess the significance of each $T_{\rm{e}}$--$T_{\rm{e}}$ relation, we calculate its p-value. A correlation is judged to be significant if it exhibits $p$ $\lesssim 10^{-3}$. With the exception of \temp{N}{2}--\temp{O}{2} and the relations involving \temp{O}{3}, the remaining $T_{\rm{e}}$--$T_{\rm{e}}$ relations are deemed significant according to their p-value. For these significant $T_{\rm{e}}$--$T_{\rm{e}}$ relations we derive the best fitting linear relation using the Bayesian linear regression tool LINMIX \footnote{https://github.com/jmeyers314/linmix}, which itself is a Python implementation of linear mixture model algorithm, LINMIX\_ERR, constructed by \cite{Kelly:2007:linmix}. The linear regression in LINMIX includes an additional term to represent the intrinsic scatter weighting each data point. In the panels of Figure \ref{fig:temps_low_gals} and Figure \ref{fig:temps_intermediate_gals} we report $\sigma_{int}$, or the median of the Normal distributed scatter around the linear regression, and the total scatter, $\sigma_{tot}$. Furthermore, to better see $T_{\rm{e}}$--$T_{\rm{e}}$ relationships in individual galaxies we show in the right panel the $T_{\rm{e}}$--$T_{\rm{e}}$ relations for temperatures binned in steps of 2000 K in the x-axis, with minimum 2 \HII\ regions per bin, for each galaxy. Alongside the binned data we show recent $T_{\rm{e}}$--$T_{\rm{e}}$ relations from \citet[hereafter CHAOS-IV]{Berg2020ApJ...893...96B}, \citet[hereafter CHAOS-VI]{Rogers:chaos:2021}, \citet[hereafter Z21]{Zurita2021MNRAS.500.2359Z}, \citet[hereafter G92]{Garnett1992AJ}, \citet[hereafter BOND]{BOND}, 
and \citet[hereafter MD23]{Delgado_Desire2023arXiv230513136M}.

\subsubsection{Low-Ionization Zone}
\label{res:low_ionization_zone_te_te}
\begin{figure*}[!h]
\centering
\includegraphics[scale=0.6]{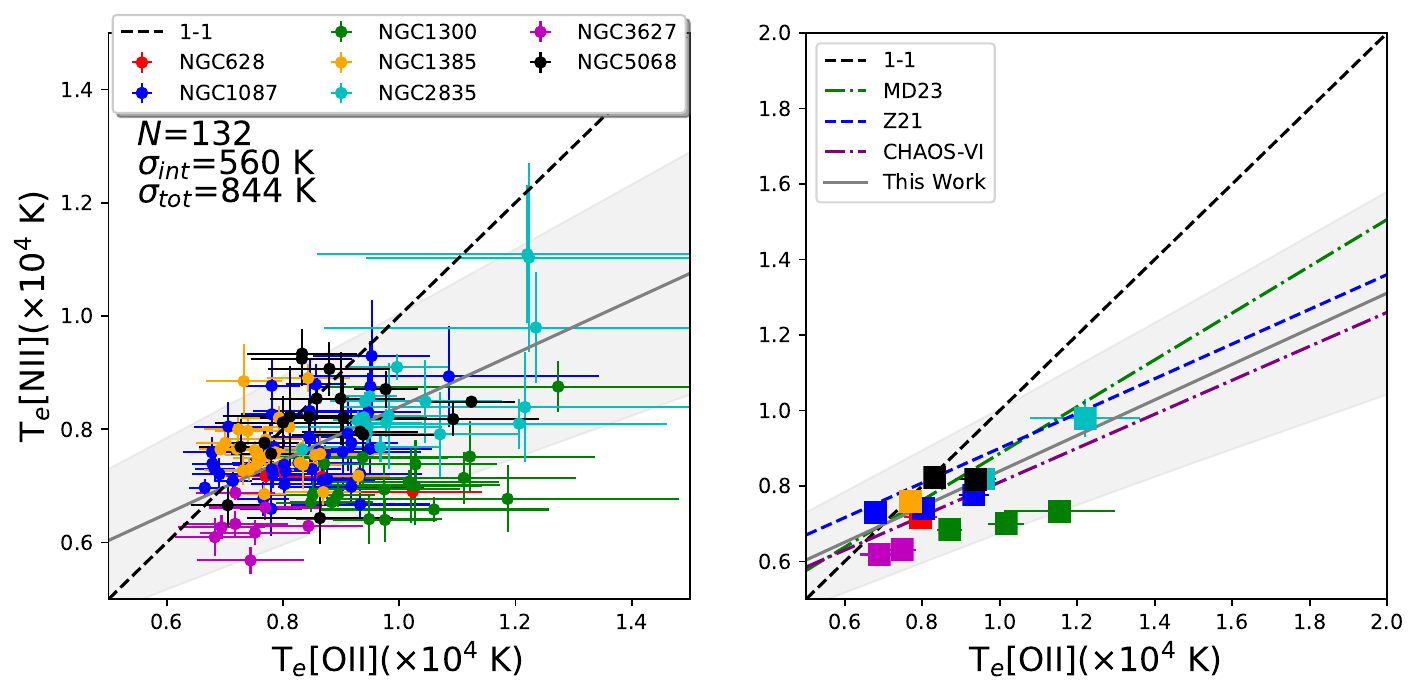}
\includegraphics[scale=0.6]{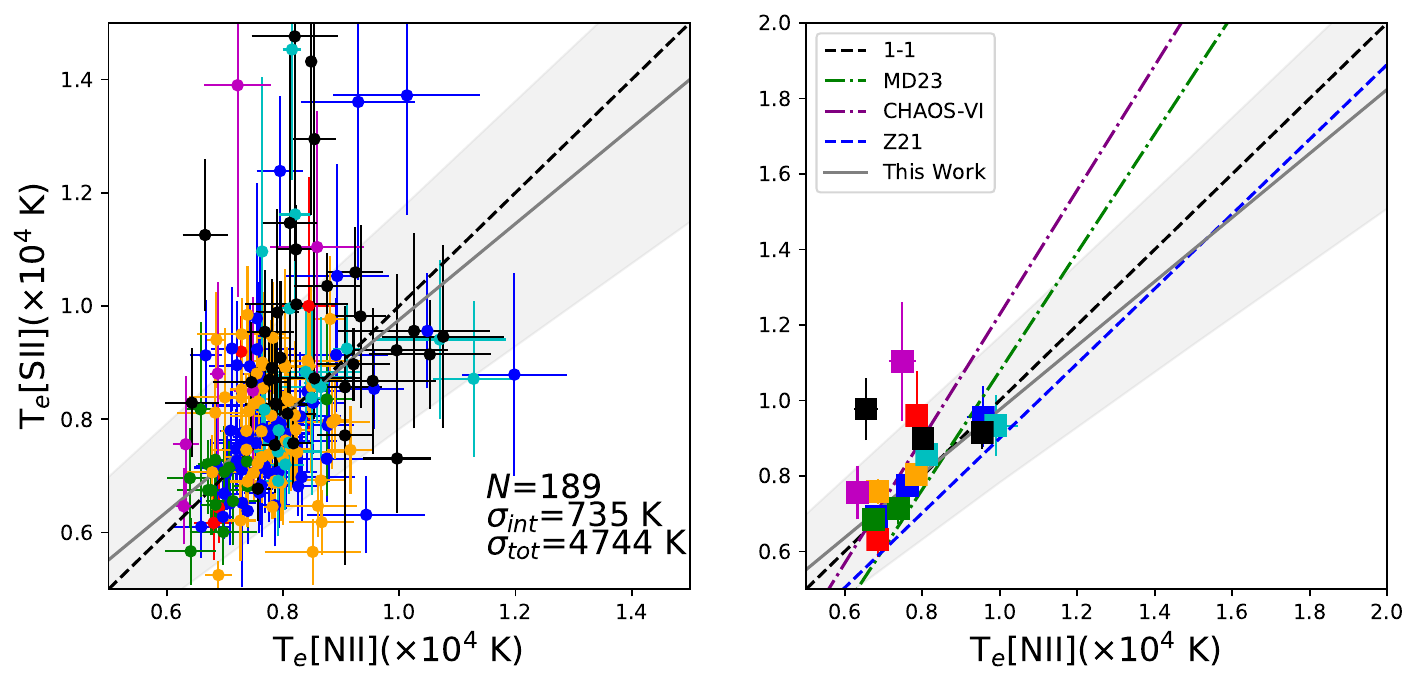}
\includegraphics[scale=0.6]{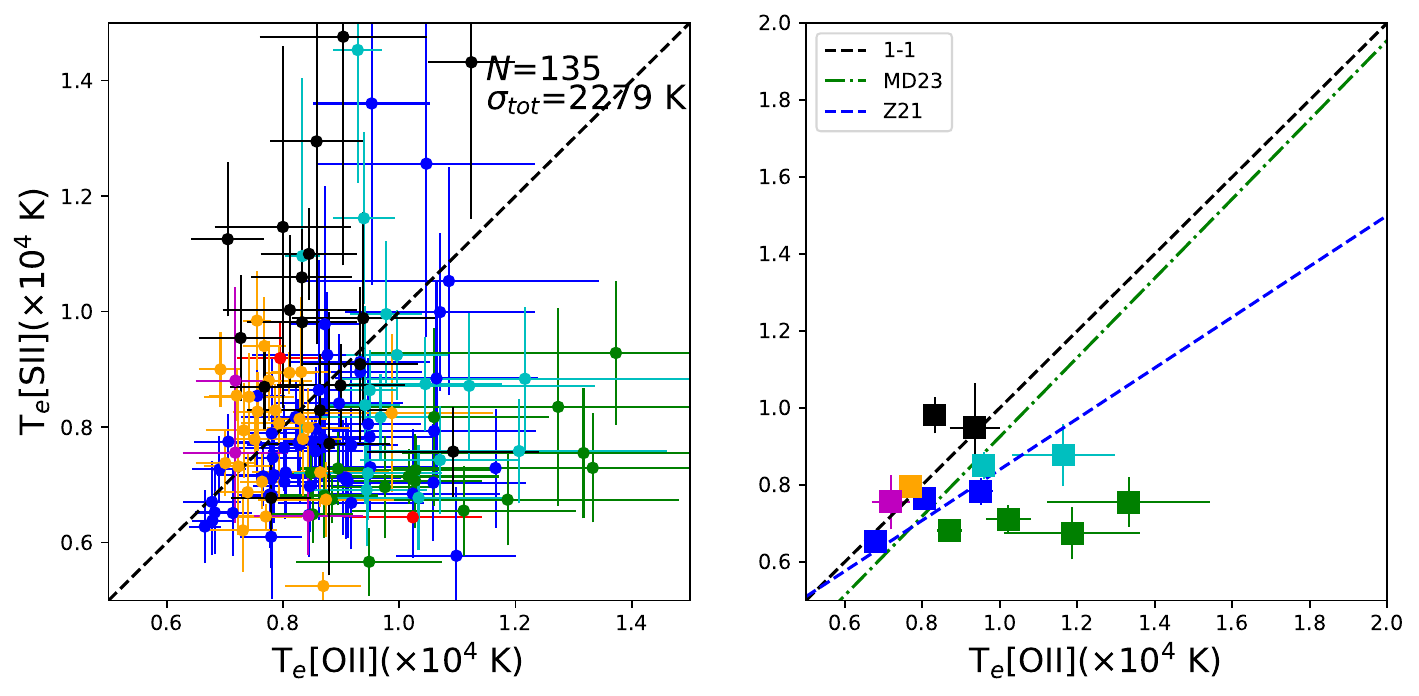}
\caption{$T_e$--$T_e$ relationships for the low-ionization zone temperatures measured from \HII\ regions in nearby galaxies. The left panel in each row displays the individual temperatures and errors with each point colored according to the host galaxy. The right panel in each row shows the $T_e$--$T_e$ relations for temperatures binned
in steps of 2000 K in the x-axis, with minimum 2 \HII\
regions per bin, for each galaxy, compared to $T_e$-$T_e$ trend lines from \citetalias{Zurita2021MNRAS.500.2359Z} (\textit{blue dash}), \citetalias{Rogers:chaos:2021} (\textit{purple dot-dash}), and \citetalias{Delgado_Desire2023arXiv230513136M} (\textit{green dot-dash}). To aid the eye, we include in (\textit{black-dash}) the 1--1 line. For comparisons with p-value $<~10^{-3}$, we include the best-fit line (\textit{grey-solid}) and 1$\sigma$ fit uncertainty (\textit{grey-shaded}).}
\label{fig:temps_low_gals}
\end{figure*}

The low-ionization zone $T_{\rm{e}}$--$T_{\rm{e}}$ relations are shown in Figure \ref{fig:temps_low_gals}. The top figure shows the \temp{N}{2}--\temp{O}{2} comparison. We observe that \temp{O}{2} gives higher values than \temp{N}{2} by 1000 K on average. The magnitude of this offset is largest when \temp{O}{2}$>1.0\times10^4$ K. 
The \temp{O}{2}$>$\temp{N}{2} inequality is also reflected in the relations from 
\citetalias{Zurita2021MNRAS.500.2359Z}, \citetalias{Rogers:chaos:2021}, and \citetalias{Delgado_Desire2023arXiv230513136M}. We also show in grey the best-fit line, and the corresponding 1$\sigma$ uncertainty, described by, \temp{N}{2}$=(0.47 \pm 0.13)\times$\temp{O}{2}$+(0.36\pm0.10)$. The \citetalias{Rogers:chaos:2021}, \citetalias{Delgado_Desire2023arXiv230513136M}, and \citetalias{Zurita2021MNRAS.500.2359Z} relations are within the uncertainties of the fit, indicating that the \temp{N}{2}--\temp{O}{2} relation is in good agreement with these studies. We also measure scatter around the trend line of $\sigma_{int}=560$~K and $\sigma_{tot}=844$~K, both of which are in good agreement with the reported values of $\sigma_{int}=588$~K and $\sigma_{tot}=810$~K by \citetalias{Rogers:chaos:2021}. 

The comparison between \temp{S}{2}--\temp{N}{2} is shown in the middle panel of Figure \ref{fig:temps_low_gals}. The best-fit line is described by \temp{S}{2}$=(0.85 \pm 0.15)\times$\temp{N}{2}$+(0.13\pm0.12)$. Only the trend from \citetalias{Zurita2021MNRAS.500.2359Z} is within the uncertainties of our fit. Both \citetalias{Rogers:chaos:2021} and \citetalias{Delgado_Desire2023arXiv230513136M} favor a steep slope (i.e. hotter \temp{S}{2}) which are outside the 1$\sigma$ bounds of our fit. We measure a low intrinsic scatter $\sigma_{int}=735$~K but a larger total scatter $\sigma_{tot}=4744$~K. For comparison, \citetalias{Rogers:chaos:2021} report $\sigma_{int}=945$~K and $\sigma_{tot}=1460$~K.

We show the temperature relation between \temp{S}{2}-\temp{O}{2} in the bottom-left panel. We observe large scatter around the line of equality, $\sigma_{tot}=2279$~K, towards hot temperatures and measure a p-value, p$=0.09$, that suggests the two temperatures are uncorrelated. We discuss a potential, physical explanation of the scatter in Section \ref{results:density_inhomogeneities}, but first we explore effects on the correlation due to low S/N detections. Low S/N detections of weak emission can have an intrinsic bias towards higher values \citep[see,][]{Rola1994A&A_SN}, which in the case of weak detections of the auroral lines of [\ion{S}{2}] and [\ion{O}{2}], would result in high temperatures. To explore how S/N changes the observed trend, we re-calculated the p-value line using only $T_{\rm{e}}$ derived from [\ion{S}{2}] and [\ion{O}{2}] auroral lines with S/N $>$ 5. Using the higher S/N threshold, the p-value returned is now, p$=0.04$, but not significant according to our criteria. Furthermore, the total scatter,  $\sigma_{tot}=1743$~K, around the 1-to-1 line is trend is high. 

We observe in the $T_{\rm{e}}$--$T_{\rm{e}}$ relations that \temp{O}{2} and \temp{S}{2} are systematically hotter than \temp{N}{2}. Based on photoionization models \citep[e.g.][]{cambell:1986,Garnett1992AJ}, the low-ionization zone temperatures are expected to be equal. There are systematic effects that could increase \temp{O}{2} and \temp{S}{2} temperatures such that \temp{O}{2}$\sim$\temp{S}{2}$>$\temp{N}{2}. The wide wavelength range between [\ion{O}{2}]$\lambda\lambda7320,7330$ and [\ion{O}{2}]$\lambda\lambda3726,3729$ as well as [\ion{S}{2}]$\lambda\lambda4068,4076$ and [\ion{S}{2}]$\lambda\lambda6716,6731$, make these ratios sensitive to the applied reddening correction. An overestimate in the extinction would lead to an underestimate of \temp{O}{2} and underestimate of \temp{S}{2}. At the same time this would leave \temp{N}{2} relatively unchanged due to the proximity in wavelength of the requisite lines. We see that both \temp{O}{2} and \temp{S}{2} are greater than \temp{N}{2} which is not compatible with the effects of overestimated extinction correction. Despite the above, our temperatures hierarchy could potentially be produced by under-estimated extinction as this would lead to over-estimates of \temp{O}{2} and \temp{S}{2} while again leaving \temp{N}{2} unchanged. We tested two different extinction prescriptions \citep{ExtinctionCorr1994ApJ...422..158O,Fitzpatrick1999PASP..111...63F} and found no change in our results. Due to the possible DIG contribution to the measured extinction, it is more likely that we could be overestimating the extinction than underestimating it \citep{Congiu:2023}.
Telluric contamination to the line emission at [\ion{O}{2}]$\lambda\lambda7320,7330$ could be an additional systematic error, but is unlikely to be significant in our data because the [\ion{O}{2}]$\lambda7320$/$\lambda7330$ ratio measured for our sample is $1.27 \pm 0.3$, in agreement with values predicted by the transition probabilities and the collisional strengths \citep{Zeippen1982,Kisielius2009} and observed in nearby \HII\ regions \citep{1957ApJ...125...66S, 1976ApJS...31..163K,Yates2020A&A...634A.107Y, mendezdelgadoHH514}. 
The [\ion{S}{2}]$\lambda\lambda4068,4076$ doublet may contain $\sim 10\%$ contamination due to \ion{O}{2} recombination emission near the location of [\ion{S}{2}]$\lambda\lambda4068,4076$ \citep{Delgado_Desire2023arXiv230513136M} that could bias \temp{S}{2} upwards by $\sim 0.2 \times 10^4$ K for \temp{S}{2} $< 2.0 \times 10^4$~K. The KCWI spectral resolution is too low to separate out any contamination in the measurements of [\ion{S}{2}]$\lambda\lambda4068,4076$. 
Lastly, \temp{O}{2} and \temp{S}{2} have a higher sensitivity to electron density inhomogeneities than \temp{N}{2} \citep{Rubin:1989,Osterbrock:2006, Delgado_Desire2023arXiv230513136M}. [\ion{O}{2}] and [\ion{S}{2}] start to be dependent on density around $n_{\rm{e}}\approx 10^3$~cm$^{-3}$ while [\ion{N}{2}] is independent of density up to $n_{\rm{e}} \approx 10^4$~cm$^{-3}$. 

Despite the all of the above factors, including S/N considerations, and given the agreement of the \temp{S}{2}--\temp{N}{2} and \temp{N}{2}--\temp{O}{2} trends with those from \citetalias{Zurita2021MNRAS.500.2359Z} using the S/N $> 3$ [\ion{O}{2}] and [\ion{S}{2}] auroral lines, we are motivated to explore potential biases due to uncertainties in the measured electron density, see Section \ref{results:density_inhomogeneities}.

\subsubsection{Intermediate Ionization Zone}
\label{res:te_te_intermediate}
\begin{figure*}[!h]
\centering
\includegraphics[scale=0.6]{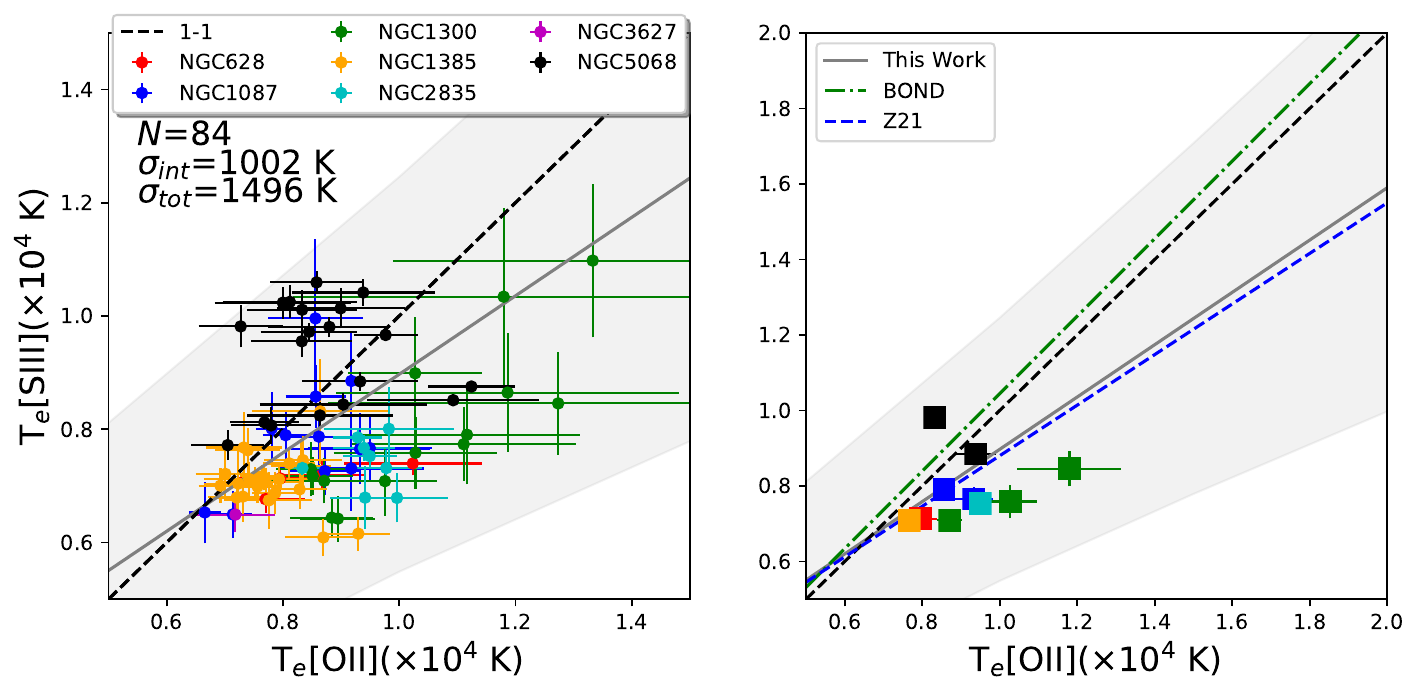}
\includegraphics[scale=0.6]{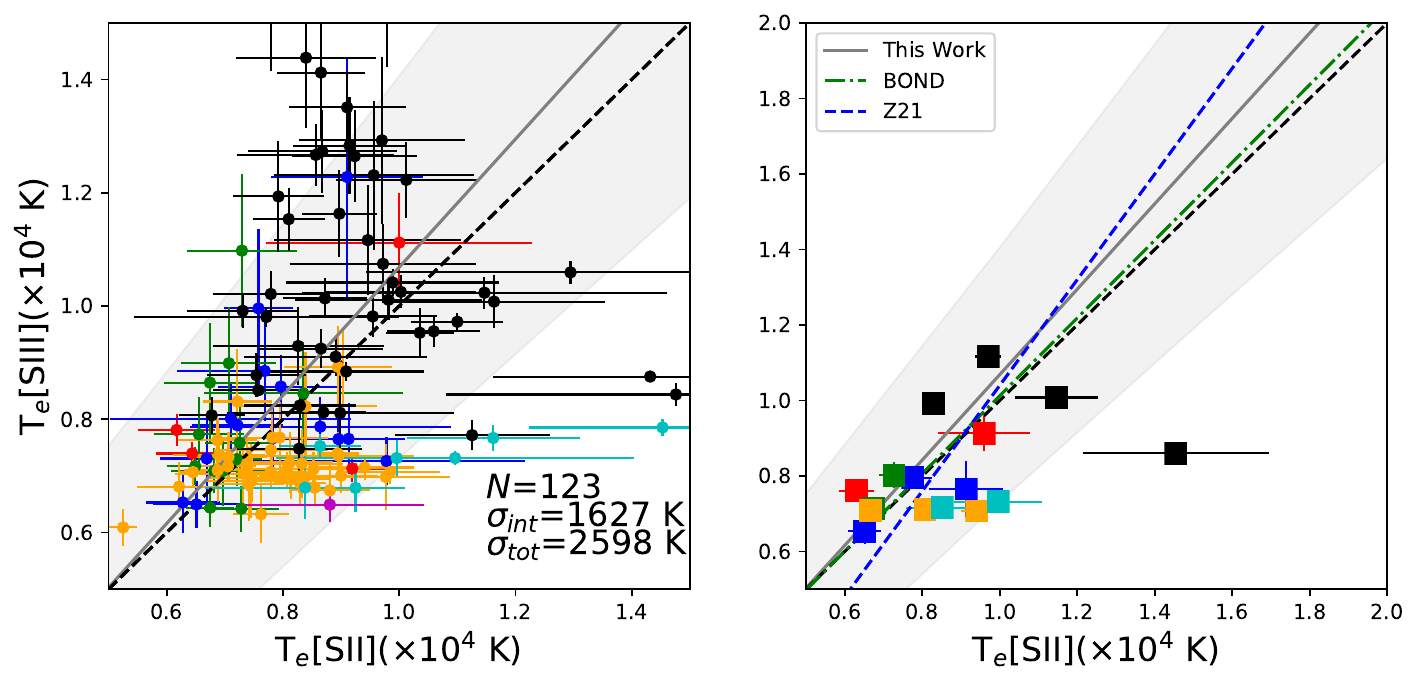}
\includegraphics[scale=0.6]{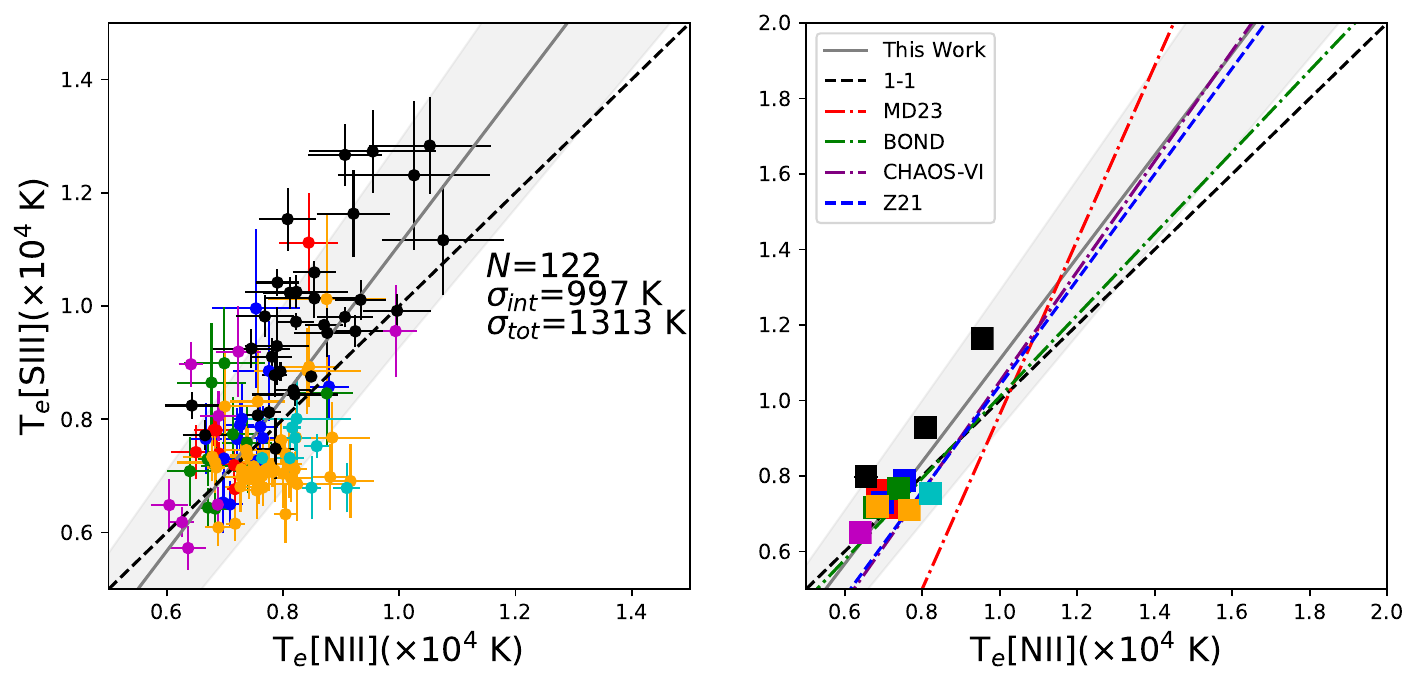}
\caption{$T_{\rm{e}}$--$T_{\rm{e}}$ relationships between the low and intermediate ionization zone temperatures measured from \HII\ regions in nearby galaxies. The temperatures measured from \HII\ regions in this work are color coded by host galaxy. The left panel in each row displays the individual temperatures and errors with each point colored according to the host galaxy. The right panel in each row show the $T_{\rm{e}}$--$T_{\rm{e}}$ relations for temperatures binned
in steps of 2000 K in the x-axis, with minimum 2 \HII\
regions per bin, for each galaxy, compared to $T_{\rm{e}}$--$T_{\rm{e}}$ trend lines from \citetalias{Zurita2021MNRAS.500.2359Z} (\textit{blue dash}), \citetalias{Rogers:chaos:2021} (\textit{purple dot-dash}), \citetalias{BOND} (\textit{green dot-dash}) and \citetalias{Delgado_Desire2023arXiv230513136M} (\textit{red dot-dash}). To aid the eye, we include in (\textit{black dash}) the 1--1 line. For comparisons with p-value $<~10^{-3}$, we include the best-fit line (\textit{grey-solid}) and 1$\sigma$ fit uncertainty (\textit{grey-shaded}).}
    \label{fig:temps_intermediate_gals}
\end{figure*}

The $T_{\rm{e}}$--$T_{\rm{e}}$ relations between the low and intermediate-ionization zone temperatures are shown in Figure \ref{fig:temps_intermediate_gals}. Given the lower ionization potentials of O$^+$, N$^+$, and S$^+$ with respect to S$^{++}$, it is possible to have differences in the temperature in the different ionization zones \citep{Garnett1992AJ}. We compare our observed $T_{\rm{e}}$--$T_{\rm{e}}$ relations to predictions from photoionization models for giant \HII\ regions produced by the Bayesian Oxygen and Nitrogen abundance project \citetalias{BOND} and found in the Mexican Million Models database \citep{MexicanMillionModels}. When available, we also compare the observations to relations from \citetalias{Rogers:chaos:2021}, \citetalias{Delgado_Desire2023arXiv230513136M}, and \citetalias{Zurita2021MNRAS.500.2359Z}.

The top panel of Figure \ref{fig:temps_intermediate_gals} shows the temperature comparisons between \temp{S}{3} and \temp{O}{2}. The best-fit line, \temp{S}{3}$=(0.67 \pm 0.27)\times$\temp{O}{2}$+(0.022\pm0.022)$, closely follows the relationship reported by \citetalias{Zurita2021MNRAS.500.2359Z}. We measure an intrinsic, $\sigma_{int}=1002$~K, and total scatter, $\sigma_{tot}=1496$~K. The \citepalias{BOND} models predict that the temperature \temp{S}{3} should be greater than \temp{O}{2} by a small constant offset across the full temperature range. However, our data suggest \temp{O}{2} rises faster than  \temp{S}{3}. Our binned data favor the empirical trend line from \citetalias{Zurita2021MNRAS.500.2359Z}. However, there is a grouping of points from NGC~5068 that show contrasting behavior in both the individual and binned data comparisons.  

The \temp{S}{3} and \temp{S}{2} comparison, comprised of 108 \HII\ regions, is shown in the middle panel of Figure \ref{fig:temps_intermediate_gals}. The best fit line for this comparisons is described by, \temp{S}{3}$=(1.11 \pm 0.25)\times$\temp{S}{2}$-(0.045\pm0.21)$. Within uncertainties, the \temp{S}{3}$--$\temp{S}{2} agree with trends observed by \citetalias{Zurita2021MNRAS.500.2359Z} and that derived from \citetalias{BOND}. However, both the intrinsic, $\sigma_{int}=1627$~K, and total scatter, $\sigma_{tot}=2598$~K, are large. The binned data do not reveal a preference for either of the literature trend.

The series of CHAOS $T_{\rm{e}}$--$T_{\rm{e}}$ comparisons \citep[see,][]{Berg2015ApJ...806...16B,Berg2020ApJ...893...96B,Rogers:chaos:2021,Rogers:chaos:2022} and \citetalias{Zurita2021MNRAS.500.2359Z} have observed a tight relationship between \temp{S}{3} and \temp{N}{2}.
We show in the bottom panel of Figure \ref{fig:temps_intermediate_gals} the \temp{S}{3} and \temp{N}{2} comparison. Satisfying the expectations driven by these past studies, we observe that the trend between \temp{S}{3} and \temp{N}{2} exhibits the smallest scatter of the $T_{\rm{e}}$--$T_{\rm{e}}$ relations presented here. The trend between these temperatures is described by, \temp{S}{3}$=(1.35 \pm 0.15)\times$\temp{N}{2}$-(0.24\pm0.11)$, and exhibit intrinsic scatter, $\sigma_{int}=997$~K, and total scatter, $\sigma_{tot}=$1313~K; both of which are larger than $\sigma_{int}=173$~K $\sigma_{tot}=$507~K reported by \citetalias{Rogers:chaos:2021}.

For the \temp{S}{3} and \temp{N}{2} comparison there are available empirical relations from \citetalias{Rogers:chaos:2021}, \citetalias{Zurita2021MNRAS.500.2359Z}, and \citetalias{Delgado_Desire2023arXiv230513136M} which we overlay in addition to the \citetalias{BOND} models. The data show a clear disagreement with the trend observed by \citetalias{Delgado_Desire2023arXiv230513136M}. We observe a large fraction of the binned data that lie near the \citetalias{Rogers:chaos:2021} model, and \citetalias{Zurita2021MNRAS.500.2359Z} relations at low \temp{N}{2}. Regions with hotter \temp{N}{2} and \temp{S}{3} would be needed to to further differentiate between the models and empirical trends.

\subsubsection{High Ionization Zone}
\begin{figure*}[!h]
\centering
\gridline{\fig{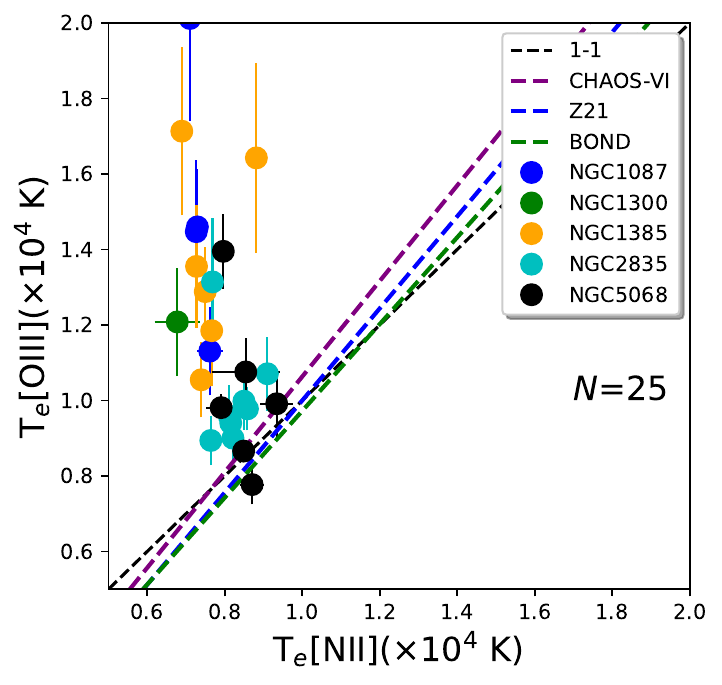}{0.49\textwidth}{}
          \fig{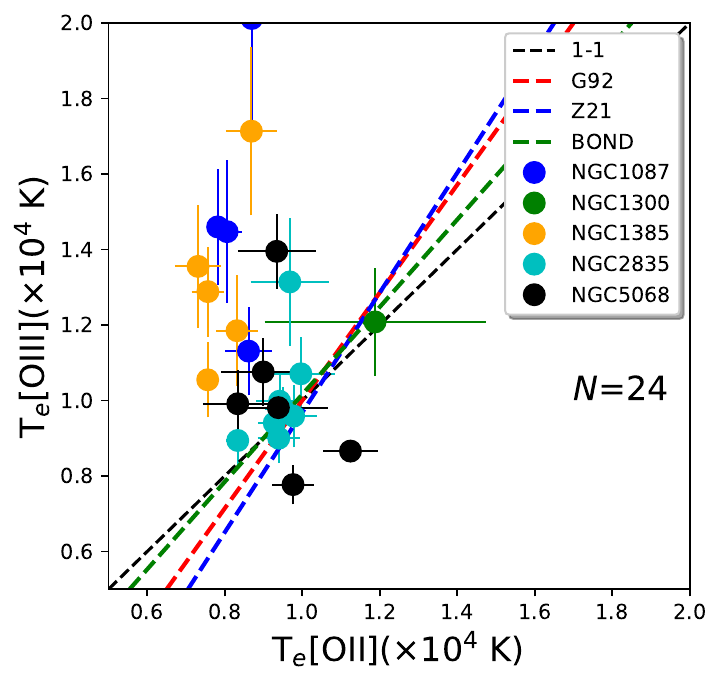}{0.49\textwidth}{}}
\gridline{\fig{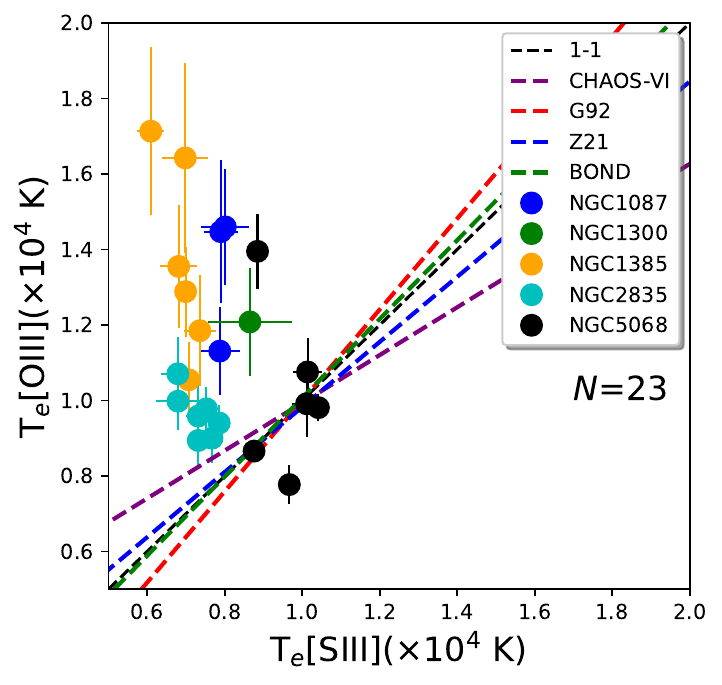}{0.49\textwidth}{} \fig{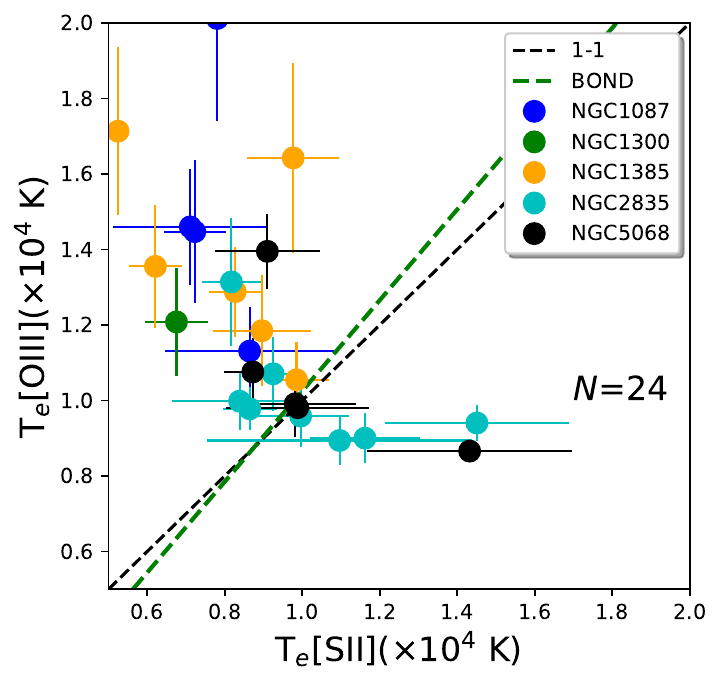}{0.49\textwidth}{}}
\vspace{-.2in}

\caption{$T_{\rm{e}}$--$T_{\rm{e}}$ relationships for the high ionization zone. The temperatures measured from \HII\ regions in this work are color coded by host galaxy. The (\textit{blue-dashed}) line show the $T_{\rm{e}}$--$T_{\rm{e}}$ relationships from \citetalias{Zurita2021MNRAS.500.2359Z}. The (\textit{red-dashed}) line show trends from \citetalias{Rogers:chaos:2021}. The (\textit{purple-dashed}) and (\textit{green-dashed}) line trends from the photoionization models of \citetalias{Garnett1992AJ} and \citetalias{BOND}. To aid the eye, we include show in \textit{black} line the 1--1 line in each panel.}
\label{fig:temps_high_gals}
\end{figure*}

We show $T_{\rm{e}}$--$T_{\rm{e}}$ relations between [\ion{O}{3}] temperatures (which trace the high ionization zone) and those from the low-intermediate ionization zones in Figure \ref{fig:temps_high_gals}. Additionally, we overlay $T_{\rm{e}}$--$T_{\rm{e}}$ relations from \citetalias{Berg2020ApJ...893...96B}, \citetalias{Zurita2021MNRAS.500.2359Z}, and \citetalias{Garnett1992AJ}. Although we observe some \HII\ regions with $T_{\rm{e}}$--$T_{\rm{e}}$ relations that agree with literature relations, we also observe numerous \HII\ regions with much higher \temp{O}{3} values of than what is predicted by models with the given low and intermediate ionization zone temperatures. The total scatter around the line of equality in \temp{O}{3} range between 3100~K and 4500~K. Given this large scatter we do not perform a linear regression analysis for these comparisons.

Based on previous findings from \citetalias{Zurita2021MNRAS.500.2359Z} and \citetalias{Rogers:chaos:2021}, the scatter towards large excess in \temp{O}{3} for regions with cooler low and intermediate ionization zone temperatures is unexpected. With IFUs we are perhaps capturing a wider range of \HII\ regions. Furthermore \citetalias{Zurita2021MNRAS.500.2359Z} and \citetalias{Rogers:chaos:2021} extend to lower metallicities, and possibly higher ionization parameters, where \temp{O}{3} may be better behaved. Nevertheless, these regions represent an extremely limited sub-set of the data and are at the boundary of significant, so are subject to higher uncertainty. For the $\sim$ Solar metallicities for our \HII\ region sample, where the relative flux [\ion{O}{3}]$\lambda4363$ is expected to be $< 10^{-2} \times$ H$\beta$ \citep{Berg2015ApJ...806...16B}, the temperature from [\ion{O}{3}]$\lambda4363$ would have to be high in order to be detected. The small number of [\ion{O}{3}]$\lambda4363$ detections reflects this. Because we do not expect to detect the line in most \HII\ regions, the ones we do detect may be unusual cases or statistical outliers, especially given that the average S/N of the [\ion{O}{3}] detections is $\sim$4. We explore S/N effects by increasing the threshold for comparison to S/N $> 5$ in [\ion{O}{3}]$\lambda4363$. This reduces the sample of regions with \temp{O}{3} measurements to 5 which is too low to confidently fit a trend, but, we measure a large total scatter, $\sigma_{tot} > 2000$~K, around the line of equality. Discussed further in Section \ref{sec:result_ip} and in \cite{Yates2020A&A...634A.107Y}, these regions exhibit low \OIIIOII\ ratios, meaning they are systems with low O$^{++}$/O$^{+}$. This would mean that small fraction of the total nebulae volume would be described by the high \temp{O}{3}. Despite this, there has been evidence \citep{Peimbert1991shocks,Binette2012_OIII_shocks} that shock excitation can preferentially enhance the high ionization zone temperature, with the highest enhancement occurring in high metallicity environments. How this scenario could apply to this small sub-set of \HII\ regions is explored in Section \ref{res:high_o3}.

\subsection{The Impact of Density Inhomogeneities on [S~II] and [O~II] Temperatures}
\label{results:density_inhomogeneities}
Recent studies have suggested that the temperatures obtained from the auroral-to-nebular lines ratios of [\ion{O}{2}] and [\ion{S}{2}] can be biased upwards due to the presence of density inhomogeneities, even while the average density is underestimated by nebular doublet line [\ion{O}{2}]$\lambda\lambda3726,3729$ and [\ion{S}{2}]$\lambda\lambda6716,6731$ diagnostics.

For example, in a sample of 190 high-signal-noise spectra of \HII\ regions and other photoionized nebulae, \cite{Delgado_Desire2023arXiv230513136M} observed systematically hotter \temp{O}{2} and \temp{S}{2} relative to \temp{N}{2}, similar to what we observe, which they attribute to the presence of density inhomogeneities. The atomic levels responsible for the nebular lines [\ion{S}{2}]$\lambda\lambda6731,6716$ and [\ion{O}{2}]$\lambda\lambda3726,3729$, listed in Table \ref{tab:atomic_references}, have critical densities of order $10^3$~cm$^{-3}$ which are at least two orders of magnitude lower than the critical densities for the nebular levels of [\ion{N}{2}]. In addition, the auroral levels of the same ions have very high critical densities. This makes the auroral-to-nebular temperature diagnostics of \temp{O}{2} and \temp{S}{2}  density sensitive above $10^3$~cm$^{-3}$, and therefore susceptible to biases if there are important contributions to the line flux from gas above that density. The auroral-to-nebular ratio of  [\ion{N}{2}], however, is not susceptible to such sensitivity until much higher densities.

Both [\ion{S}{2}]$\lambda\lambda6731,6716$ and [\ion{O}{2}]$\lambda\lambda3726,3729$ nebular line doublet ratios serve as density diagnostics for densities $10^2$~cm$^{-3}<$ $n_{\rm{e}}$ $<$ $10^{3.5}$~cm$^{-3}$. Furthermore, because of the bias described above, \citet{Delgado_Desire2023arXiv230513136M} show that at fixed temperature the auroral-to-nebular line ratios for [\ion{O}{2}] and [\ion{S}{2}] can serve as a density diagnostic over a large range of electron density, $10^2$~cm$^{-3}<$ $n_{\rm{e}}$ $<$ $10^{6}$~cm$^{-3}$. For a uniform density \HII\ region, the $n_{\rm{e}}$ returned from both of these diagnostics should be identical, as long as $n_{\rm{e}}$ is within the sensitivity range of the diagnostics. However, in the presence of density inhomogeneities, different density diagnostics can return conflicting values. Even if high density gas clumps make up a small fraction of the gas, such regions can continue to contribute to the auroral line emission while no longer contributing significantly to the nebular lines, since the effects of collisional de-excitation on the nebular lines will reduce their emissivities relative to the auroral lines \citep{Rubin:1989}. Because of this, the nebular [\ion{S}{2}] and [\ion{O}{2}] lines can reflect the dominant contribution of low-density gas, while the auroral [\ion{S}{2}] and [\ion{O}{2}] lines will be sensitive to volume of high-density gas  \citep{Peimbert:1971,Rubin:1989,Delgado_Desire2023arXiv230513136M}.
 
\begin{figure*}
    \centering
    \includegraphics[scale=0.6]{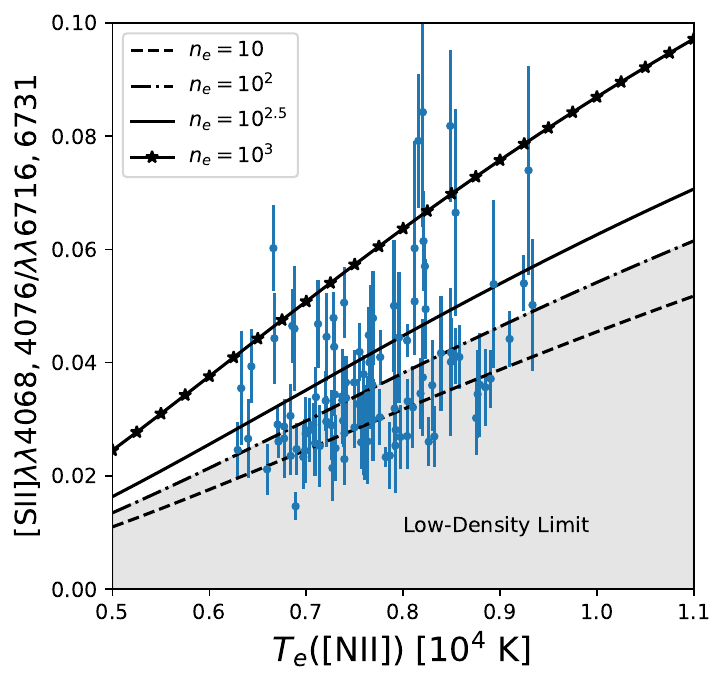}
    \includegraphics[scale=0.6]{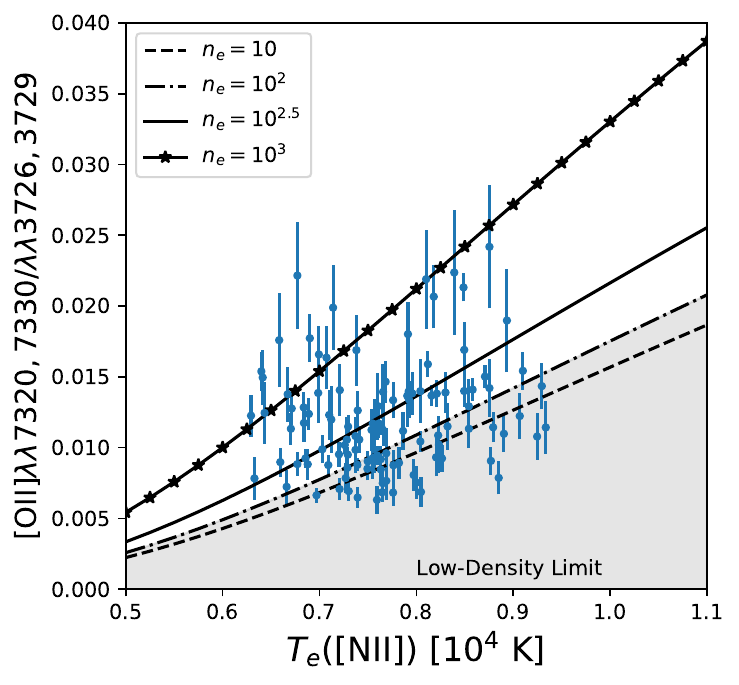}
    \includegraphics[scale=0.6]{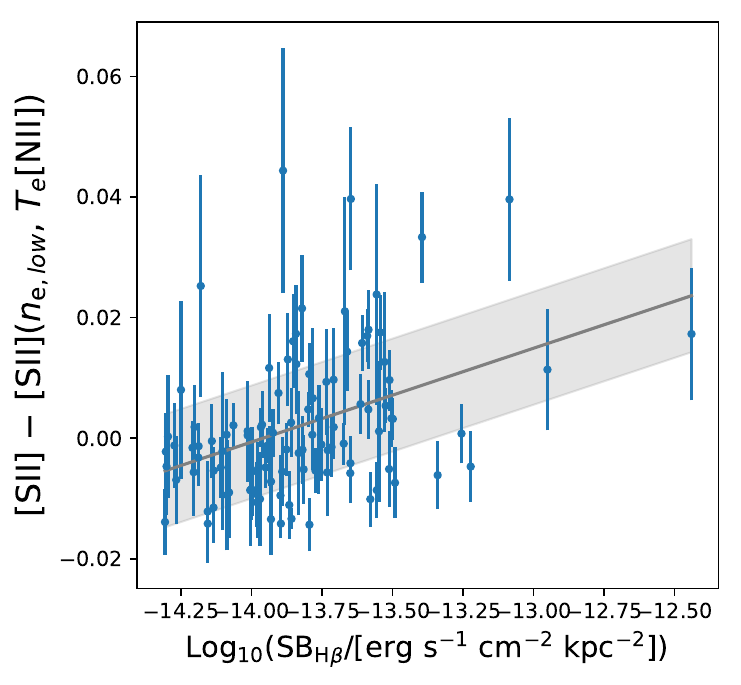}
    \includegraphics[scale=0.6]{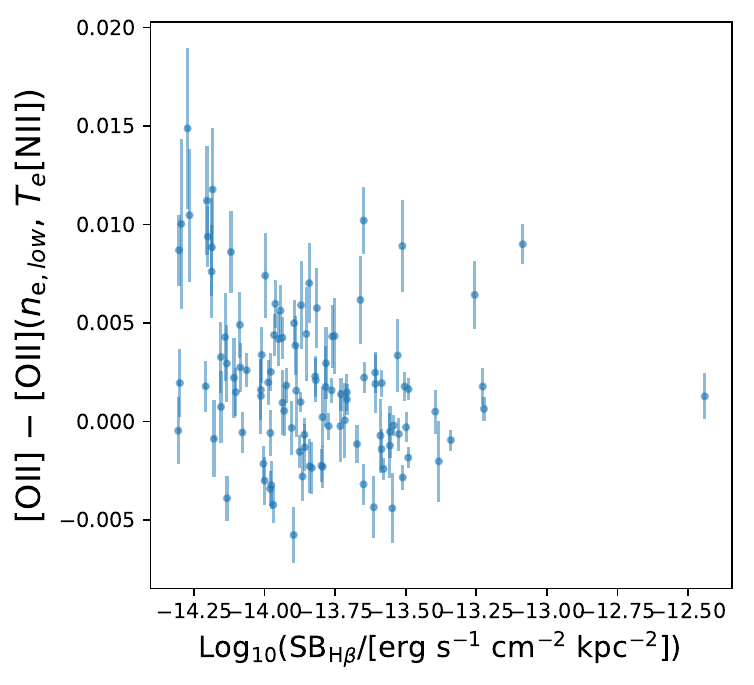} 
    \caption{In the top panels we show auroral-to-nebular line ratios measured from [\ion{S}{2}] and [\ion{O}{2}] against their measured \temp{N}{2}. We assume equality between the low- ionization zone temperatures and overlay lines of predicted [\ion{S}{2}] and [\ion{O}{2}] auroral-to-nebular line ratios for fixed electron densities $n_e=10\ \mathrm{cm}^{-3}, n_e=10^2\ \mathrm{cm}^{-3}$, $10^{2.5}\ \mathrm{cm}^{-3}$, and $10^{3}\ \mathrm{cm}^{-3}$, vs. \temp{N}{2}. The (\textit{grey-shaded}) region, in the top panels, show regime where auroral-to-nebular lines ratios for densities below the low-density limits for the [\ion{S}{2}] and [\ion{O}{2}] density diagnostics. In the bottom panels we show the departure of the measured [\ion{S}{2}] and [\ion{O}{2}] auroral-to-nebular line ratios from the theoretical auroral-to-nebular line ratio predicted using $n_{\rm{e}}=100 \rm{cm}^{-3}$ and \temp{N}{2} against the the H$\beta$ surface brightness. For the bottom-left [\ion{S}{2}] panel  we show the best-fit line (\textit{grey-solid}) and 1$\sigma$ fit uncertainty (\textit{grey-shaded}).}
    \label{fig:ne_result}
\end{figure*}

To investigate if the presence of density inhomogeneities could bias our measured \temp{S}{2} and \temp{O}{2} we compare the observed auroral-to-nebular line ratios [\ion{O}{2}] and [\ion{S}{2}] to those predicted using fixed \temp{N}{2}. For this comparison we use the regions with auroral line detections for all three low-ionization zone ions. We show in Figure \ref{fig:ne_result} the measured [\ion{O}{2}]$\lambda\lambda7320,7330$/$\lambda\lambda3726,3729$ and [\ion{S}{2}]$\lambda\lambda4069,4076$/$\lambda\lambda6716,6731$ line ratios versus the region's \temp{N}{2}. We overlay the predicted trends of auroral-to-nebular line ratios calculated using $n_{\rm{e}}=10^2~\mathrm{cm}^{-3}$, $10^{2.5}~\mathrm{cm}^{-3}$, and $10^{3}~\mathrm{cm}^{-3}$. We see in Figure \ref{fig:ne_result} that under the assumption that \temp{N}{2}$=$\temp{S}{2}$=$\temp{O}{2}, the largest measured auroral-to-nebular line ratios could be consistent with $T_{\rm{e}}$ traced by \temp{N}{2} but with a higher electron density than that returned by [\ion{S}{2}]$\lambda\lambda6731,6716$ in the low-density limit. This suggests that under inhomogenous conditions, underestimated contributions from $>10^3$~cm$^{-3}$ gas to the nebular [\ion{O}{2}] and [\ion{S}{2}] lines could bias the density diagnostics and then the use of underestimated densities in temperature calculations for ions with low critical densities like  [\ion{O}{2}] and [\ion{S}{2}] could lead to hotter estimated temperatures inferred from auroral-to-nebular ratios. 

At the same time, Figure \ref{fig:ne_result} also shows that the lowest measured auroral-to-nebular ratios lie below the theoretical curves for ratios with densities equal to the low-density limit. It may be the case that, under our assumption that the \HII\ region low-ionization zone $T_{\rm{e}}$ is described by \temp{N}{2}, these regions physically exhibit volumes of gas with $n_{\rm{e}} \lesssim 10~\rm{cm}^{-3}$ or lower \citep{Kennicutt1984ApJ...287..116K}. However, we are unable to verify this using the available diagnostics because densities returned by either the [\ion{O}{2}] and [\ion{S}{2}] doublet or their auroral-to-nebular line ratios are uncertain in this regime. For these reason we do not pursue further interpretation of these points.

We can further explore potential density inhomogeneities by comparing the measured auroral-to-nebular lines to the regions H$\beta$ surface brightness. The emission from any recombination lines is proportional to $n_{\rm{e}}n({\rm{X}^+})\alpha_{eff}$, where $n({\rm{X}^+})$ is the number density of the emitting ion and $\alpha_{eff}$ is the effective recombination coefficient \citep{Piembert2017PASP..129h2001P}. For H$\beta$, $n({\rm{X}^+})\propto n_{\rm{e}}$ because 90\% of free electrons will come from the photoionization of H which represents $\sim$90\% of all the gas. For a uniform density \HII\ region, the H$\beta$ surface brightness, SB$_{\rm{H}\beta}$, would be proportional to $n_{\rm{e}}^2$. In the case of \HII\ regions with high density inclusions, the variance of $n_{\rm{e}}$ would be expected to rise due to the increase in the average of the density squared (i.e. $<n_{\rm{e}}^2>$). Under such conditions, it would be reasonable to expect that H$\beta$ surface brightness would increase as SB$_{\rm{H}\beta} \propto <n_{\rm{e}}^2>$.

If we interpret the departure of the measured auroral-to-nebular lines ratios from the theoretical ratios calculated by fixing the electron density at the low-density limit, $n_{\rm{e}}=100~\rm{cm}^{-3}$ and $T_{\rm{e}}=$\temp{N}{2}, then we could expect that this deviation would correlate with the H$\beta$ surface brightness. We show in the bottom panels of Figure \ref{fig:ne_result} the degree of inhomogeneities, measured by [\ion{S}{2}]-[\ion{S}{2}]($n_{\rm{e}}=100$, \temp{N}{2}) and  [\ion{O}{2}]-[\ion{O}{2}]($n_{\rm{e}}=100$, \temp{N}{2}), against the regions H$\beta$ surface brightness.

We find a significant correlation, p-value~$<10^{-3}$, between [\ion{S}{2}]-[\ion{S}{2}]($n_{\rm{e}}=100$, \temp{N}{2}) and SB$_{\rm{H}\beta}$, which suggests that the regions with large deviations from the predicted low-density limit auroral-to-nebular line ratios are consistent with density inhomogeneities. The best-fit line is described by [\ion{S}{2}]-[\ion{S}{2}]($n_{\rm{e}}=100$, \temp{N}{2}) $=0.0147(\pm 0.004)\times\rm{SB}_{\rm{H}\beta}+0.20(\pm0.05)$. To account for any uncertainties in the atomic data, we vary the absolute value of the [\ion{S}{2}]($n_{\rm{e}}=100$, \temp{N}{2}) curve by $\pm$ 10\% \citep{Mendoza_atomic_comparisons} and find no change in correlation strength. While this correlation is suggestive of density inhomogeneities, we acknowledge that the \HII\ regions are not fully resolved which means we are measuring a PSF-averaged surface brightness. Because of this, it is unclear if this correlation can be fully link to density inhomogeneities. 

For the comparison involving [\ion{O}{2}]-[\ion{O}{2}]($n_{\rm{e}}=100$, \temp{N}{2}), we find no significant correlation with H$\beta$ surface brightness, even when using S/N$>5$ auroral lines. While the exact reason for the non-correlation between [\ion{O}{2}]-[\ion{O}{2}]($n_{\rm{e}}=100$, \temp{N}{2}) and H$\beta$ surface brightness is unknown it is important to note that the ionization potential of [\ion{S}{2}] is less than both $H\beta$ and [\ion{O}{2}]. This difference in ionization potential means that \HII\ region [\ion{S}{2}] and [\ion{O}{2}] may not be co-spatial, and may have different sensitivities as tracers of high density inclusions. However, given that these regions have measurements of all three low-ionization zone auroral lines and survive a DIG contrast constraint, we find that the correlation between [\ion{S}{2}]-[\ion{S}{2}]($n_{\rm{e}}=100$ and SB$_{\rm{H}\beta}$ suggests that density inhomogeneities may be affecting the low-ionization zone temperatures. Future studies with multiple density diagnostics and high spatial resolution will be valuable to exploring the potential impact inhomogeneous conditions have on these diagnostics.

\subsection{Temperature Differences Compared to H~II Region Ionized Gas, Stellar Population, and Molecular Gas Properties.}
\label{res:differences}
Studies have shown that \HII\ region temperatures for different ionization zones can be differently impacted by properties of the ISM. Temperature comparisons presented in \cite{Berg2020ApJ...893...96B} using \HII\ regions observed in four nearby galaxies revealed that the dispersion around low-intermediate and intermediate-high $T_{\rm{e}}$--$T_{\rm{e}}$ relation ships increased (or decreased) with the ionization parameter. Another trend with ionization parameter was observed by \cite{Yates2020A&A...634A.107Y}. They found that systems with low ionization, or larger ratios of O$^{+}$/O$^{2+}$ parameter would exhibit systematically hotter \temp{O}{3}. Though \cite{Arellano2020} argue this could be explained by increased iron contamination to [\ion{O}{3}]$\lambda4363$.  Discussed in Section \ref{results:density_inhomogeneities}, density fluctuations can also bias the temperatures for the low-ionization zone due to the sensitivity of [\ion{O}{2}] and [\ion{S}{2}] with density \citep{Delgado2023arXiv230511578M,Delgado_Desire2023arXiv230513136M}.

Given that stars are the primary source of ionizing photons, it is reasonable to suspect that the properties of the stellar population ionizing the \HII\ region can potentially play a role in setting the $T_{\rm{e}}$ structure of \HII\ regions. Another potential factor on the $T_{\rm{e}}$ structure, and traced by its effects on the surrounding molecular gas, is the degree of stellar feedback within \HII\ regions. Although very important to our understanding, the physical processes that impact the $T_{\rm{e}}$ structure in \HII\ regions remain uncertain \citep{Garnett1991ApJ...373..458G,Nicholls2020PASP..132c3001N}.

The KCWI+MUSE \HII\ regions combined with the PHANGS-HST and PHANGS-ALMA observations allow us to investigate how $T_{\rm{e}}$ is impacted by different \HII\ ISM, stellar and molecular gas properties.
We compare temperature differences, $\Delta$($T_{\mathrm{ion,1}}$,$T_{\mathrm{ion,2}}$)=$T_{\mathrm{ion,1}}$$-$$T_{\mathrm{ion,2}}$, between the low, intermediate, and high-ionization zone temperatures. with \HII\ region properties derived from emission line diagnostics; with the properties of the surrounding molecular gas  measured from ALMA \citep{2021Leroy_data}; and with  stellar population masses/ages from SED fitting to HST photometry \citep{2022PHANGS_HST,ThilkerCompactClusters,Larson2023}. 

To gauge the significance and monotonicity of each comparison we calculate the Pearson correlation coefficient (i.e. p-value or $p$), as well as the Spearman Rank Correlation Coefficient, $\rho$. A correlation is judged to be significant if it exhibits $p$ $\lesssim 10^{-3}$. The strength of the correlation is separated into the following regimes: $1 > |\rho| > 0.8 $ corresponds to a \textit{strong} correlation, $0.8 > |\rho| > 0.4 $ corresponds to a \textit{moderate} correlation and $0.4 > |\rho| > 0 $ identifies a \textit{weak} or no correlation.

As expected, we observe strong correlations between temperatures differences with ionization parameter. We do not report any significant correlations between $\Delta T_{\rm{e}}$ with any of the following properties: integrated CO intensity, CO peak temperature, molecular gas velocity dispersion, cluster mass, cluster age, association age, association age. Figures showing the $\Delta T_{\rm{e}}$ comparisons to these parameters are shown in Appendix \ref{appn:uncorrelated_properties}. Despite no significant correlations, we do find interesting behavior between \tempDiff{N}{2}{S}{3} with association mass, and excess \temp{O}{3} with molecular gas velocity dispersion. We discuss these special cases and the details of the comparisons in the following subsections. The statistics of the correlations with \HII\ region ionization parameter are summarized in Table \ref{tab:summary_table_1}.

\begin{deluxetable*}{cDDD}
\tablecaption{Summary of the p-values and Spearman Rank coefficients for comparisons between $\Delta T_{\rm{e}}$ with the ionization and radiation softness parameter of the \HII\ regions.}
\label{tab:summary_table_1}
\tablehead{\colhead{$\Delta T_{\rm{e}}$} & \multicolumn2c{\SIIISII} & \multicolumn2c{\OIIIOII} & \multicolumn2c{$\eta$}  \\ \colhead{} & \multicolumn2c{($\rho$, p)} & \multicolumn2c{($\rho$, p)} & \multicolumn2c{($\rho$, p)} }
\decimals
\startdata
\tempDiff{O}{2}{N}{2} & - & - & - \\
\tempDiff{N}{2}{S}{2} & (-0.27,\ $<10^{-3}$) & - & - \\
\tempDiff{S}{2}{O}{2} & (0.32,\ $<10^{-3}$)  & - &   - \\
\tempDiff{O}{2}{S}{3} & - & - & - \\
\tempDiff{N}{2}{S}{3} & - & - & - \\
\tempDiff{S}{2}{S}{3} & (0.53,\ $<10^{-3}$)  & - & (0.46,\ $<10^{-3}$) \\
\tempDiff{O}{2}{O}{3} & - & (0.85,\ $<10^{-3}$) & -  \\
\tempDiff{N}{2}{O}{3} & (0.72,\ $<10^{-3}$)  & (0.82,\ $<10^{-3}$)  & - \\
\tempDiff{S}{2}{O}{3} & (0.75,\ $<10^{-3}$)  & (0.75,\ $<10^{-3}$) &  - \\
\tempDiff{S}{3}{O}{3} &  - & (0.82,\ $<10^{-3}$)  & - 
\enddata
\end{deluxetable*}

\subsubsection{Correlations between Temperature Differences, Ionization Parameter, and Radiation Softness Parameter}
\label{sec:result_ip}
\begin{figure*}[!h]
    \includegraphics[scale=0.8]{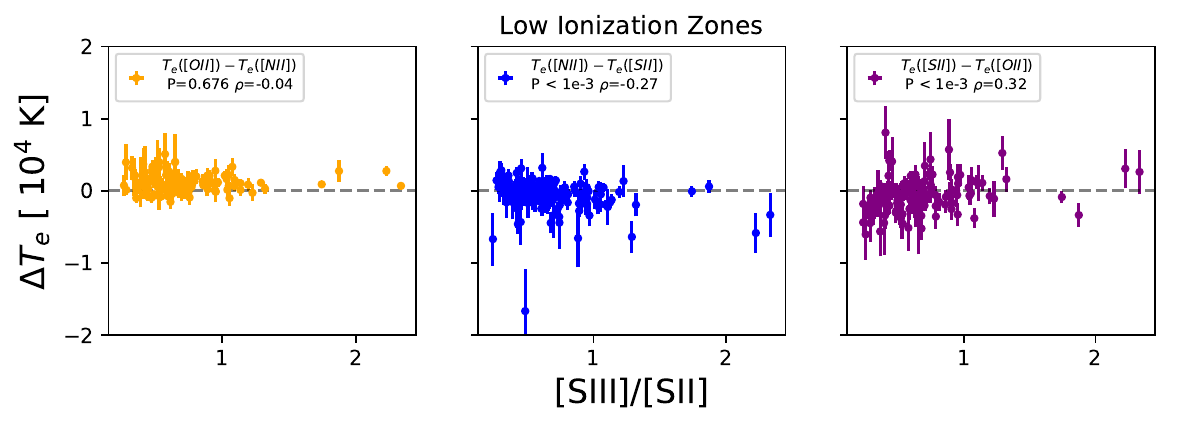}
    \includegraphics[scale=0.8]{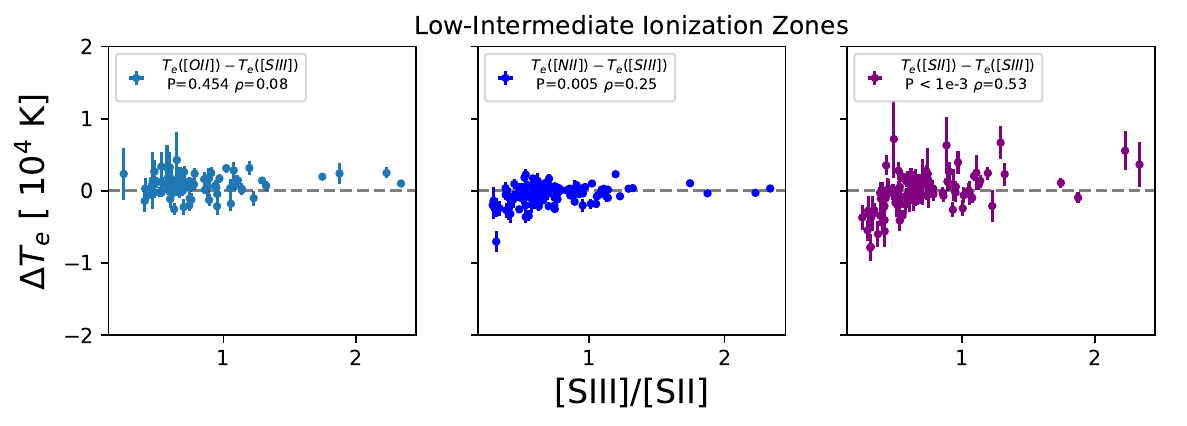}
    \includegraphics[scale=0.8]{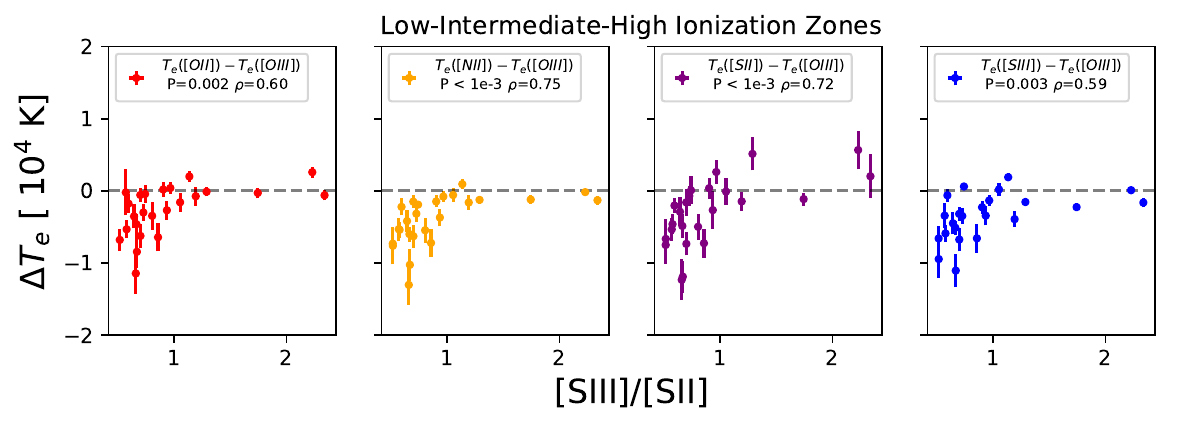}
    \caption{Electron temperature differences compared to the \HII\ region ionization parameter, $U$, traced by \SIIISII. Top: The $\Delta T_e$'s between the low ionization zone temperatures. Middle: The $\Delta T_e$'s between the low and intermediate ionization zone temperatures. Bottom: The $\Delta T_e$'s between the low, intermediate and high ionization zone temperatures. We observe significant correlations with $U$, traced by \SIIISII, between the low ionization zone temperatures differences \tempDiff{N}{2}{S}{2} and \tempDiff{S}{2}{O}{2}; between the low and intermediate ionization zones \tempDiff{S}{2}{S}{3}; and the high ionization \tempDiff{N}{2}{S}{3} and \tempDiff{S}{2}{O}{3}.}  
    \label{fig:ip_S_comparisons}
\end{figure*}

\begin{figure*}[!h]
    \includegraphics[scale=0.8]{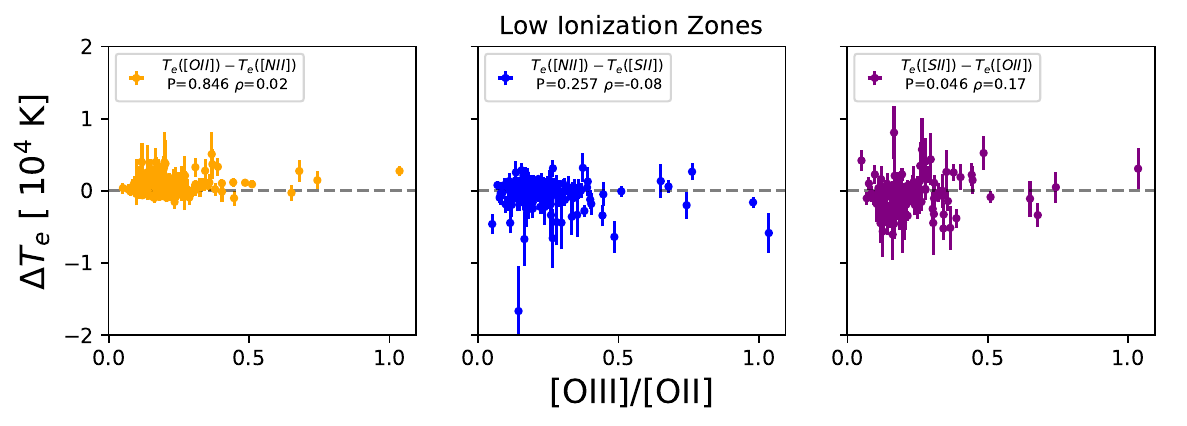}
    \includegraphics[scale=0.8]{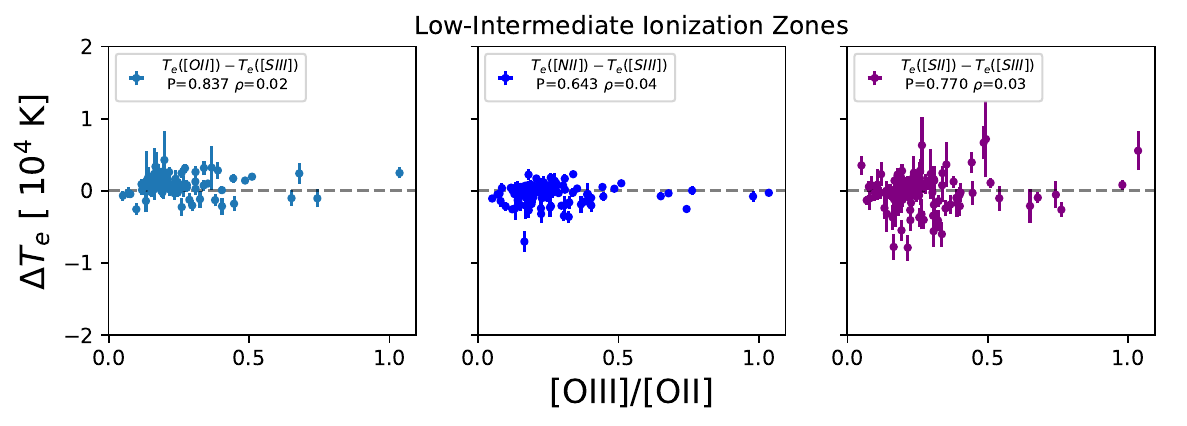}
    \includegraphics[scale=0.8]{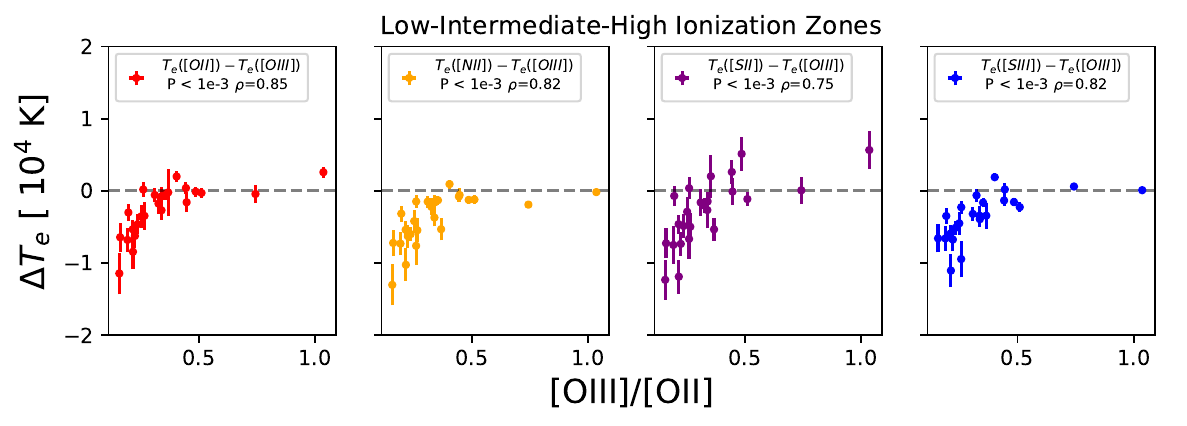}
    \caption{Electron temperature differences compared to the \HII\ region ionization parameter, $U$, traced by \OIIIOII. The ionization zones depicted in each row follow those in Figure \ref{fig:ip_S_comparisons}. We observe strong correlations between all the $\Delta T_{\rm{e}}$ involving \temp{O}{3}. The correlations between the low and intermediate ionization zones and $U$, traced by with \SIIISII\ and shown in Figure \ref{fig:ip_S_comparisons}, do not appear when tracing $U$ with \OIIIOII.}  
    \label{fig:ip_O_comparisons}
\end{figure*}

We show in Figures \ref{fig:ip_S_comparisons} and \ref{fig:ip_O_comparisons} the comparisons between temperature differences and the ionization parameter, $U$, traced with both \SIIISII\ and \OIIIOII. In both comparisons the low-ionization zones show no correlations, but there is a moderate correlation the low and intermediate-ionization zone, traced by \tempDiff{S}{2}{S}{3}, with \SIIISII. The comparisons of the low-intermediate and high-ionization zones have high correlation with both \SIIISII\ and \OIIIOII. 

We observe correlations between \tempDiff{S}{2}{S}{3} with \SIIISII. For all temperature differences with \temp{O}{3} we also observe strong correlations between $\Delta T_{\rm{e}}$ and \SIIISII. The largest temperature differences are associated with the smallest values of \SIIISII. As \SIIISII\ increases, $\Delta T_{\rm{e}}$ converges to $\Delta T_{\rm{e}}=0$. Shown in Figure \ref{fig:ip_O_comparisons}, we present the same temperature differences but using \OIIIOII\ as a tracer for $U$. All the correlations with ionization parameter that do not involve \temp{O}{3} disappear when using \OIIIOII\ as a tracer. The remaining $\Delta T_{\rm{e}}$ that include \temp{O}{3} show similar correlations as before with ionization parameter when using \OIIIOII. In both cases, the largest temperature differences occur at the lowest values of ionization parameter, traced by either \SIIISII\ or \OIIIOII.

Our correlations with ionization parameter tracers are similar to the results presented in \cite{Yates2020A&A...634A.107Y}. When comparing the temperatures of the low and high ionization zone for oxygen, \cite{Yates2020A&A...634A.107Y} observed an increase in the \temp{O}{3}/\temp{O}{2} ratio that is anti-correlated with the ratio of $\rm{O}^{2+}/\rm{O}^{+}$ (which closely follows the ionization parameter traced by \OIIIOII). Our \HII\ regions are all likely to be in the relatively high metallicity regime, where $\rm{O}^{+}$ should be the dominant ionization state of oxygen. Here, the average electron temperature will be best described by the auroral-to-nebular line ratio of [\ion{O}{2}] (with caution about density inhomogeneities, as previously noted). Nevertheless, emission from [\ion{O}{3}]$\lambda4363$ can still be produced, albeit more weakly, and given the exponential temperature dependence, it will be biased towards hotter gas. Therefore, if there are temperature inhomogenieities, the [\ion{O}{3}] temperatures  may reflect a small amount of hot, high ionization gas and may not agree with the auroral-to-nebular line ratio of [\ion{O}{2}].  \cite{Yates2020A&A...634A.107Y} predict that regions with \temp{O}{3} $>$ \temp{O}{2} will be $\rm{O}^{+}$ dominant, i.e.\ $\rm{O}^{+}/\rm{H}^{+}> \rm{O}^{2+}/\rm{H}^{+}$ . Because \OIIIOII\ $\propto$  $\rm{O}^{2+}/\rm{O}^{+}$, we would expect to see the largest deviations in \temp{O}{3} at the lowest \OIIIOII. To summarize, \cite{Yates2020A&A...634A.107Y} postulate that the differences between hotter high ionization zone temperatures, \temp{O}{3} and the low ionization zone temperature, \temp{O}{2} will increase with decreasing ionization parameter. These trends should in theory also be evident for sulfur, although these correlations were not explored by \cite{Yates2020A&A...634A.107Y}. Similar trends of temperature differences associated with different ionization states of the gas have been discussed by \cite{Berg2020ApJ...893...96B}.

\begin{figure*}[ht]
    \centering
    \includegraphics[scale=0.65]{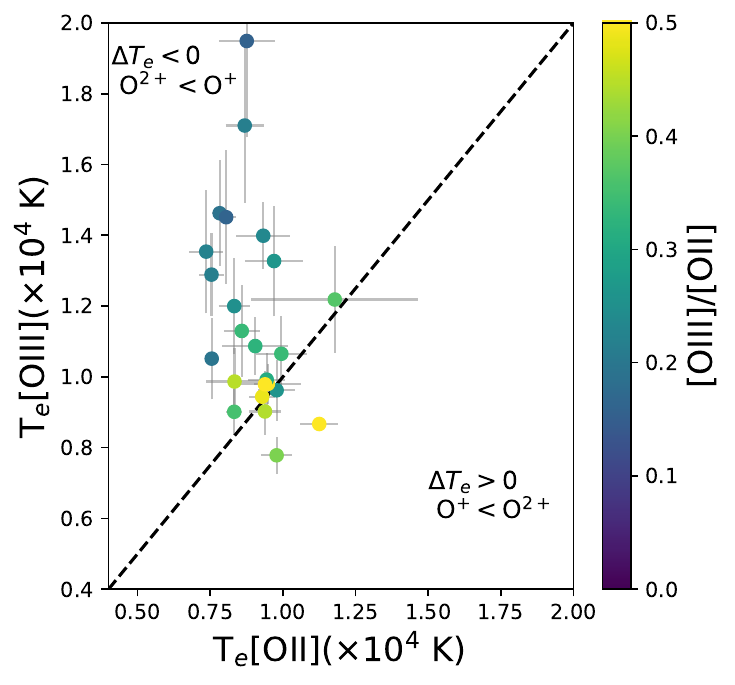}
    \includegraphics[scale=0.65]{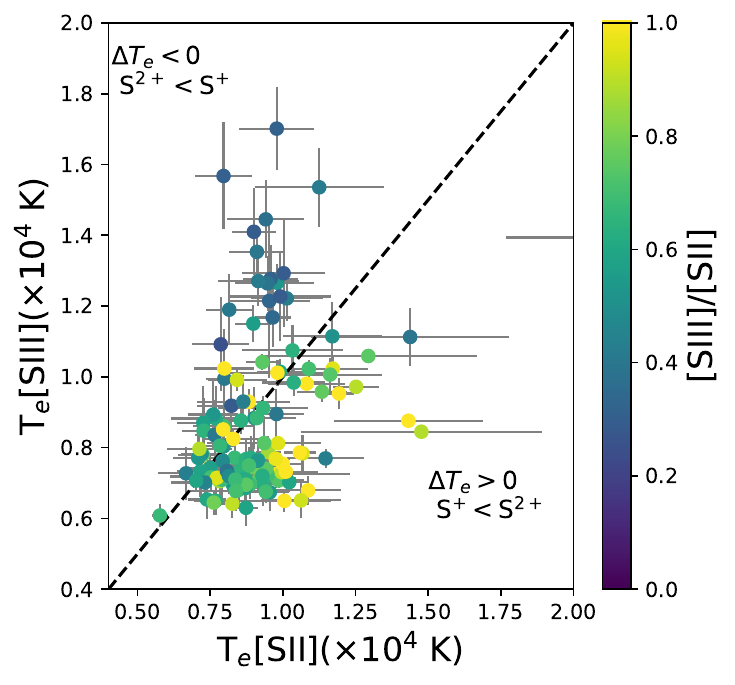}
    \caption{The $T_{\rm{e}}$--$T_{\rm{e}}$ relation between (left) \temp{O}{3}--\temp{O}{2} and (right) \temp{S}{3}--\temp{S}{2} colored by the value of \OIIIOII\ and \SIIISII. In both panels, the largest ratios of \temp{X}{3}/\temp{X}{2}, where X is either S or O, occur for the lowest values of ionization parameter, as indicated by the value of the line ratio, as expected according to the results of \cite{Yates2020A&A...634A.107Y}.}
    \label{fig:te_yates_plot}
\end{figure*}

In Figure \ref{fig:te_yates_plot}, we plot the $T_{\rm{e}}$--$T_{\rm{e}}$ between \temp{O}{3}--\temp{O}{2} and \temp{S}{3}--\temp{S}{2} colored by the value of \OIIIOII\ and \SIIISII. We also annotate the plot according to the schematics from Figure 3 of \cite{Yates2020A&A...634A.107Y}. For both oxygen and sulfur, the largest ratios of \temp{X}{3}/\temp{X}{2}, where X is either S or O, occur for the lowest values of ionization parameter, as indicated by the value of the line ratio. This suggests we are observing a similar correlation with ionization parameter as postulated by \cite{Yates2020A&A...634A.107Y}. 

If we consider the similarity between our results and \cite{Yates2020A&A...634A.107Y} as evidence for correlations between $\Delta T_{\rm{e}}$ and $U$, then why is it that the correlation between \tempDiff{S}{2}{S}{3} and $U$ traced by \SIIISII\ is not evident when using \OIIIOII\ as a tracer for $U$? It might be the case that \SIIISII\ and \OIIIOII\ do not change with $U$ in similar ways, as [\ion{S}{3}] and [\ion{O}{3}] arise from different ionization zones and conditions. It is possible that density inhomogeneities may be playing a role, both [\ion{S}{2}] and [\ion{O}{2}] have similar low critical densities, unlike either [\ion{S}{3}] and [\ion{O}{3}]. It remains unclear why some  $\Delta T_{\rm{e}}$ vs.\ $U$ trends disappear. 

We explore the correlations with $\Delta T_{\rm{e}}$ using a combination of both \SIIISII\ and \OIIIOII. \cite{Vilchez:1988} define the ``radiation softness" parameter, $\eta=$([\ion{O}{2}]/[\ion{O}{3}])/([\ion{S}{2}]/[\ion{S}{3}]), as a diagnostic of the effective temperature of the ionizing stars. Shown in Appendix \ref{appn:uncorrelated_properties} Figure \ref{fig:eta_comparisons}, we find a correlation between \tempDiff{S}{2}{S}{3} and $\eta$ with similar $p$ and $\rho$ values as the correlation with $U$. There are no correlations between temperature differences involving \temp{O}{3} with $\eta$. Because $\eta$ is a measure of ionizing properties of the ionizing stars, the correlation between $\eta$ and \tempDiff{S}{2}{S}{3} may suggest that $T_{\rm{e}}$'s derived from sulfur lines are sensitive to the stellar population while $T_{\rm{e}}$'s from oxygen are more sensitive to the physical conditions of the ionized gas. However, a future comparison with a larger sample of \temp{O}{3}'s would be beneficial in solidifying such an interpretation.

\subsubsection{Temperatures Differences with Stellar Mass and Age}
Next we compare temperature differences to stellar mass and ages from compact stellar clusters and associations matched to our \HII\ regions. As presented in Appendix~\ref{appn:comparison_of_muse_and_kcwi_hii_regions} Figures~\ref{fig:age_cluster_comparisons}--\ref{fig:age_associations_comparisons}, we find no correlations between $\Delta T_{\rm{e}}$ with cluster mass, cluster age, or with association mass and age.

\begin{figure*}[h]
    \centering
    \includegraphics[scale=0.8]{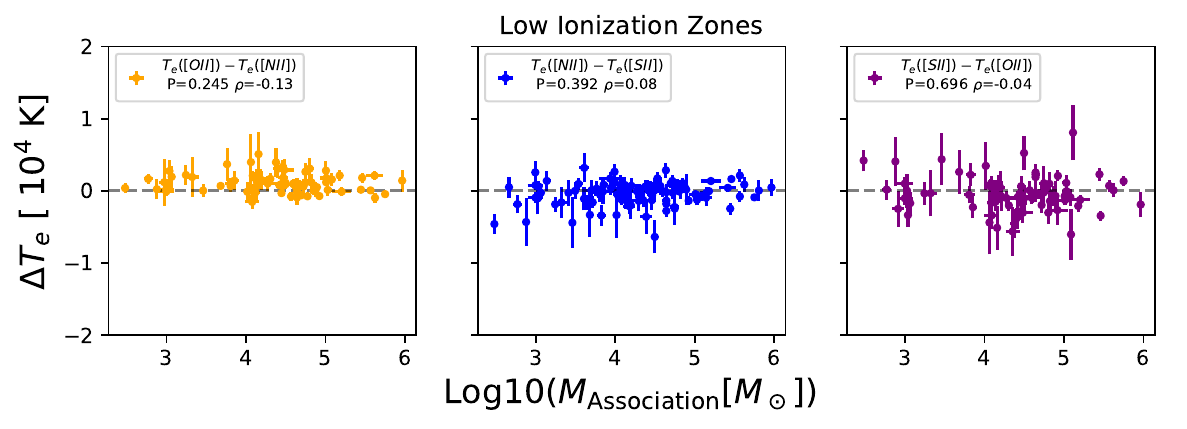}
    \includegraphics[scale=0.8]{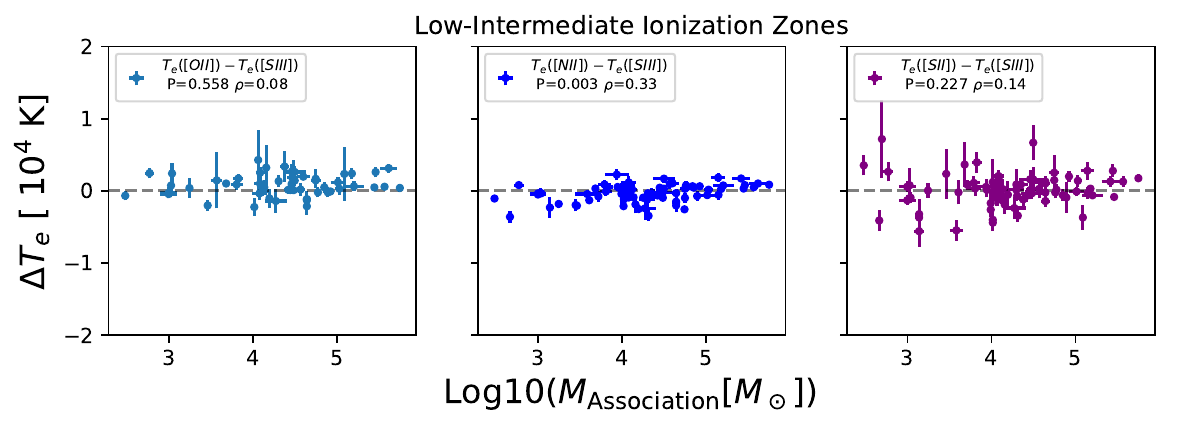}
    \includegraphics[scale=0.8]{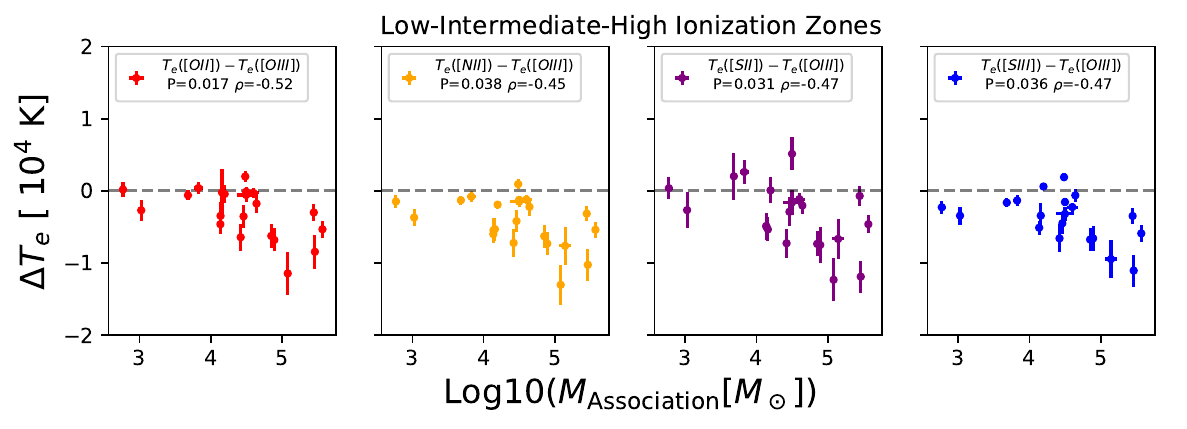}
    \caption{Electron temperature differences compared to stellar association mass. The ionization zones depicted in each row follow those in Figure \ref{fig:ip_S_comparisons}. Although the correlations are insignificant according to their p-value, we observe potential, weak, correlations between \tempDiff{N}{2}{S}{3}, and all $\Delta T_{\rm{e}}$ involving \temp{O}{3}, with stellar association mass.}
    \label{fig:mass_associations_comparisons}
\end{figure*}

\begin{figure}[!h]
\centering
\includegraphics[scale=0.6]{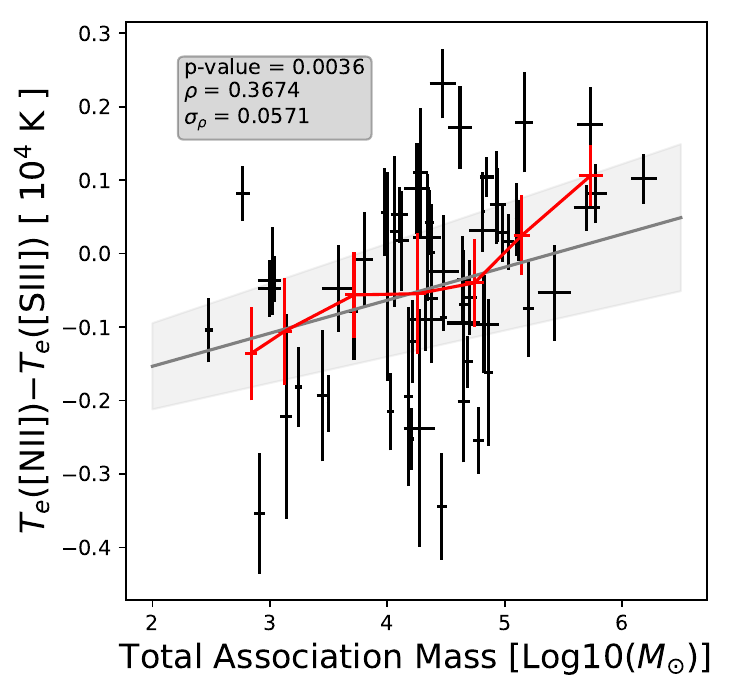}
\caption{The \temp{N}{2}$-$\temp{S}{3} temperature difference versus the total stellar association mass. The \textit{black-points} are individual \HII\ region association masses. In \textit{red} we show the average temperature differences and \HII\ region association masses calculated in Log10($M_{\odot}$)=0.5 bins. We also show a fit and $1\sigma$ fit in \textit{grey}. Compared to using the largest mass measurement, see middle row Figure \ref{fig:mass_associations_comparisons}, the p-value using total stellar mass p-value is higher, indicating a less statistically significant correlation.}
    \label{fig:total_mass}
\end{figure}

Although the correlation is not statistically significant according to our criteria, we speculate on a possible positive correlation, between the most reliable $\Delta T_{\rm{e}}$ indicator without density inhomogeneity issues, \tempDiff{N}{2}{S}{3} with the association mass. Figure \ref{fig:mass_associations_comparisons} shows the comparisons between \tempDiff{S}{2}{S}{3}, \tempDiff{O}{2}{S}{3}, and \tempDiff{O}{2}{S}{3}. While the scatter for both \tempDiff{S}{2}{S}{3} and  \tempDiff{O}{2}{S}{3} are centered around zero, it appears to be the case that \temp{N}{2} is cooler than \temp{S}{3} towards the low-mass end and vice-versa on the high-mass end. As discussed in Section \ref{sec:hst_matching}, we assigned the largest mass stellar population to the \HII\ region if the region was matched to more than one association. To see if this choice has any impact on the strength of the correlation, we plot in Figure \ref{fig:total_mass} the \tempDiff{N}{2}{S}{3} against the sum of the matched association masses. Using the total masses, we observe no change in the strength of the correlation. 

It is possible that the correlation between \tempDiff{N}{2}{S}{3} and association mass may result from biases introduced by an under-sampling of initial mass function. Similar to our study, \cite{Scheuermann2023MNRAS.522.2369S} matched the stellar association catalog \citep{Larson2023} to \HII\ regions in the Nebular catalog \citep{Kreckel2019ApJ...887...80K, Groves2023NebCat}. Instead of including all masses measured for the associations, which we do in this work, \cite{Scheuermann2023MNRAS.522.2369S} implement a cut-off and assume masses $< 10^4 M_{\odot}$, as masses below the threshold do not sample the IMF \citep{DaSilva:2012}. We perform no such cutoff in this study. The correlation between \tempDiff{N}{2}{S}{3} and $M_{\mathrm{Association}}$ correlation includes many regions with masses $< 10^4 M_{\odot}$, where the IMF may not be fully sampled. Despite this, we do not fully dismiss this correlation; however, the effects that undersampling the IMF could have on the correlation with \tempDiff{N}{2}{S}{3} warrants further investigation with a larger sample size on order to expand the dynamic range of association mass. 

\subsubsection{Temperatures Differences and Molecular Gas Properties}
We compare temperature differences to the properties of the molecular gas derived from CO emission measured within the projected boundaries of our sample of \HII\ regions. For the comparisons to $I_{\rm CO}$, shown in Figure \ref{fig:ico_comparisons} of Appendix \ref{appn:uncorrelated_properties} , and comparisons to $T_{\rm{peak}}$, shown in Figure \ref{fig:ptemp_comparisons} of Appendix \ref{appn:uncorrelated_properties}, we observe no correlations with temperature differences between the low, intermediate, and high ionization zones.

\begin{figure*}[h]
    \centering
    \includegraphics[scale=0.8]{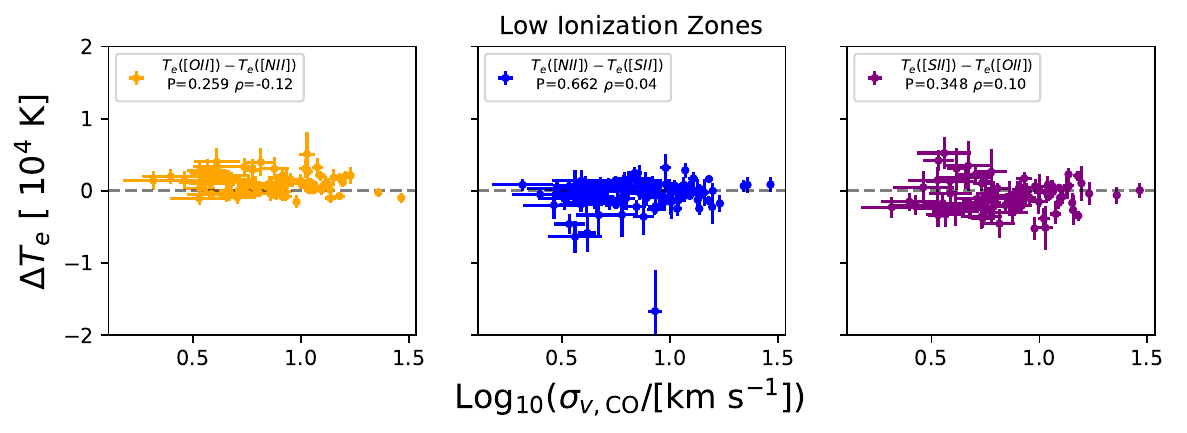}
    \includegraphics[scale=0.8]{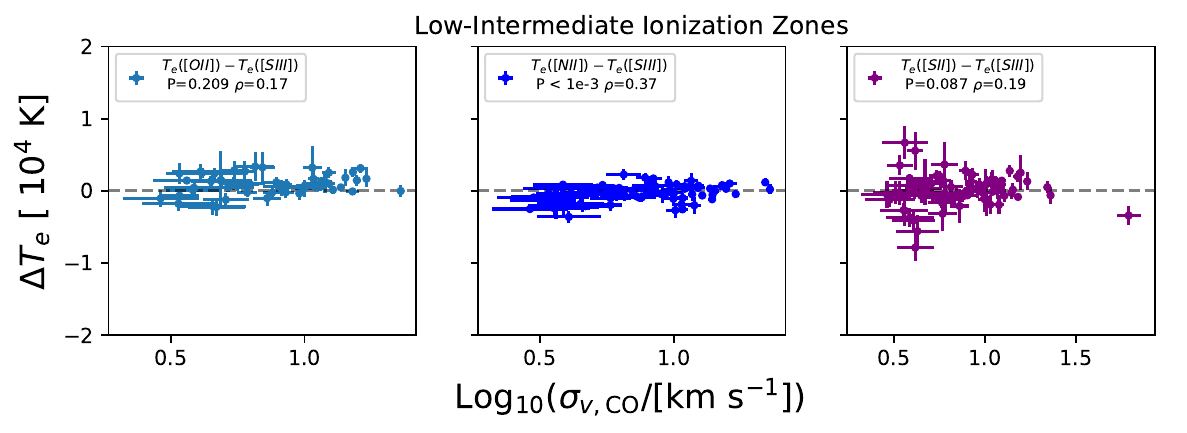}
    \includegraphics[scale=0.8]{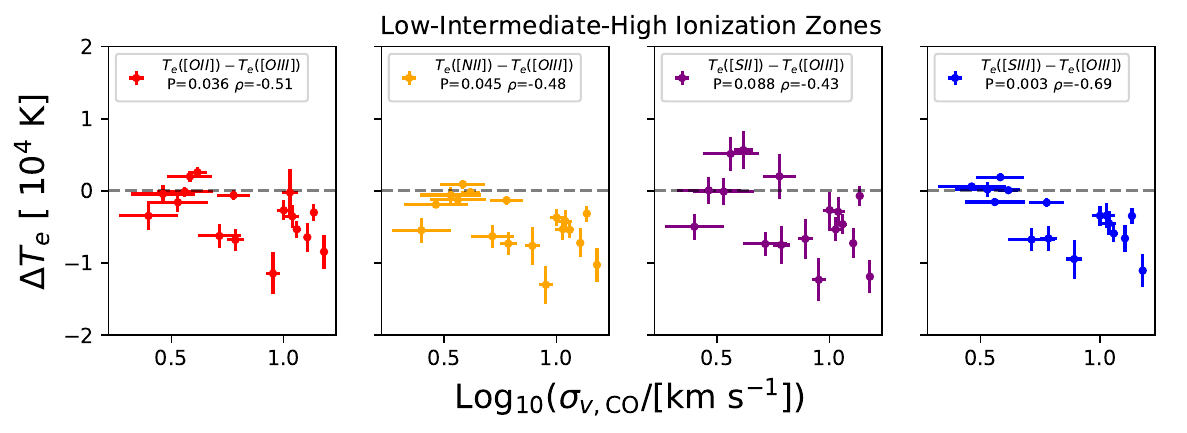}
    \caption{Electron temperature differences compared to the CO velocity dispersion, $\sigma_{v,\rm{CO}}$. The order of the panels follow those in Figure \ref{fig:ip_S_comparisons}.  We observe a weak correlation between \tempDiff{N}{2}{S}{3}. Although insignificant according to their p-values, we also observe that highest \temp{O}{3} values are associated with the highest $\sigma_{v,\rm{CO}}$.}
    \label{fig:rms_comparisons}
\end{figure*}

We show in Figure \ref{fig:rms_comparisons} the temperature differences compared to the CO velocity dispersion ($\sigma_{v, \mathrm{CO}}$). We observe a moderate correlation between $\sigma_{v, \mathrm{CO}}$ and \tempDiff{N}{2}{S}{3}. The values of \tempDiff{N}{2}{S}{3} over the range of $\sigma_{v, \mathrm{CO}}$ is small, only encompassing \tempDiff{N}{2}{S}{3} from $-$1000~ K to 2000~K. Given the possible correlation with association mass, and this correlation with $\sigma_{v, \mathrm{CO}}$, we find it intriguing that these correlations between the low and intermediate ionization zone are only seen in \tempDiff{N}{2}{S}{3}. It is possible that the low scatter in $T_{\rm{e}}$--$T_{\rm{e}}$ relationship between \temp{S}{3} and \temp{N}{2} in the presence of density inhomogeneities, contrary to the temperatures from [\ion{S}{2}] and [\ion{O}{2}], allow for better insight into these underlying trends.

We observe that all the correlations of $\Delta T_{\rm{e}}$ involving \temp{O}{3} and $\sigma_{v, \mathrm{CO}}$ have large, negative, $\rho$ values, but are all insignificant according to their p-values. The high Spearman rank values for the comparisons between $\Delta T_{\rm{e}}$ and $\sigma_{v, \mathrm{CO}}$ appears to be largely driven by the highest values of \temp{O}{3}. Despite the fact that the correlations are not strong, it is clear that the high \temp{O}{3} regions go along with high CO velocity dispersion.

The CO velocity dispersion can be enhanced by low-velocity shocks originating from the interaction of molecular gas with late-time-expansion of supernovae remnants \citep[see][]{Koo:2001_SN_CO,Zhou:2023arXiv}, as well as from interaction with low-velocity shocks from pressure-radiation driven \HII\ region expansion \citep{Hill1978HII_Shocks,Kothes:2002,Watkins2023CO_bubbles}. These low-velocity shocks are also predicted to enhance \temp{O}{3}. Shock modeling has shown that outward expanding low-velocity shocks can create conditions, such as pockets of high post-shock temperature, where the strength of [\ion{O}{3}]$\lambda4363$ emission will be enhanced compared to the little to no increase in emission from [\ion{O}{3}]$\lambda5007$  \citep{Peimbert1991shocks, Binette2012_OIII_shocks,MendezDelgado2021HerbigHaro}. The combination of large temperature differences with \temp{O}{3} and high $\sigma_{v, \mathrm{CO}}$ suggests that we may be observing the effects of low-velocity shocks. Motivated by this scenario, we search the \HII\ regions for evidence of low-velocity shocks in the following section.

\subsubsection{Investigating H~II Regions for Presence of Low-Velocity Shocks}
\label{res:high_o3}
The correlations of $\Delta T_{\rm{e}}$ involving the high ionization zone and \HII\ region properties: \SIIISII, \OIIIOII, and the CO velocity dispersion appears to be driven by the presence of regions with high \temp{O}{3}, high $\sigma_{\mathrm{v,CO}}$ and low $U$. One potential explanation is the presence of low velocity shocks enhancing \temp{O}{3} \citep{Peimbert1991shocks, Binette2012_OIII_shocks,MendezDelgado2021HerbigHaro}. To test this explanation, we search for evidence of shocks in enhanced optical line ratios and line broadening. 

\begin{figure}[h]
\centering
\includegraphics[scale=0.6]{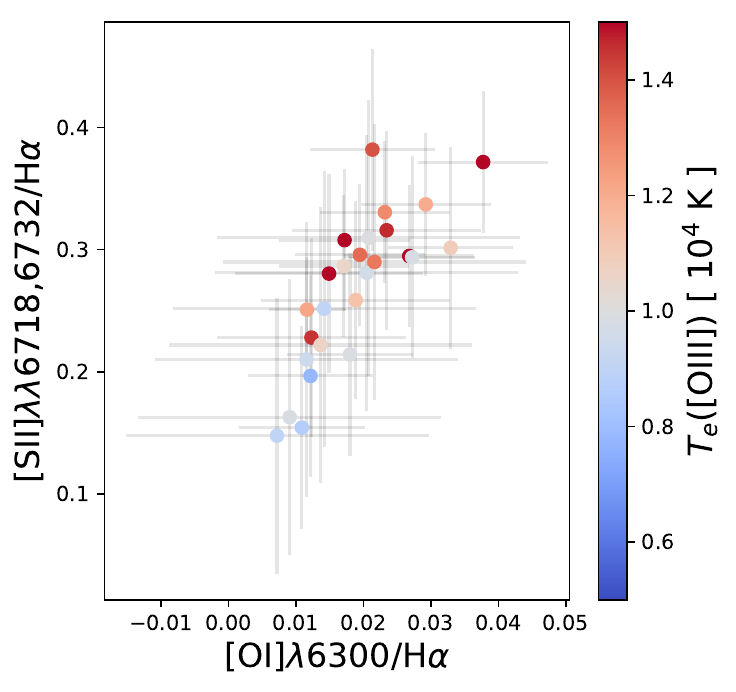}
\caption{We compare the relative strength of the low-ionization species [\ion{O}{1}] and [\ion{S}{3}] to Balmer emission. We color-code each data point by the value of its measured [\ion{O}{3}] temperature. The hottest \temp{O}{3} regions also have the highest line ratios. The higher line ratios are indicative of the existence of a partially-ionized zone that may be due to the presence of shocks or harder photons such as X-rays \citep{Grise:2008}.}
\label{fig:te_shocks}
\end{figure}

When shocks collide with and compress gas, the ionization parameter of the gas is reduced leading to partially ionized zones of enhanced nebular emission of low-ionization species such as S$^{+}$ and O relative to H$\beta$ and/or H$\alpha$ \citep{Dopita1996ApJS..102..161D,Allen2008ApJS..178...20A}. In Figure \ref{fig:te_shocks}, we show the [\ion{S}{2}]/H$\alpha$ vs. [\ion{O}{1}]/H$\alpha$ ratios, color-coded by \temp{O}{3}, for regions with measured [\ion{O}{3}]$\lambda4363$. We find that \HII\ regions with hotter \temp{O}{3} tend to populate a region with enhanced [\ion{S}{2}]/H$\alpha$ and [\ion{O}{1}]/H$\alpha$ ratios which suggests that these regions may host a partially ionized zone due to shocks. Between the two line ratios, \temp{O}{3} is better correlated with [\ion{S}{2}]/H$\alpha$, p-value$=0.0004$, than [\ion{O}{1}]/H$\alpha$, p-value$=0.0016$. We note here that it is possible that harder photons and X-rays produced by X-ray binaries also enhance [\ion{S}{2}] and [\ion{O}{1}] relative to the Balmer emission \citep{Abolmasov:2007, Grise:2008}. Furthermore, X-rays would provide high energy photons able to boost [\ion{O}{3}]$\lambda4363$. However, lacking the high spatial resolution X-ray imaging of these \HII\ regions, exploration of X-ray contributions to [\ion{O}{3}]$\lambda4363$ emission is beyond the scope of this paper.

Another tracer that may indicate presence of shocks is the \ion{He}{2} $\lambda4686$/H$\beta$ ratio \citep{Allen2008ApJS..178...20A} though it is also sensitive to the shape of the Lyman continuum below 228 \AA\  \citep{Garnett1991ApJ...373..458G, Guseva2000ApJ...531..776G, Allen2008ApJS..178...20A}. Wolf-Rayet (WR) stars are capable of releasing photons able to produce \ion{He}{2} $\lambda4686$. WR stars host stellar winds and can be a source of high energy photons, E$> 54$ eV, capable of doubly ionizing Helium. We measured the \ion{He}{2}$ \lambda4686$/H$\beta$ in \HII\ regions with \temp{O}{3} detections and found that these regions exhibit an average \ion{He}{2} $\lambda4686$/H$\beta = 4.1\pm1.6\%$. This value is within the expected range of \ion{He}{2} $\lambda4686$/H$\beta$ values for \HII\ regions with WR, 0.04\%-7\% \citep{Guseva2000ApJ...531..776G,Thuan2005ApJS..161..240T,Mayya2023MUSE}, and 100~km s$^{-1}$ shocks, 4\%-6\% \citep{Allen2008ApJS..178...20A}. We visually inspected the spectra of regions with measured \temp{O}{3} for the characteristic red/blue bump associated with the presence of WR stars. We found the blue bump in only 2 regions with measurable [\ion{O}{3}]$\lambda4363$, both of which have \temp{O}{3} $< 10^4$~K.
Nevertheless, the \ion{He}{2} $\lambda4686$/H$\beta$ for the high \temp{O}{3} regions lends evidence that these regions may host shocks or undetected WR stars.

We also searched for kinematic signatures of shocks, but find no clear kinematic evidence. Shocks can imprint asymmetries and/or broad emission near the base of an emission line. We inspected the fit residuals of [\ion{O}{3}]$\lambda5007$ in \HII\ regions with measured \temp{O}{3} and found no evidence of line broadening due to shocks. Next, we inspected the measured line widths of [\ion{O}{3}]$\lambda5007$ in these regions. We compared the line-spread-function corrected [\ion{O}{3}]$\lambda5007$, as measured by MUSE, velocity dispersion, $\sigma_{v,\lambda5007}$ vs. the CO velocity dispersion,  $\sigma_{v, \mathrm{CO}}$. The line-widths of the optical and CO emission for regions with high \temp{O}{3} are comparable to regions with low, or not detected \temp{O}{3}.  Despite this, the absence of these features may only exclude the presence of high-velocity shocks. 

If the regions do host low-velocity shocks, then their impact on the line width of the optical emission may be too small to be resolved given the MUSE resolution; $\sim 70$~km s$^{-1}$ at $\lambda=5007$ \AA. The high velocity resolution, $2.5~\mathrm{km} \mathrm{s}^{-1}$ of the PHANGS-ALMA data, are more sensitive than MUSE to low-velocity shocks, and are the strongest evidence for their presence in the high \temp{O}{3} \HII\ regions.

\section{Discussion}
\label{sec:DISCUSSION}
\subsection{Electron Density Inhomogeneities}
Similar to the results of \cite{Delgado_Desire2023arXiv230513136M}, described in Section \ref{results:density_inhomogeneities}, we have found that the presence of electron density inhomogeneities may cause the temperatures measured from [\ion{O}{2}] and [\ion{S}{2}] to be biased hotter compared to those measured using the [\ion{N}{2}] auroral lines. 

The critical densities of the nebular lines of [\ion{S}{2}]$\lambda\lambda6716,6731$ and more so [\ion{O}{2}]$\lambda\lambda3726,3729$ are low enough that at densities $n_{\rm{e}} > 10^3\ \mathrm{cm}^{-3}$ the lines will undergo increased collisional de-excitation. In the presence of density inhomogoneities above this value, the low critical densities will reduce the emissivity of the nebular emission lines from [\ion{S}{2}] and [\ion{O}{2}]. Because of this, the nebular diagnostic lines of [\ion{S}{2}] and [\ion{O}{2}] will mainly describe gas component with $n_{\rm{e}} < 10^3\ \mathrm{cm}^{-3}$. This biases the [\ion{S}{2}] and [\ion{O}{2}] density diagnostic to return electron densities that are lower than the true average density of the \HII\ region. Since the auroral line critical densities are far higher, this also makes the measured [\ion{S}{2}] and [\ion{O}{2}] temperatures appear to reflect hotter values.

This effect has been observed in many studies. Densities measured in Milky Way \HII\ regions (the Orion Nebula, NGC\,3604 and NGC\,3576) using [\ion{Cl}{3}], which is sensitive to greater densities than [\ion{S}{2}], routinely show that the [\ion{S}{2}] diagnostic returns lower values than [\ion{Cl}{3}].
\citep{PoggeOrion, GarciaRojas2007ApJ...670..457G,Nunez2013hsa7ORION, Weilbacher2015_MUSE_ORION}. Although densities derived from the [\ion{S}{2}] doublet are commonly used in the literature due to their strengths and their insensitivity to dust extinction, the [\ion{S}{2}]$\lambda6731/\lambda6716$ ratio is less sensitive to density than [\ion{Cl}{3}], [\ion{Ar}{4}], and [\ion{Fe}{3}] diagnostics when $n_{\rm{e}} > 10^3$ cm$^{-3}$ \citep[see Figure 2 in][]{Delgado_Desire2023arXiv230513136M}. If the equality \temp{O}{2}$\approx$\temp{S}{2}$\approx$\temp{N}{2} predicted by photoionization models is true, then the auroral-to-nebular line ratios would suggest a factor of 10 higher density than the [\ion{S}{2}] doublet, but neither may well represent the true average density of the region. 

Electron density variations may arise from shocks, turbulence, and pre-existing non-uniform structure in the ISM \citep{Hill1978HII_Shocks, Dopita1996ApJS..102..161D,Allen2008ApJS..178...20A}. \citet{JinFractal2022} have extended photoionization modeling of ionized nebulae to more complex geometries. Starting with an initial clumpy ISM, ionizing photons will pass through diffuse regions more readily than denser clumps. The resulting photoionized region will exhibit fluctuations in density and irregular geometry as opposed to a uniform density and spherical morphology. The ionization parameter in the dense clumps will be relatively lower than other regions of the nebulae due to the increased density. At these locations, the emission of low-ionization species, including [\ion{S}{2}] and [\ion{O}{2}], will be enhanced compared to higher ionization species. Due to the higher critical densities of the auroral lines of these ions, the emissivity of the auroral lines will be greater than those of the nebular lines in the high-density portion the nebula. In this scenario, the nebular density diagnostics can return an average density that traces the low-density portion of the nebula. In doing so, the value of $n_{\rm{e}}$ returned by the nebular diagnostics may inaccurately describe the ionizing conditions of the high-density clumps where the auroral line emissivities are greater than the nebular lines, and overestimate [\ion{S}{2}] temperatures. 

Density inhomogeneities have been reported in studies of highly resolved local HII regions like Orion where the inhomogeneities can be spatially resolved \citep{Baldwin_Orion:1991,PoggeOrion,Weilbacher2015_MUSE_ORION, McLeod2016Orioin, ODell_Orion:2017}.  \citet{Weilbacher2015_MUSE_ORION} mapped the spatial variation of density in the Orion nebulae and found variations of density between 500~cm$^{-3}$ and in excess of 10,000 cm$^{-3}$. A maximum of 25,000~cm$^{-3}$ is measured using [\ion{Cl}{3}] at the location of the ionization front in the ``Orion S'' area. Density inhomogeneities, associated with turbulence driven velocity fluctuations, have been invoked as one mechanism to generate surface brightness fluctuations within the Orion Nebula \citep{Kainulainen2017ORION}. 

The Orion Nebula is not particularly comparable to the \HII\ regions studied here due to the difference in scales (i.e. Orion is more compact) and resolution. 
A subset of our \HII\ regions, as shown in the comparison between the MUSE \HII\ regions mask in Appendix \ref{appn:comparison_of_muse_and_kcwi_hii_regions}, are unresolved clusters of individual regions.
Measurements of density inhomogeneities using density diagnostics besides the nebular [\ion{O}{2}] and [\ion{S}{2}] doublets for extragalactic \HII\ regions that more closely match our sample are rare and require deep, high $S/N$ spectra \citep{Delgado_Desire2023arXiv230513136M}. One consequence of the lack of different diagnostics is that many studies will often assume a fixed density of $100$ cm$^{-3}$ when either [\ion{S}{2}], or [\ion{O}{2}], return a density in the low-density limit \citep[e.g.][]{Kreckel2019ApJ...887...80K}. Studies using mid-infrared observations have shown this latter assumption could be incorrect, as densities up to $1000~\mathrm{cm}^{-3}$ have been measured using the [\ion{S}{3}]$\lambda 18.7/33.5\ \mu$m density diagnostic \citep[see][]{Rubin:2016}, and indicate that density inhomogeneities are present in extra-galactic \HII\ regions.

The consistency of auroral-to-nebular ratios with $n_{\rm{e}}\sim1000$~cm$^{-3}$ assuming $T_{\rm{e}}=$ \temp{N}{2} in Figure \ref{fig:ne_result}, as well as the correlation between degree of inhomogeneities with H$\beta$ surface brightness observed for [\ion{S}{2}], supports this picture that inhomogeneities must be considered when deriving temperatures from the [\ion{O}{2}] and [\ion{S}{2}]. As a result, we consider \temp{N}{2} a more reliable indicator of the low-ionization zone temperature due to its relative insensitivity to density.

\subsection{\temp{N}{2} and \temp{S}{3} as Accurate Tracers of HII Region Temperatures}
\label{dis:nii_siii}
Within the set of $T_{\rm{e}}$--$T_{\rm{e}}$ between the low and intermediate ionization zones, we observed that the comparisons between \temp{S}{3}--\temp{O}{2} and \temp{S}{3}--\temp{S}{2} largely agree with those from \cite{Zurita2021MNRAS.500.2359Z}.
However, similar to recent studies \cite{Berg2015ApJ...806...16B,Berg2020ApJ...893...96B,Zurita2021MNRAS.500.2359Z,Rogers:chaos:2021} and \cite{Rogers:chaos:2022}, we observed that the $T_{\rm{e}}$--$T_{\rm{e}}$ trend between \temp{N}{2} and \temp{S}{3} exhibits the lowest scatter and agrees with many of the literature trends presented in Figure \ref{fig:temps_intermediate_gals}. These results suggest \temp{N}{2} and \temp{S}{3} temperatures are optimal tracers of \HII\ region temperatures, more so than \temp{S}{2}, \temp{O}{2} and \temp{O}{3}. 

Judged from the Oxygen CEL and RL temperatures, the high ionization zone is expected to be most affected by temperature fluctuations due to its proximity to sources of feedback \citep{Delgado2023arXiv230511578M}. To what degree the intermediate ionization zone temperatures are affected is unclear \citep{Delgado_Desire2023arXiv230513136M}. In Figure \ref{fig:temps_intermediate_gals}, we showed that our best-fit trend, including those from the literature, between \temp{N}{2} and \temp{S}{3} ``generally'' (within 2$\sigma$) follow trends predicted from the \citetalias{BOND} photoionization models with temperature fluctuations set to zero. 
This suggests that \temp{S}{3} may be minimally affected by temperature inhomogeneities. \cite{Diaz2022MNRAS_sulphur} provide additional arguments for the use of \temp{S}{3} over \temp{O}{3}, including: 1) the emission lines of [\ion{S}{3}] have a lower exponential dependence on electron temperature and 2) because [\ion{S}{3}] overlaps gas volumes containing both $\mathrm{O}^{2+}$ and $\mathrm{O}^{+}$ that \temp{S}{3} can be representative of the entire \HII\ region.

We also showed that \tempDiff{N}{2}{S}{3} is stable around the zero line for across the range of molecular gas velocity dispersion and the low values of ionization parameter, traced by both \OIIIOII\ and \SIIISII, observed in this work. This result follows the observations of \cite{Berg2020ApJ...893...96B}, whom find that agreement between these two temperature tracers become more uncertain in high ionization parameter, traced by \OIIIOII, \HII\ regions. Due to stable behavior of between \temp{N}{2} and \temp{S}{3} across the multiple \HII\ region properties observed in this work, and following the suggestions of \cite{Berg2015ApJ...806...16B,Berg2020ApJ...893...96B} and \cite{Rogers:chaos:2021}, we will compare ``direct'' metallicities derived prioritizing [\ion{N}{2}] and [\ion{S}{3}] temperatures to several calibrated methods (Rickards Vaught et al., in prep).

\subsection{The High-Ionization Zone Temperature Excess}
\label{disc:o3_te_excess}
Within the sample of \HII\ regions with measured [\ion{O}{3}]$\lambda4363$, there are a small number of \HII\ regions with high \temp{O}{3}, enhanced velocity dispersion in the surrounding molecular gas, and low-ionization parameter. We investigated these \HII\ regions for enhanced low-ionization species emission line ratios [\ion{S}{2}]/H$\alpha$, [\ion{O}{1}]/H$\alpha$ and the high-ionization ratio \ion{He}{2}/H$\beta$. We found that the high \temp{O}{3} regions exhibited enhanced low-ionization line ratios suggestive of shock ionization (or possibly X-ray ionization). We also found that the \ion{He}{2}/H$\beta$ ratios for these regions are within the range of those expected from shock velocities $<100$ km s$^{-1}$ \citep{Allen2008ApJS..178...20A} and/or WR stars \citep{Guseva2000ApJ...531..776G,Thuan2005ApJS..161..240T,Mayya2023MUSE}, though we only find characteristic red/blue bumps associated with the presence of WR stars, in 2 \HII\ regions neither of which had elevated \temp{O}{3}. Absent the WR signatures, we are motivated to explore shocks as enhancers of the [\ion{O}{3}] temperature.

We did not find any kinematic signatures of shocks in the optical emission line profiles. We determine that if shocks are present and broadening the CO emission, then the shock velocities are too low to be resolved by the MUSE spectral resolution. Either way, high-velocity shocks are not expected to effectively boost \temp{O}{3} \citep{MendezDelgado2021HerbigHaro}.
Despite the uncertainty in whether shocks are present in the high [\ion{O}{3}] temperature regions, we can discuss the plausibility that low-velocity shocks are the cause for excess \temp{O}{3}. 

\subsubsection{Shock-enhanced [\ion{O}{3}] Temperature}
Low velocity, $<$ 100 km s$^{-1}$, shocks can increase [\ion{O}{3}]$\lambda4363$ while leaving [\ion{O}{3}]$\lambda5007$ unchanged \citep{Peimbert1991shocks,Binette2012_OIII_shocks,MendezDelgado2021HerbigHaro}. For a sample of giant \HII\ regions, \cite{Binette2012_OIII_shocks} measured \temp{O}{3} up to 6000 K higher than \temp{S}{3} in regions with \temp{S}{3} $< 10^4$~K. To explore if shocks were boosting their [\ion{O}{3}] temperatures relative to [\ion{S}{3}], they model outward expanding shocks, mimicking those generated by stellar winds, by combining shock$+$photoionization models, with increasing shock velocities (analogous to increasing the post-shock temperature from 1.6$\times 10^4$~K to 7.2$\times 10^4$~K) between 20 and 60~km s$^{-1}$. Additionally, the models span 5 different metallicities between Z$=$0.01~Z$_{\odot}$ to Z$=$1.6~Z$_{\odot}$. Comparing the average properties of between the lowest and highest shock velocity models, \cite{Binette2012_OIII_shocks} found that mean doubly ionized fraction of oxygen, $\mathrm{O}^{2+}$/$\mathrm{O}$ , decreases while at the same time leaving the doubly ionized fraction of sulfur, $\mathrm{S}^{2+}$/$\mathrm{S}$, unchanged. The imbalance of the ionization fraction between oxygen and sulfur means that hotter, post-shock gas contributes proportionally more to the observed [\ion{O}{3}] emission than [\ion{S}{3}]. Because of the exponential sensitivity to temperature, the [\ion{O}{3}] auroral line will be enhanced, tracing the hotter post-shock temperature rather than the local photoionized nebula temperature returned by \temp{S}{3}. Furthermore, the highest metallicity models show the largest temperature differences between \temp{S}{3} and \temp{O}{3}, up to $\sim 7000$~K. Although this difference is 3000~K lower than our largest measured \tempDiff{S}{3}{O}{3}, a complete understanding of the degree of enhancement of [\ion{O}{3}]$\lambda4363$ with shocks require more complex 3D hydro-dynamical simulations \citep{Binette2012_OIII_shocks}. Despite this, one extreme example of shock impact on electron temperatures was observed in the outflow of the \HII\ region Sh 2-129 \citep{Corradi2014Shocks_outflows}. This outflow, with velocity $\approx 100$ $\mathrm{km} \mathrm{s}^{-1}$, exhibits \temp{O}{3} $= 55,000$ K and \temp{O}{2} $\sim 20,000$ K. Much larger than $\Delta T_{\rm{e}}$ observed in this study.

It has also been shown that shocks driven by pressure--radiation \HII\ region expansion, and SNR, impact the surrounding cold molecular and ionized gas. If the velocity of the expansion is greater than the sound speed ionized gas $\sim 10$ km s$^{-1}$, a layer of shocked H gas will form in-between the expanding ionization front and a surrounding molecular gas \citep{ Hill1978HII_Shocks,Kothes:2002, Tremblin:2014,Watkins2023CO_bubbles}. The impact of shock interaction with molecular gas has been studied in 18 galaxies observed as part of PHANGS-ALMA \citep{2021Leroy_sample, 2021Leroy_data}, where \cite{Watkins2023CO_bubbles} identified hundreds of super bubbles (i.e. pockets of expanding gas arising as byproducts of feedback). The superbubbles were identified using spatial correspondence between CO shells and stellar populations contained in PHANGS-HST catalogs \citep{2022PHANGS_HST, ThilkerCompactClusters, Larson2023}. Due to the ALMA spatial resolution (50--150 pc), \cite{Watkins2023CO_bubbles} measured the expansion velocity only for the largest superbubbles. Assuming the CO expansion velocity is equal to the shock velocity, \cite{Watkins2023CO_bubbles} measure the velocity of the approaching/receding CO shells and determine an average, line-of-sight, expansion velocity of $v_{\rm exp}=$9.8 $\pm$ 4.3~km~s$^{-1}$. Although this value is similar to the sound speed of the ionized gas, superbubbles exhibit asymmetries in their morphology, and the velocity can potentially reach up to a few tens of km s$^{-1}$ depending on the conditions of the gas, source or energy injection, and age \citep{Watkins2023CO_bubbles}. 1D models indicate that, at minimum, \HII\ regions exhibit expansion speeds of a $\sim$few~km~s$^{-1}$ \citep{Tremblin:2014}. This suggests that \HII\ regions expand with a large range of velocities.

As for ionized gas, \cite{Egorov2023} identify in the PHANGS-MUSE galaxies more than 1400 regions of ionized gas with elevated intrinsic H$\alpha$ velocity dispersions $> 45$~km~s$^{-1}$, and, under the assumption that these regions are undergoing expansion, \cite{Egorov2023} infer expansion velocities between $v_{\rm{ exp}}=$ 10--40 km s$^{-1}$ \citep[see also][]{Egorov:2014,Egorov:2017, Cosens2022ApJ...929...74C}. The ubiquity of  \HII\ region expansion, as well as their effects on the surrounding molecular gas, make them good candidates for drivers of low-velocity shocks capable of boosting [\ion{O}{3}] temperatures.

\subsubsection{Temperature Inhomogeneities}
Temperature fluctuations within \HII\ regions are another potential explanation for the high [\ion{O}{3}] temperature. As discussed in Section \ref{sec:intro}, RL emissivities have a linear sensitivity to temperature rather than the exponential sensitivity of CELs. In the presence of temperature fluctuations CELs will return higher estimates of temperature. \cite{Delgado2023arXiv230511578M} have recently shown that differences between [\ion{O}{3}] and [\ion{N}{2}] temperatures  are strongly correlated with the temperature fluctuations parameter, $t^2$, of the highly ionized gas. This suggests that \temp{O}{3} is likely to overestimate the representative \HII\ region temperature. The observed excess in \temp{O}{3} are likely to be produced by phenomena other than those commonly observed in \HII\ regions. Simple models from \cite{Binette2012_OIII_shocks} and the observations from \cite{Delgado2023arXiv230511578M} show that the effect from $t^2>0$ is more pronounced in lower-metallicity/high ionization parameter regions. Furthermore, the temperature excess \temp{O}{3} relative to the other ionization zones is too large in our observations to be caused solely by inhomogeneities. Using our measured \tempDiff{N}{2}{O}{3}, we can infer from \cite{Delgado2023arXiv230511578M} that $t^2 > 0.2$ for \tempDiff{N}{2}{O}{3} $>5000$~K. These values are much higher than what has been observed for nearby \HII\ regions \citep{PPP2012_t2, Binette2012_OIII_shocks, Delgado2023arXiv230511578M}. Even in the presence of classical temperature inhomogeneities, a secondary effect would also need to be included to explain our [\ion{O}{3}] temperatures.

\subsubsection{Potential Observation Bias}
Another possibility is that the high [\ion{O}{3}] temperatures are statistical outliers. The galaxies in our sample exhibit strong line oxygen abundances $\gtrsim$  8.3 \citep{Kreckel2019ApJ...887...80K}. Because electron temperature is anti-correlated to the metallicity of the gas, the [\ion{O}{3}]$\lambda4363$ temperatures for these galaxies are expected to be low. For [\ion{O}{3}]$\lambda4363$ to be detectable, the temperature would need to be high, otherwise, we would likely not detect the auroral line. [\ion{O}{3}] temperatures from a lower-metallicity sample may be compatible with photoionization models and $T_{\rm{e}}$--$T_{\rm{e}}$ relations. Finally, \cite{Rola1994A&A_SN} have shown that emission line measurements with S/N $< 5$ can potentially overestimate the true intensity by $80\%$. The average S/N [\ion{O}{3}]$\lambda4363$ measured from our sample is $\sim 4.5$. Nevertheless, why we would measure high [\ion{O}{3}]$\lambda4363$, even after removal of [\ion{Fe}{2}]$\lambda4360$ contamination, that also exhibit enhanced molecular gas velocity dispersions is difficult to explain purely with statistical outliers. 

\section{Conclusions}
We presented combined KCWI and MUSE observations of the [\ion{N}{2}]$\lambda5756$, [\ion{O}{2}]$\lambda\lambda7320,7330$, [\ion{S}{2}]$\lambda\lambda4069,4076$, [\ion{O}{3}]$\lambda4363$, and [\ion{S}{3}]$\lambda6312$ auroral lines in a sample of 421 \HII\ regions in 7 nearby galaxies. We compared the derived electron temperatures and temperature differences between multiple \HII\ region ionization zones to several \HII\ region properties such as electron density, ionization parameter, molecular gas velocity dispersion, stellar mass, and age obtained from PHANGS observations. We found that:

\begin{itemize}
    \item Similar to the results from \cite{Delgado_Desire2023arXiv230513136M}, temperatures obtained from [\ion{S}{2}] and [\ion{O}{2}] are consistent with being overestimated due to the presence of density inhomogeneities in the \HII\ regions. Because of these potential biases, we recommend the use [\ion{N}{2}] temperatures to trace the low ionization zone.
    
    \item In addition to previous studies: \cite{Berg2015ApJ...806...16B,Berg2020ApJ...893...96B,Rogers:chaos:2021}, and \cite{Zurita2021MNRAS.500.2359Z}, we found that the [\ion{N}{2}] and [\ion{S}{3}] temperatures exhibited the lowest scatter of the $T_{\rm{e}}$-- $T_{\rm{e}}$ relations and follow trends predicted from photoionization models. The well-behaved relationship between [\ion{N}{2}] and [\ion{S}{3}], even in potentially inhomogeneous conditions, may be better tracing the underlying \HII\ region temperatures. This result, and those from the above studies, further stress the prioritization of [\ion{N}{2}] and [\ion{S}{3}] temperatures for metallicity determinations.

    \item We observed a subset of \HII\ regions with high [\ion{O}{3}] temperatures that do not agree with the cooler temperatures measured in the low and intermediate ionization zones. We found that the regions with high [\ion{O}{3}] temperature tended to have enhanced molecular gas velocity dispersion and lower ionization parameter than those regions with [\ion{O}{3}] temperatures that were in better agreement with other ionization zones. These regions also showed enhanced [\ion{S}{2}]/H$\alpha$, [\ion{O}{1}]/H$\alpha$, and \ion{He}{2}/H$\beta$ ratios indicating the presence of secondary ionization sources (e.g. shocks, Wolf-Rayet and X-ray binary stars). Absent direct detection of shocks, we explored whether or not shocks are able to both enhance \temp{O}{3} and CO velocity dispersion. We found that low-velocity shocks are a plausible explanation for the observed  [\ion{O}{3}] temperatures and CO velocity dispersions. However, disentangling the effects of shocks from possible contributions to [\ion{O}{3}] temperatures by harder ionization sources such as Wolf-Rayet stars or X-ray binaries require further investigation. 

    \item We also explored temperature inhomogenities and observational uncertainties as causes for high [\ion{O}{3}] temperatures measured for a small sub-sample of H~II regions. We found that the degree of temperature inhomogeneity that would be required to produce the difference between high [\ion{O}{3}] temperatures and those of the low and intermediate ionization zone are larger than what has been observed in most star-forming regions. Furthermore, if the regions with high [\ion{O}{3}] temperatures are statistical outliers leading to overestimated temperatures, we lack an explanation as to why these temperatures would correlate with high molecular gas velocity dispersion. 
\end{itemize}

In a follow-up paper, Rickards Vaught et al. (in prep), We will test temperature recommendations for measuring ``direct" metallicities using our full set of measured auroral line temperatures. This work, along with the PHANGS-MUSE survey, demonstrate the power of integral field spectrographs on 10m class telescopes for measuring faint auroral emission lines from large samples of \HII\ regions in nearby galaxies.  Future efforts with deeper observations or expanded samples will be critical for further elucidating the temperature and ionization behavior of these regions, particularly as [\ion{O}{3}]$\lambda4363$ and other faint lines are now being routinely detected in galaxies at high redshift with JWST and used in metallicity determinations.

\section*{Acknowledgements}
The authors thank the referee for a very thorough report which significantly improved the analysis presented here. The data presented herein were obtained at the W. M. Keck Observatory, which is operated as a scientific partnership among the California Institute of Technology, the University of California and the National Aeronautics and Space Administration. The Observatory was made possible by the generous financial support of the W. M. Keck Foundation. The authors wish to recognize and acknowledge the very significant cultural role and reverence that the summit of Maunakea has always had within the indigenous Hawaiian community.  We are most fortunate to have the opportunity to conduct observations from this mountain. We also wish to thank all the Keck Observatory staff, including Gregg Doppman, Luca Rizzi, Sheery Yeh, and Rosalie McGurk for observational support. 

This research is also based on observations collected at the European Southern Observatory under ESO programmes 094.C-0623 (PI: Kreckel), 095.C-0473,  098.C-0484 (PI: Blanc), 1100.B-0651 (PHANGS-MUSE; PI: Schinnerer), as well as 094.B-0321 (MAGNUM; PI: Marconi), 099.B-0242, 0100.B-0116, 098.B-0551 (MAD; PI: Carollo) and 097.B-0640 (TIMER; PI: Gadotti). 

RRV and KS acknowledge funding support from National Science Foundation Award No.\ 1816462.
FB acknowledges funding from the INAF Fundamental Astrophysics 2022 program. 
RRV wishes to thank Jonah Gannon, and Maren Consens for fruitful discussions on reducing the KCWI data. RRV would also like to acknowledge discussions with JEM-D, KK, OVE which greatly improved the manuscript.
KK, JEM-D and OVE gratefully acknowledge funding from the Deutsche Forschungsgemeinschaft (DFG, German Research Foundation) in the form of an Emmy Noether Research Group (grant number KR4598/2-1, PI Kreckel).
KG is supported by the Australian Research Council through the Discovery Early Career Researcher Award (DECRA) Fellowship (project number DE220100766) funded by the Australian Government. 
KG is supported by the Australian Research Council Centre of Excellence for All Sky Astrophysics in 3 Dimensions (ASTRO~3D), through project number CE170100013. 
JN acknowledges funding from the European Research Council (ERC) under the European Union’s Horizon 2020 research and innovation programme (grant agreement No. 694343).
RSK acknowledges funding from the European Research Council via the ERC Synergy Grant ``ECOGAL'' (project ID 855130), from the German Excellence Strategy via the Heidelberg Cluster of Excellence (EXC 2181 - 390900948) ``STRUCTURES'', and from the German Ministry for Economic Affairs and Climate Action in project ``MAINN'' (funding ID 50OO2206). He also thanks for computing resources provided by {\em The L\"{a}nd} and DFG through grant INST 35/1134-1 FUGG and for data storage at SDS@hd through grant INST 35/1314-1 FUGG.
G.A.B. acknowledges the support from the ANID Basal project FB210003. 

This research made use of Montage. It is funded by the National Science Foundation under Grant Number ACI-1440620, and was previously funded by the National Aeronautics and Space Administration's Earth Science Technology Office, Computation Technologies Project, under Cooperative Agreement Number NCC5-626 between NASA and the California Institute of Technology. This research has made use of NASA's Astrophysics Data System. This research made use of Astropy, a community-developed core Python package for Astronomy \citep{Astropy}, as well as substantial use of the nebular diagnostics toolkit Pyneb \citep{Luridiana2015A&A...573A..42L}.

The distances in Table \ref{tab:sample_properties} were
compiled by \citet{Anand2021MNRAS.501.3621A} and are based on \cite{Freedman:2001}; \cite{Nugent:2006}; \cite{Jacobs:2009}; \cite{Kourkchi:2017}; \cite{Shaya:2017}; \cite{Kourkchi:2020}; \cite{Anand2021MNRAS.501.3621A}; \cite{Scheuermann:2022}

\bibliography{main}{}
\bibliographystyle{aasjournal}


\appendix

\section{KCWI Seeing FWHM from Standard Star Observations and Table of Observations}
\label{appn:observations}
We summarize the details of all of the KCWI observations and standard star observations in Table \ref{tab:obs} and \ref{tab:std_obs}. Additionally, we describe below the measurement of seeing from the set of standard stars.

To measure seeing, we fit a 2D Gaussian to each standard star observation. We find an average FWHM of $\sim 1.2$\arcsec, however, at some points during the nights of 10-17-2018, 03-27-2019, and 03-28-2019 the seeing was poorer with values between 1.6\arcsec-2\arcsec. Aside from these portions of the nights, the seeing was stable near the average FWHM. The FWHM measurements for each standard star observation are summarized in Table \ref{tab:std_obs}.

\startlongtable
\begin{deluxetable*}{cccccccccc}
\tablecaption{\textit{Summary of KCWI Observations}. This table reports the the galaxy and the field observed (Field). Note the repeated fields are offset by 1/2 slice width, with the exception of all the fields in NGC3627. The table also reports: the original coordinates of the field center in right ascension and declination (RA, DEC), the applied astrometric correction in right ascension and declination (RA$_{\mathrm{corr}}$, DEC$_{\mathrm{corr}}$) determined from the image registration, the UTC start-time and total integration the exposure in seconds (Date, Exposure) as well as the airmass (Airmass), position angle (Angle) and the angular FWHM of the pointing.}
\label{tab:obs}
\tablehead{\colhead{Field\tablenotemark{a}} & \colhead{RA}  & \colhead{RA$_{\mathrm{corr}}$\tablenotemark{b}} &   \colhead{Dec}  & \colhead{Dec$_{\mathrm{corr}}$\tablenotemark{b}} & \colhead{Date} & \colhead{Exposure} & \colhead{Airmass} & \colhead{PA} & \colhead{FWHM\tablenotemark{c}} \\ 
 \colhead{} & \colhead{($^{\circ}$)} & \colhead{($\arcsec$)}  & \colhead{($^{\circ}$)} & \colhead{($\arcsec$)} & \colhead{(UTC)} & \colhead{(s)} & \colhead{} & \colhead{($^{\circ}$)} & \colhead{($''$)} }
\startdata
NGC0628F17 &  24.188826 &    0.16 &  15.771491 &    -1.75 &  2018-10-17T07:00:46 &    1200.0 &     1.46 &        45.0 &    2.14 \\
 NGC0628F17 &  24.189230 &   -1.30 &  15.771380 &    -1.35 &  2018-10-17T07:35:05 &    1200.0 &     1.28 &        45.0 &    1.79 \\
 NGC0628F17 &  24.189230 &   -1.30 &  15.771380 &    -1.35 &  2018-10-17T07:56:17 &    1200.0 &     1.20 &        45.0 &    1.97 \\
 NGC0628F18 &  24.189057 &   -0.68 &  15.771546 &    -1.95 &  2018-10-17T08:29:45 &    1200.0 &     1.11 &        45.0 &    2.49 \\
 NGC0628F18 &  24.188941 &   -0.26 &  15.771657 &    -2.35 &  2018-10-17T08:51:03 &    1200.0 &     1.07 &        45.0 &    2.49 \\
 NGC0628F22 &  24.189576 &   -2.55 &  15.771268 &    -0.95 &  2018-10-17T09:29:26 &    1200.0 &     1.02 &        45.0 &    2.20 \\
 NGC0628F22 &  24.189461 &   -2.13 &  15.771380 &    -1.35 &  2018-10-17T09:50:42 &    1200.0 &     1.01 &        45.0 &    2.49 \\
 NGC0628F23 &  24.189519 &   -2.34 &  15.770380 &     2.25 &  2018-10-17T10:24:20 &    1200.0 &     1.00 &        45.0 &    1.49 \\
 NGC0628F23 &  24.189634 &   -2.75 &  15.770324 &     2.45 &  2018-10-17T10:45:51 &    1200.0 &     1.01 &        45.0 &    1.60 \\
  NGC0628F04 &  24.190558 &   -6.08 &  15.770213 &     2.85 &  2018-10-16T10:08:04 &    1200.0 &     1.00 &        45.0 &    2.48 \\
 \hline
 NGC1087F10 &  41.606106 &   -1.05 & -0.494058 &    -0.15 &  2018-10-17T11:23:24 &    1200.0 &     1.07 &        90.0 &    1.26 \\
 NGC1087F10 &  41.606106 &   -1.05 & -0.494058 &    -0.15 &  2018-10-17T11:44:36 &    1200.0 &     1.07 &        90.0 &    1.29 \\
 NGC1087F09 &  41.605773 &    0.15 & -0.494446 &     1.25 &  2018-10-17T12:17:46 &    1200.0 &     1.09 &        90.0 &    1.19 \\
 NGC1087F09 &  41.605773 &    0.15 & -0.494391 &     1.05 &  2018-10-17T12:38:58 &    1200.0 &     1.12 &        90.0 &    1.15 \\
 NGC1087F08 &  41.605884 &   -0.25 & -0.494391 &     1.05 &  2018-10-17T13:12:00 &    1200.0 &     1.19 &        90.0 &    1.16 \\
 NGC1087F08 &  41.605884 &   -0.25 & -0.494446 &     1.25 &  2018-10-17T13:33:09 &    1200.0 &     1.25 &        90.0 &    1.27 \\
 NGC1087F04 &  41.606050 &   -0.85 & -0.494169 &     0.25 &  2018-10-17T14:06:31 &    1200.0 &     1.39 &        90.0 &    1.21 \\
 NGC1087F04 &  41.606050 &   -0.85 & -0.494113 &     0.05 &  2018-10-17T14:27:51 &    1200.0 &     1.52 &        90.0 &    1.31 \\
 NGC1087F11 &  41.605884 &   -0.25 & -0.494391 &     1.05 &  2018-10-16T13:30:01 &    1200.0 &     1.23 &        90.0 &    1.24 \\
 NGC1087F11 &  41.605773 &    0.15 & -0.494391 &     1.05 &  2018-10-16T13:08:26 &    1200.0 &     1.17 &        90.0 &    1.16 \\
 NGC1087F12 &  41.606939 &   -4.05 & -0.495113 &     3.65 &  2018-10-16T11:36:44 &    1200.0 &     1.07 &        90.0 &    1.07 \\
 NGC1087F12 &  41.606050 &   -0.85 & -0.494724 &     2.25 &  2018-10-16T12:11:21 &    1200.0 &     1.08 &        90.0 &    1.19 \\
 NGC1087F12 &  41.606050 &   -0.85 & -0.494669 &     2.05 &  2018-10-16T12:32:55 &    1200.0 &     1.11 &        90.0 &    1.19 \\
 \hline 
 NGC1300F1 &  49.903140 &   -2.39 & -19.400675 &     1.25 &  2021-10-07T10:58:14 &    1200.0 &     1.44 &        90.0 &    1.14 \\
 NGC1300F1 &  49.903140 &   -2.39 & -19.400786 &     1.65 &  2021-10-07T11:19:16 &    1200.0 &     1.38 &        90.0 &    1.50 \\
 NGC1300F2 &  49.903022 &   -1.96 & -19.401008 &     2.45 &  2021-10-07T11:52:07 &    1200.0 &     1.32 &        90.0 &    1.14 \\
 NGC1300F2 &  49.903022 &   -1.96 & -19.400953 &     2.25 &  2021-10-07T12:13:05 &    1200.0 &     1.30 &        90.0 &    1.33 \\
 NGC1300F3 &  49.902963 &   -1.75 & -19.401619 &     4.65 &  2021-10-07T12:45:25 &    1200.0 &     1.29 &        90.0 &    1.34 \\
 NGC1300F3 &  49.902963 &   -1.75 & -19.401508 &     4.25 &  2021-10-07T13:06:26 &    1200.0 &     1.30 &        90.0 &    1.20 \\
 NGC1300F4 &  49.903788 &   -4.72 & -19.401341 &     3.65 &  2021-10-07T13:42:40 &    1200.0 &     1.35 &        90.0 &    1.04 \\
 NGC1300F4 &  49.903729 &   -4.51 & -19.401341 &     3.65 &  2021-10-07T14:03:37 &    1200.0 &     1.40 &        90.0 &    1.03 \\
 NGC1300F5 &  49.903611 &   -4.08 & -19.401730 &     5.05 &  2021-10-07T14:36:25 &    1200.0 &     1.51 &         0.0 &    1.33 \\
 NGC1300F5 &  49.903552 &   -3.87 & -19.401730 &     5.05 &  2021-10-07T14:57:23 &    1200.0 &     1.62 &         0.0 &    1.34 \\
 \hline
 NGC1385F1 &  54.372099 &   -2.25 & -24.496321 &     1.65 &  2021-10-06T10:52:05 &    1200.0 &     1.70 &        90.0 &    1.52 \\
 NGC1385F1 &  54.372099 &   -2.25 & -24.496321 &     1.65 &  2021-10-06T11:13:46 &    1200.0 &     1.59 &        90.0 &    1.43 \\
 NGC1385F2 &  54.372038 &   -2.03 & -24.496265 &     1.45 &  2021-10-06T11:50:22 &    1200.0 &     1.47 &        90.0 &    1.26 \\
 NGC1385F2 &  54.372038 &   -2.03 & -24.496321 &     1.65 &  2021-10-06T12:11:22 &    1200.0 &     1.43 &        90.0 &    1.20 \\
 NGC1385F3 &  54.371916 &   -1.59 & -24.496432 &     2.05 &  2021-10-06T12:44:09 &    1200.0 &     1.40 &        90.0 &    1.06 \\
 NGC1385F3 &  54.371916 &   -1.59 & -24.496487 &     2.25 &  2021-10-06T13:05:09 &    1200.0 &     1.39 &        90.0 &    1.35 \\
 NGC1385F4 &  54.371794 &   -1.15 & -24.496265 &     1.45 &  2021-10-06T13:38:12 &    1200.0 &     1.42 &        90.0 &    1.16 \\
 NGC1385F4 &  54.371794 &   -1.15 & -24.496265 &     1.45 &  2021-10-06T13:59:12 &    1200.0 &     1.45 &        90.0 &    1.06 \\
 NGC1385F5 &  54.371611 &   -0.49 & -24.495710 &    -0.55 &  2021-10-06T14:41:06 &    1200.0 &     1.57 &         0.0 &    1.21 \\
 \hline
NGC2835F1 &  139.471355 &   -3.51 & -22.341874 &     0.05 &  2019-03-27T05:57:41 &    1200.0 &     1.47 &        90.0 &    1.42 \\
 NGC2835F1 &  139.471355 &   -3.51 & -22.341874 &     0.05 &  2019-03-27T06:18:44 &    1200.0 &     1.42 &        90.0 &    1.52 \\
 NGC2835F2 &  139.471475 &   -3.95 & -22.342430 &     2.05 &  2019-03-27T06:52:51 &    1200.0 &     1.36 &        90.0 &    1.51 \\
 NGC2835F2 &  139.471475 &   -3.95 & -22.342430 &     2.05 &  2019-03-27T07:13:56 &    1200.0 &     1.35 &        90.0 &    1.42 \\
 NGC2835F3 &  139.471114 &   -2.65 & -22.342874 &     3.65 &  2019-03-27T07:47:08 &    1200.0 &     1.36 &        90.0 &    1.39 \\
 NGC2835F3 &  139.471054 &   -2.43 & -22.342874 &     3.65 &  2019-03-27T08:08:11 &    1200.0 &     1.38 &        90.0 &    1.52 \\
 NGC2835F4 &  139.470694 &   -1.14 & -22.343041 &     4.25 &  2019-03-27T08:42:01 &    1200.0 &     1.45 &        90.0 &    1.26 \\
 NGC2835F4 &  139.470634 &   -0.92 & -22.342985 &     4.05 &  2019-03-27T09:03:11 &    1200.0 &     1.51 &        90.0 &    1.17 \\
 \hline
NGC3627F1 &  170.056873 &   -2.92 &  12.998914 &     1.45 &  2017-12-24T12:43:33 &     120.0 &     1.32 &        90.0 &    1.27 \\
 NGC3627F1 &  170.056873 &   -2.92 &  12.998858 &     1.65 &  2017-12-24T12:46:31 &     120.0 &     1.30 &        90.0 &    1.27 \\
 NGC3627F1 &  170.056873 &   -2.92 &  12.998858 &     1.65 &  2017-12-24T12:49:28 &     120.0 &     1.29 &        90.0 &    1.26 \\
 NGC3627F1 &  170.056873 &   -2.92 &  12.998858 &     1.65 &  2017-12-24T12:52:26 &     120.0 &     1.28 &        90.0 &    1.27 \\
 NGC3627F1 &  170.056873 &   -2.92 &  12.998858 &     1.65 &  2017-12-24T12:55:24 &     120.0 &     1.26 &        90.0 &    1.27 \\
 NGC3627F1 &  170.056132 &   -0.26 &  12.999192 &     0.45 &  2017-12-24T15:33:08 &     120.0 &     1.01 &        90.0 &    1.16 \\
 NGC3627F1 &  170.056132 &   -0.26 &  12.999192 &     0.45 &  2017-12-24T15:36:06 &     120.0 &     1.01 &        90.0 &    1.22 \\
 NGC3627F1 &  170.056132 &   -0.26 &  12.999192 &     0.45 &  2017-12-24T15:39:04 &     120.0 &     1.01 &        90.0 &    1.21 \\
 NGC3627F1 &  170.056132 &   -0.26 &  12.999192 &     0.45 &  2017-12-24T15:42:02 &     120.0 &     1.01 &        90.0 &    1.20 \\
 NGC3627F2 &  170.057215 &   -4.16 &  12.998747 &     2.05 &  2017-12-24T13:02:38 &     120.0 &     1.24 &        90.0 &    1.21 \\
 NGC3627F2 &  170.057215 &   -4.16 &  12.998747 &     2.05 &  2017-12-24T13:05:35 &     120.0 &     1.23 &        90.0 &    1.23 \\
 NGC3627F2 &  170.057215 &   -4.16 &  12.998747 &     2.05 &  2017-12-24T13:08:33 &     120.0 &     1.22 &        90.0 &    1.22 \\
 NGC3627F2 &  170.057215 &   -4.16 &  12.998747 &     2.05 &  2017-12-24T13:11:31 &     120.0 &     1.21 &        90.0 &    1.19 \\
 NGC3627F2 &  170.057215 &   -4.16 &  12.998747 &     2.05 &  2017-12-24T13:14:29 &     120.0 &     1.20 &        90.0 &    1.23 \\
 NGC3627F3 &  170.057215 &   -4.16 &  12.998747 &     2.05 &  2017-12-24T13:20:54 &     120.0 &     1.18 &        90.0 &    1.13 \\
 NGC3627F3 &  170.057215 &   -4.16 &  12.998747 &     2.05 &  2017-12-24T13:23:52 &     120.0 &     1.17 &        90.0 &    1.12 \\
 NGC3627F3 &  170.057215 &   -4.16 &  12.998747 &     2.05 &  2017-12-24T13:26:50 &     120.0 &     1.16 &        90.0 &    1.12 \\
 NGC3627F3 &  170.057215 &   -4.16 &  12.998747 &     2.05 &  2017-12-24T13:29:48 &     120.0 &     1.15 &        90.0 &    1.12 \\
 NGC3627F3 &  170.057215 &   -4.16 &  12.998747 &     2.05 &  2017-12-24T13:32:45 &     120.0 &     1.14 &        90.0 &    1.14 \\
 NGC3627F4 &  170.057215 &   -4.16 &  12.998580 &     2.65 &  2017-12-24T13:40:31 &     120.0 &     1.12 &        90.0 &    1.14 \\
 NGC3627F4 &  170.057215 &   -4.16 &  12.998580 &     2.65 &  2017-12-24T13:43:29 &     120.0 &     1.12 &        90.0 &    1.14 \\
 NGC3627F4 &  170.057215 &   -4.16 &  12.998580 &     2.65 &  2017-12-24T13:46:27 &     120.0 &     1.11 &        90.0 &    1.14 \\
 NGC3627F4 &  170.057215 &   -4.16 &  12.998580 &     2.65 &  2017-12-24T13:49:25 &     120.0 &     1.10 &        90.0 &    1.15 \\
 NGC3627F4 &  170.057215 &   -4.16 &  12.998580 &     2.65 &  2017-12-24T13:52:23 &     120.0 &     1.10 &        90.0 &    1.15 \\
 NGC3627F5 &  170.057215 &   -4.16 &  12.998525 &     2.85 &  2017-12-24T13:58:52 &     120.0 &     1.09 &        90.0 &    1.06 \\
 NGC3627F5 &  170.057158 &   -3.95 &  12.998525 &     2.85 &  2017-12-24T14:01:50 &     120.0 &     1.08 &        90.0 &    1.07 \\
 NGC3627F5 &  170.057158 &   -3.95 &  12.998525 &     2.85 &  2017-12-24T14:04:48 &     120.0 &     1.08 &        90.0 &    1.07 \\
 NGC3627F5 &  170.057158 &   -3.95 &  12.998525 &     2.85 &  2017-12-24T14:07:46 &     120.0 &     1.07 &        90.0 &    1.07 \\
 NGC3627F5 &  170.057158 &   -3.95 &  12.998525 &     2.85 &  2017-12-24T14:10:44 &     120.0 &     1.07 &        90.0 &    1.07 \\
 NGC3627F6 &  170.056246 &   -0.67 &  12.999136 &     0.65 &  2017-12-24T14:24:16 &     120.0 &     1.05 &        90.0 &    1.12 \\
 NGC3627F6 &  170.056189 &   -0.46 &  12.999136 &     0.65 &  2017-12-24T14:27:14 &     120.0 &     1.04 &        90.0 &    1.15 \\
 NGC3627F6 &  170.056189 &   -0.46 &  12.999136 &     0.65 &  2017-12-24T14:30:11 &     120.0 &     1.04 &        90.0 &    1.17 \\
 NGC3627F6 &  170.056189 &   -0.46 &  12.999136 &     0.65 &  2017-12-24T14:33:10 &     120.0 &     1.04 &        90.0 &    1.17 \\
 NGC3627F6 &  170.056189 &   -0.46 &  12.999136 &     0.65 &  2017-12-24T14:36:07 &     120.0 &     1.03 &        90.0 &    1.15 \\
 NGC3627F7 &  170.056303 &   -0.87 &  12.999136 &     0.65 &  2017-12-24T14:42:14 &     120.0 &     1.03 &        90.0 &    1.16 \\
 NGC3627F7 &  170.056303 &   -0.87 &  12.999136 &     0.65 &  2017-12-24T14:45:11 &     120.0 &     1.03 &        90.0 &    1.16 \\
 NGC3627F7 &  170.056303 &   -0.87 &  12.999136 &     0.65 &  2017-12-24T14:48:09 &     120.0 &     1.02 &        90.0 &    1.17 \\
 NGC3627F7 &  170.056303 &   -0.87 &  12.999136 &     0.65 &  2017-12-24T14:51:07 &     120.0 &     1.02 &        90.0 &    1.18 \\
 NGC3627F7 &  170.056303 &   -0.87 &  12.999136 &     0.65 &  2017-12-24T14:54:05 &     120.0 &     1.02 &        90.0 &    1.20 \\
 NGC3627F8 &  170.056303 &   -0.87 &  12.998914 &     1.45 &  2017-12-24T15:00:09 &     120.0 &     1.02 &        90.0 &    1.05 \\
 NGC3627F8 &  170.056303 &   -0.87 &  12.998914 &     1.45 &  2017-12-24T15:03:07 &     120.0 &     1.01 &        90.0 &    1.05 \\
 NGC3627F8 &  170.056246 &   -0.67 &  12.998914 &     1.45 &  2017-12-24T15:06:05 &     120.0 &     1.01 &        90.0 &    1.05 \\
 NGC3627F8 &  170.056246 &   -0.67 &  12.998914 &     1.45 &  2017-12-24T15:09:03 &     120.0 &     1.01 &        90.0 &    1.06 \\
 NGC3627F8 &  170.056246 &   -0.67 &  12.998914 &     1.45 &  2017-12-24T15:12:00 &     120.0 &     1.01 &        90.0 &    1.05 \\
 NGC3627F9 &  170.056018 &    0.15 &  12.998803 &     1.85 &  2017-12-24T15:15:05 &     120.0 &     1.01 &        90.0 &    1.12 \\
 NGC3627F9 &  170.056018 &    0.15 &  12.998803 &     1.85 &  2017-12-24T15:18:03 &     120.0 &     1.01 &        90.0 &    1.12 \\
 NGC3627F9 &  170.056018 &    0.15 &  12.998803 &     1.85 &  2017-12-24T15:21:01 &     120.0 &     1.01 &        90.0 &    1.12 \\
 NGC3627F9 &  170.056018 &    0.15 &  12.998803 &     1.85 &  2017-12-24T15:23:59 &     120.0 &     1.01 &        90.0 &    1.11 \\
NGC3627F9 &  170.056018 &    0.15 &  12.998803 &     1.85 &  2017-12-24T15:26:57 &     120.0 &     1.01 &        90.0 &    1.11 \\
 \hline
NGC5068F1 &  199.704057 &   -2.20 & -21.023638 &     0.85 &  2019-03-27T09:49:19 &    1200.0 &     1.47 &        90.0 &    1.51 \\
 NGC5068F1 &  199.704057 &   -2.20 & -21.023694 &     1.05 &  2019-03-27T10:10:22 &    1200.0 &     1.41 &        90.0 &    1.33 \\
 NGC5068F2 &  199.703819 &   -1.34 & -21.023638 &     0.85 &  2019-03-27T10:44:05 &    1200.0 &     1.35 &        90.0 &    1.40 \\
 NGC5068F3 &  199.703938 &   -1.77 & -21.023971 &     2.05 &  2019-03-27T11:38:24 &    1200.0 &     1.32 &        90.0 &    2.49 \\
 NGC5068F3 &  199.703938 &   -1.77 & -21.023971 &     2.05 &  2019-03-27T11:59:28 &    1200.0 &     1.34 &        90.0 &    2.49 \\
 NGC5068F4 &  199.703700 &   -0.91 & -21.024138 &     2.65 &  2019-03-27T12:32:51 &    1200.0 &     1.39 &        90.0 &    1.55 \\
 NGC5068F4 &  199.703700 &   -0.91 & -21.024194 &     2.85 &  2019-03-27T12:53:55 &    1200.0 &     1.45 &        90.0 &    1.88 \\
 NGC5068F5 &  199.703403 &    0.16 & -21.023860 &     1.65 &  2019-03-27T13:28:21 &    1200.0 &     1.58 &        90.0 &    1.97 \\
 NGC5068F5 &  199.703998 &   -1.98 & -21.023694 &     1.05 &  2019-03-28T10:19:33 &    1200.0 &     1.38 &        90.0 &    1.69 \\
 NGC5068F7 &  199.704057 &   -2.20 & -21.023694 &     1.05 &  2019-03-28T10:54:58 &    1200.0 &     1.33 &        90.0 &    1.68 \\
 NGC5068F7 &  199.703998 &   -1.98 & -21.023694 &     1.05 &  2019-03-28T11:16:01 &    1200.0 &     1.32 &        90.0 &    1.67 \\
 NGC5068F8 &  199.703879 &   -1.55 & -21.024082 &     2.45 &  2019-03-28T11:49:43 &    1200.0 &     1.33 &        90.0 &    1.86 \\
 NGC5068F8 &  199.703819 &   -1.34 & -21.024027 &     2.25 &  2019-03-28T12:10:49 &    1200.0 &     1.36 &        90.0 &    1.87 \\
 NGC5068F9 &  199.703403 &    0.16 & -21.024138 &     2.65 &  2019-03-28T12:44:48 &    1200.0 &     1.43 &        90.0 &    1.67 \\
 NGC5068F9 &  199.703343 &    0.37 & -21.024138 &     2.65 &  2019-03-28T13:05:52 &    1200.0 &     1.50 &        90.0 &    1.65 \\
 NGC5068F9 &  199.703700 &   -0.91 & -21.023138 &    -0.95 &  2022-02-24T14:57:54 &    1200.0 &     1.46 &         0.0 &    1.39 \\
 NGC5068F9 &  199.703641 &   -0.70 & -21.023138 &    -0.95 &  2022-02-24T15:08:56 &    600.0 &     1.50 &         0.0 &    1.46 \\
 NGC5068F10 &  199.703700 &   -0.91 & -21.023527 &     0.45 &  2022-02-24T15:30:23 &    800.0 &     1.59 &        90.0 &    1.17 \\
 NGC5068F10 &  199.703700 &   -0.91 & -21.023471 &     0.25 &  2022-02-24T15:38:22 &    420.0 &     1.63 &        90.0 &    1.13 
 \enddata
 \tablenotetext{a}{The field numbering is from internal lists of potential pointings which were not performed in numerical order, therefore the field number does not reflect the sequence of observations.}
\tablenotetext{b}{The astrometric correction is described in Section \ref{sec:registration}.}
\tablenotetext{c}{The measured angular FWHM for the pointing, see Section \ref{sec:psf+per_pointing}.}
\end{deluxetable*}

\startlongtable
\begin{deluxetable*}{ccccccc}
\tablecaption{Summary of KCWI Standard Star Observations.}
\label{tab:std_obs}
\tablehead{\colhead{Star\tablenotemark{a}} & \colhead{RA} & \colhead{Dec} & \colhead{Date} & \colhead{Exposure} & \colhead{Airmass} & \colhead{FWHM\tablenotemark{b}} \\ 
\colhead{} & \colhead{} & \colhead{} & \colhead{(UTC)} & \colhead{(s)} & \colhead{} & \colhead{(\arcsec)}}
\startdata
He3 &  06:47:37.99 & +37:30:57.1 &  2018-10-16T15:00:13 &      20.0 &     1.06 &    1.32 \\
Feige15 & 01:49:09.49 &  +13:33:11.7 &  2018-10-17T09:04:57 &      10.0 &     1.07 & 1.69 \\
Feige15 & 01:49:09.49 &  +13:33:11.7 &  2018-10-17T09:06:11 &       1.0 &     1.07 & 1.12 \\
Feige15 &   01:49:09.49 &  +13:33:11.7 &  2018-10-16T10:44:02 &      10.0 &   1.01 &    1.04 \\
Feige34 & 10:39:36.71 &  +43:06:10.1 &  2019-03-27T09:21:06 &       5.0 &     1.10 &    1.28 \\
Feige34 & 10:39:36.71 &  +43:06:10.1 &  2019-03-27T09:22:26 &       1.0 &     1.10 &    1.21 \\
Feige66 & 12:37:23.52 &  +25:03:59.3 &  2019-03-27T13:45:25 &       1.0 &     1.35 &    1.99 \\
Feige66 & 12:37:23.52 &  +25:03:59.3 &  2019-03-28T13:19:36 &       1.0 &     1.26 &    1.83 \\
Feige66 & 12:37:23.52 &  +25:03:59.3 &  2019-03-28T13:20:57 &       5.0 &     1.26 &    1.83 \\
GD50 &   03:48:50.31  & -00:58:35.8 &  2021-10-06T10:17:51 &       1.0 &     1.47 &    1.13 \\
GD50 &   03:48:50.31  & -00:58:35.8 &  2021-10-06T10:19:37 &       1.0 &     1.46 &    1.13 \\
GD50 &   03:48:50.31  & -00:58:35.8 &  2021-10-06T10:20:53 &      10.0 &     1.45 &    1.18 \\
GD50 &   03:48:50.31  & -00:58:35.8 &  2021-10-06T10:22:37 &      15.0 &     1.44 &    1.22 \\
GD50 &   03:48:50.31  & -00:58:35.8 &  2021-10-06T14:13:09 &      10.0 &     1.11 &    1.01 \\
GD50 &   03:48:50.31  & -00:58:35.8 &  2021-10-06T14:14:54 &      15.0 &     1.11 &    1.04 \\
GD50 &   03:48:50.31  & -00:58:35.8 &  2021-10-06T14:16:28 &       1.0 &     1.11 &    1.06 \\
GD50 &   03:48:50.31  & -00:58:35.8 &  2021-10-07T10:29:40 &       1.0 &     1.38 &    1.07 \\
GD50 &   03:48:50.31  & -00:58:35.8 &  2021-10-07T10:30:56 &      10.0 &     1.38 &    1.11 \\
GD50 &   03:48:50.31  & -00:58:35.8 &  2021-10-07T10:32:17 &      15.0 &     1.37 &    1.10 \\
GD50 &   03:48:50.31  & -00:58:35.8 &  2021-10-07T13:19:16 &      10.0 &     1.07 &    1.11 \\
GD50 &   03:48:50.31  & -00:58:35.8 &  2021-10-07T13:20:43 &      15.0 &     1.07 &    1.18 
\enddata
\tablenotetext{a}{The name of each standard star as displayed in the KDRP list of standards.}
\tablenotetext{b}{The angular FWHM was measured from a 2D Gaussian fit to a white-light image of the standard star.}
\end{deluxetable*}
\newpage

\section{KCWI Line Spread Function}
\label{sec:lsf}
\begin{figure*}[h]
    \centering
    \includegraphics[width=0.45\textwidth]{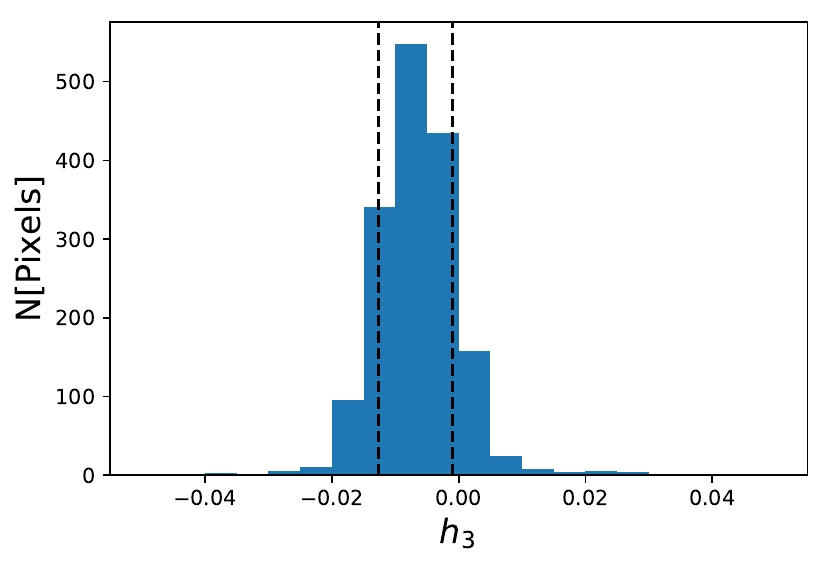}
    \includegraphics[width=0.45\textwidth]{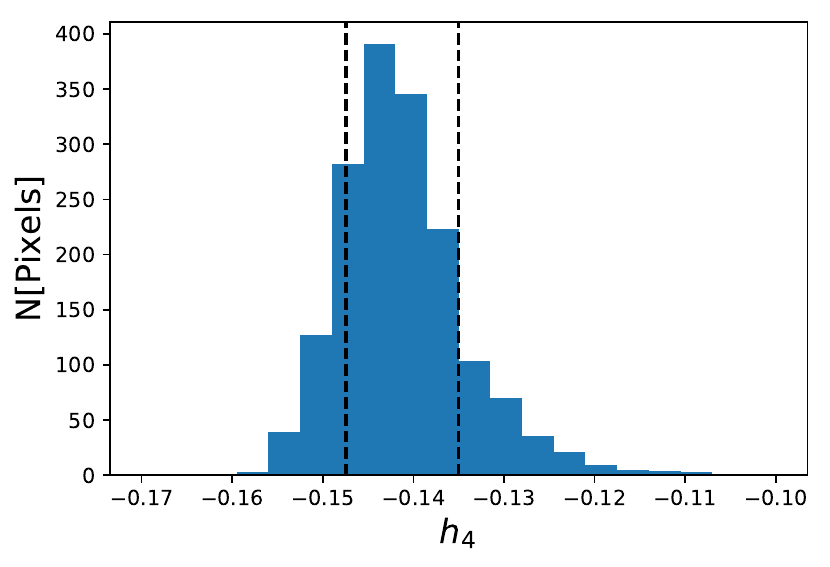}
    \caption{Histograms of the the Gauss-Hermite parameters $h_3$, shown in the left panel, and $h_4$, shown in the right panel. The above histograms show the distribution of values for the fitted constants $h_3$ (anti-symmetric) and $h_4$ (symmetric) from all pixels in the datacube. A Gaussian profile would exhibit $h_3=h_4=0$, however, the above distributions show that the line profile of the instrument exhibits symmetric deviations. The broken black vertical lines represent the position of $\pm 1 \sigma$.}
    \label{fig:hermite_maps}
\end{figure*}

The KCWI line spread function in the large slicer is non-Gaussian. Deviations from a Gaussian profile have been observed in KCWI data before. For example, using a combination of the medium slicer and grating \cite{vanDokkum_guass_hermite}. Following \cite{vanderMarel_gauss_hermite}, in order to parameterize the degree of deviation from a Gaussian profile, we fit the spectra from a pipeline reduced arc lamp exposure a Gauss-Hermite function of the form:
\begin{eqnarray}
    \label{eq:gh}
   \nonumber G(X)=\frac{\gamma}{\sigma\sqrt{2\pi}}\exp{(-X^2/2)}\bigg[1+\ldots \\  h_3\frac{X(2X^2-3)}{\sqrt{3}} +  h_4\frac{4(X^2-3)X^2+3}{\sqrt{24}} \bigg],
\end{eqnarray}
with $X=(\lambda -\lambda_0)/\sigma$, amplitude ($\gamma$) and spectral width ($\sigma$). The anti-symmetric and symmetric deviations from pure Gaussian profiles are captured by the constants, $h_3$ and $h_4$. A Gaussian function is recovered by setting $h_3=h_4=0$. In Figure \ref{fig:hermite_maps} we show histograms of the fitted values from $h_3$ and $h_4$ after fitting Eq. \ref{eq:gh} to approximately $20$ isolated arc lamp emission lines. We find that the emission lines in the arc lamp images are consistently flat topped, with an average value of $h_4=-0.14 \pm 0.01$ across all spatial pixels. The degree that the line profile is non-Gaussian due to asymmetric deviations is small compared to symmetric deviations with an average value of $h_3=-0.007\pm 0.01$. 

The authors state that the root cause of the deviation from a Gaussian profile is due to the slit width-limited resolution of the medium slicer \citep[see,][]{slit_limiting}. We posit that a similar limitation for the large slicer is responsible for the deviations measured in this work.
\newpage

\section{Comparison of \HII\ Region Identification Between KCWI and MUSE}
\label{appn:comparison_of_muse_and_kcwi_hii_regions}
We identify potential \HII\ regions with H$\beta$ emission maps constructed from the KCWI galaxy mosaics using \texttt{HIIPhot}. \HII\ regions, for the same galaxies, in the PHANGS Nebular Catalog were identified with \texttt{HIIPhot} and MUSE H$\alpha$ maps \citep{Kreckel2019ApJ...887...80K, Groves2023NebCat}. Given the higher resolution of the MUSE imaging, as well as the three-fold brightness increase of H$\alpha$ relative to H$\beta$, we expect our \HII\ region catalog to be less sensitive to the faintest and smallest \HII\ regions. 

\begin{figure}[h]
    \centering
    \includegraphics[width=0.4\textwidth]{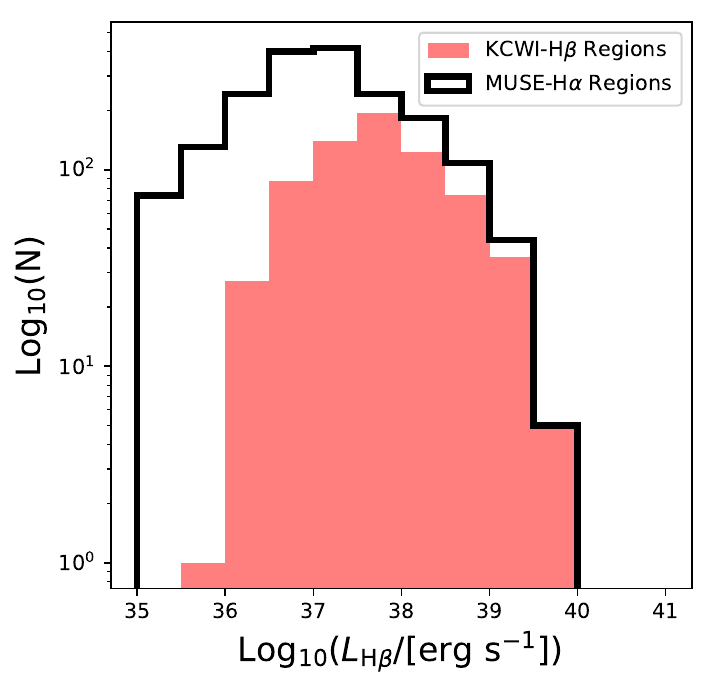}
    \includegraphics[width=0.4\textwidth]{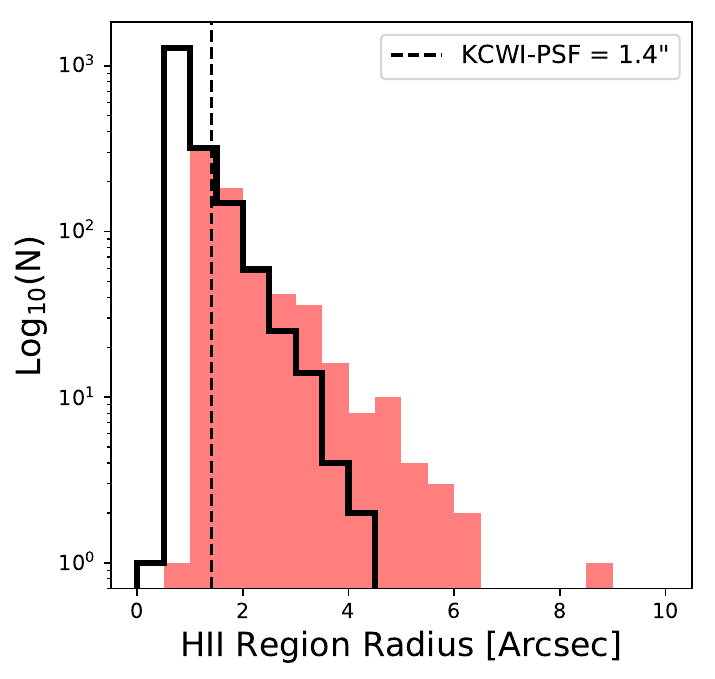}
    \caption{Histogram of the dust-corrected H$\beta$ luminosity for regions identified by \texttt{HIIPhot} using KCWI--H$\beta$ maps (\textit{red}) and regions in the Nebular Catalog identified using MUSE--H$\alpha$ maps  (\textit{black}). The number of faint regions detected using the KCWI--H$\beta$ map is set by the limiting sensitivity, Log$_{10}(L_{\rm{H}\beta}/[\rm{erg}\ \rm{s}^{-1}]$) $<$ 37, and angular resolution, FWHM=1.4\arcsec. }
    \label{fig:lum}
\end{figure}

In Figure \ref{fig:lum} we show both the distribution of dust-corrected H$\beta$ luminosity, $L_{\rm{H}\beta}$, and radii, for regions identified by \textit{HIIPhot} using KCWI H$\beta$ maps, ``KCWI--H$\beta$ regions'', and MUSE H$\alpha$ maps or ``MUSE--H$\alpha$ regions'' within the KCWI mosaic footprint. For KCWI--H$\beta$ regions we measure a median Log$_{10}(L_{\rm{H}\beta}/[\rm{erg}\ \rm{s}^{-1}]$) of 37.7$^{+0.9}_{-0.7}$ while for MUSE--H$\alpha$ regions the median is 37.1$^{+0.8}_{-1.7}$. This comparison shows that the KCWI H$\beta$ maps is less sensitive to regions with Log$_{10}(L_{\rm{H}\beta}/[\rm{erg}\ \rm{s}^{-1}]$) $<$ 37.

\begin{figure*}[h]
    \centering
    \includegraphics[scale=0.7]{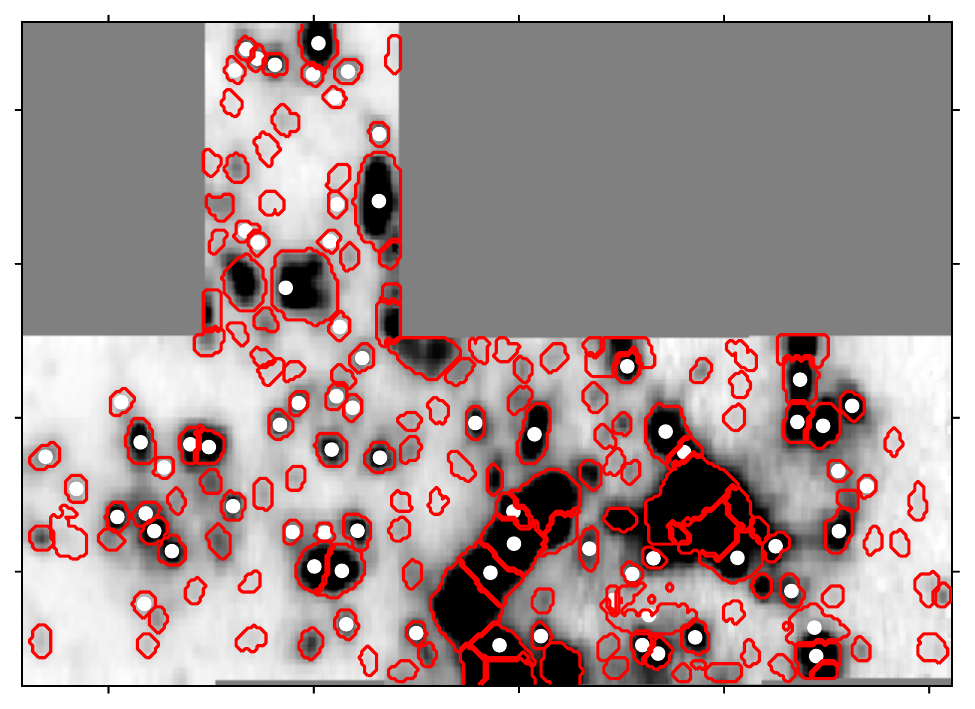}
    \includegraphics[scale=0.7]{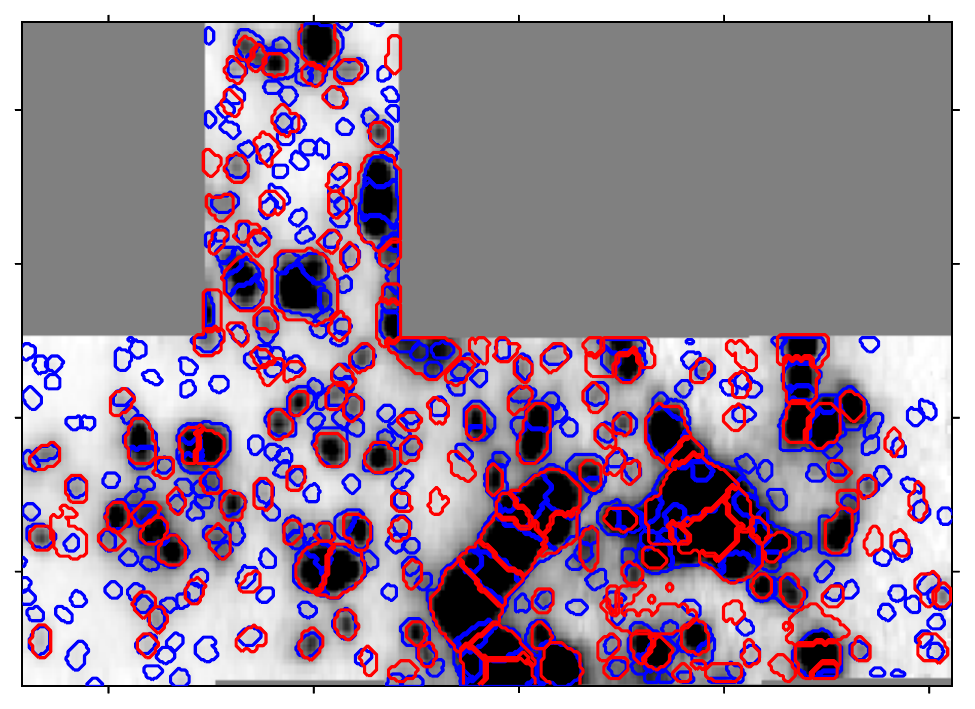}
    \caption{Comparison of region boundaries identified using \texttt{HIIPhot} and either KCWI--H$\beta$ or MUSE--H$\alpha$ emission line maps for the galaxy NGC\,1087. The KCWI--H$\beta$ emission line map is shown in both panels. We overlay in \textit{red} the morphology of regions identified by \texttt{HIIPhot} using the KCWI--H$\beta$ emission line map. A white marker indicates a region with significant auroral line detections in 2 or more auroral lines. In \textit{blue} we overlay region boundaries from \texttt{HIIPhot} using the MUSE H$\alpha$ emission line maps \citep{Kreckel2019ApJ...887...80K, Groves2023NebCat}.}
    \label{fig:hii_masks_1087}
\end{figure*}

\begin{figure*}[h]
    \centering
    \includegraphics[scale=0.7]{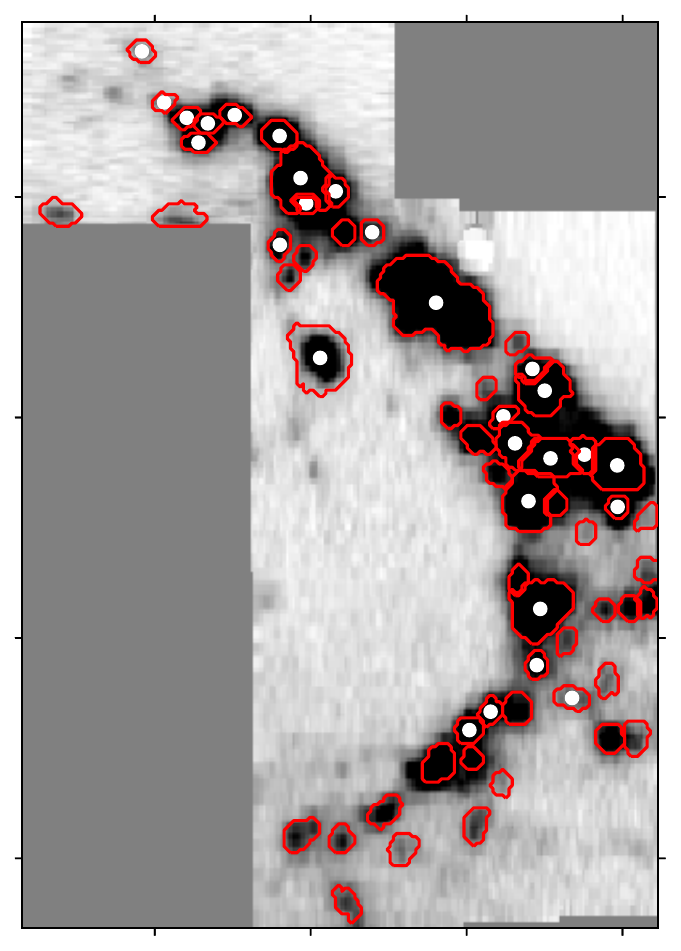}
    \includegraphics[scale=0.7]{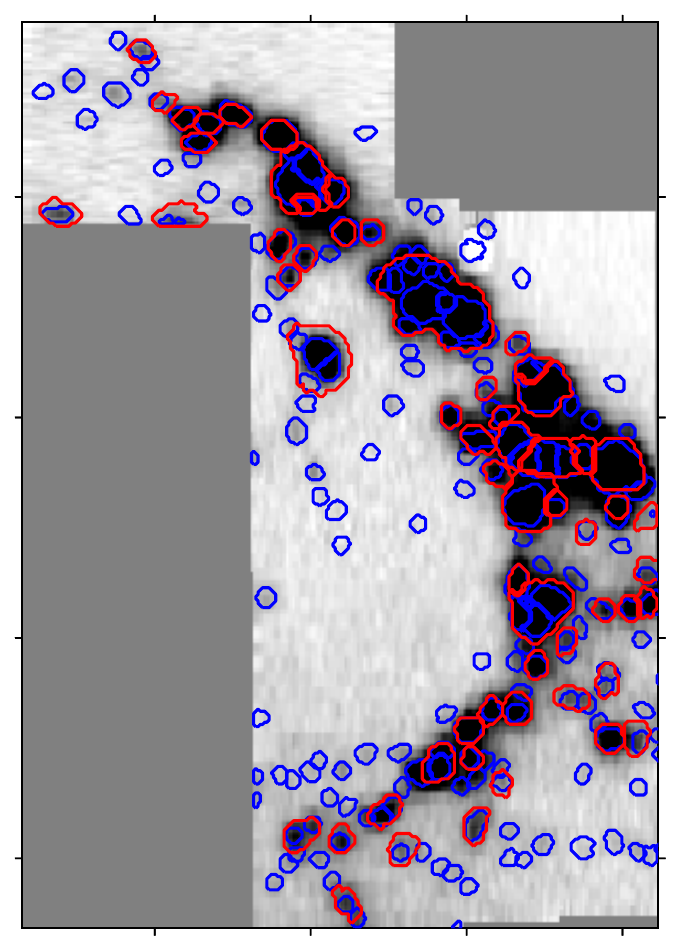}
    \caption{Comparison of \HII\ region boundaries in NGC\,1300, following Figure \ref{fig:hii_masks_1087}.}
    \label{fig:hii_masks_1300}
\end{figure*}

\begin{figure*}[h]
    \centering
    \includegraphics[scale=0.7]{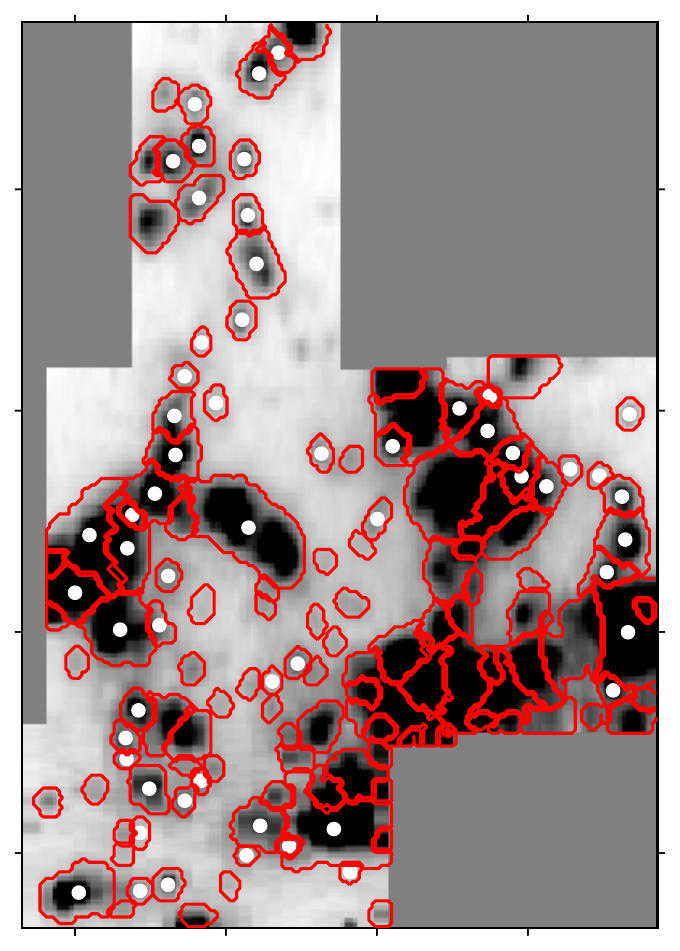}
    \includegraphics[scale=0.7]{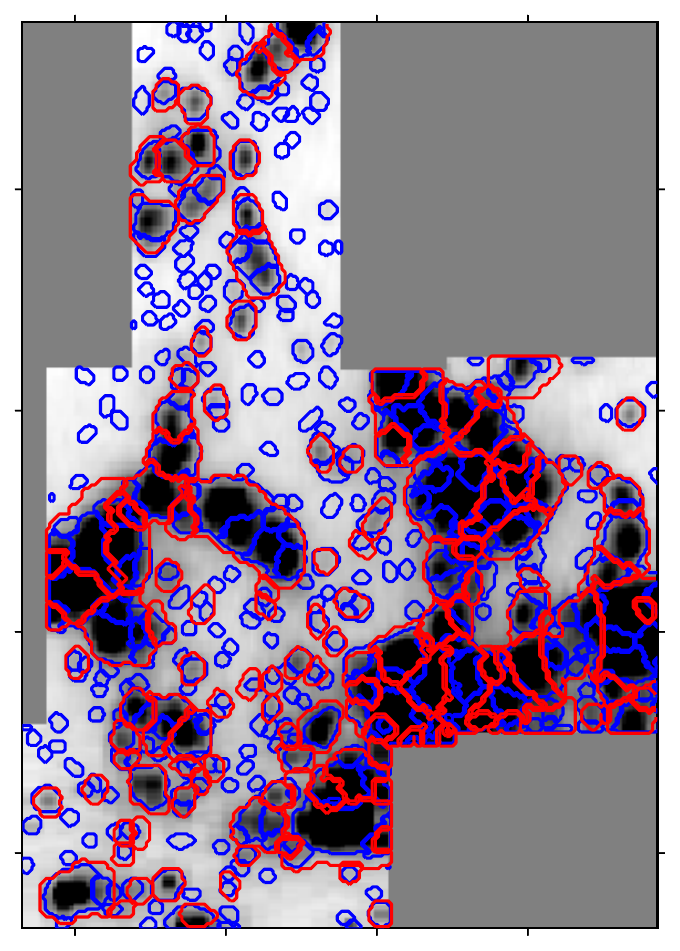}
    \caption{Comparison of \HII\ region boundaries in NGC\,1385, following Figure \ref{fig:hii_masks_1087}.}
    \label{fig:hii_masks_1385}
\end{figure*}

\begin{figure*}[h]
    \centering
    \includegraphics[scale=0.6]{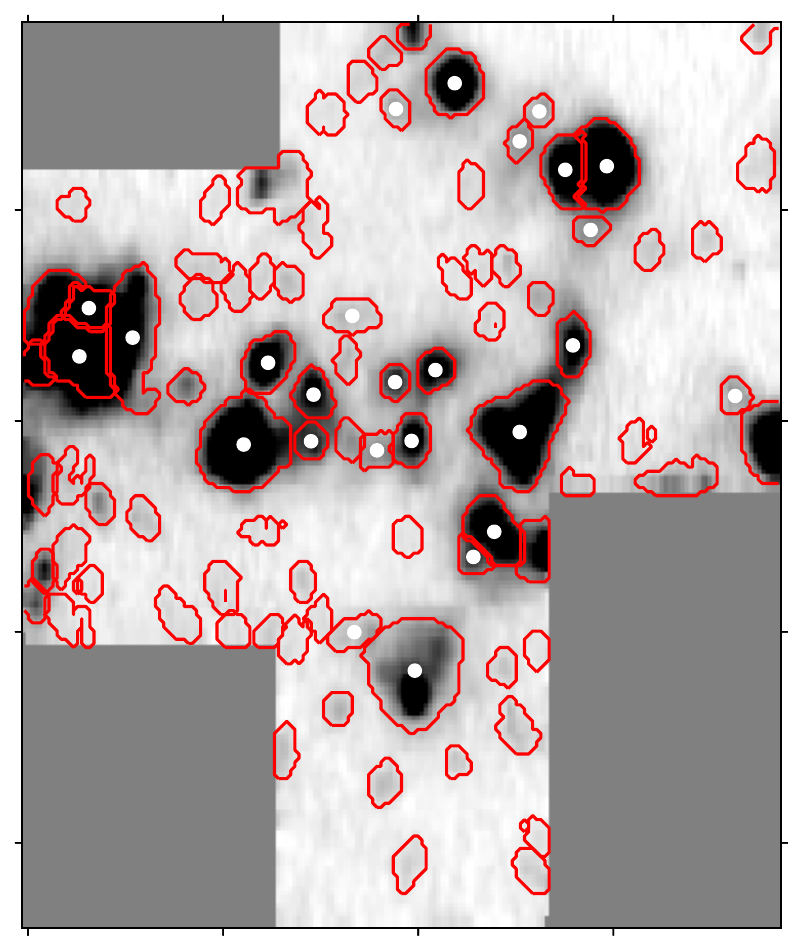}
    \includegraphics[scale=0.6]{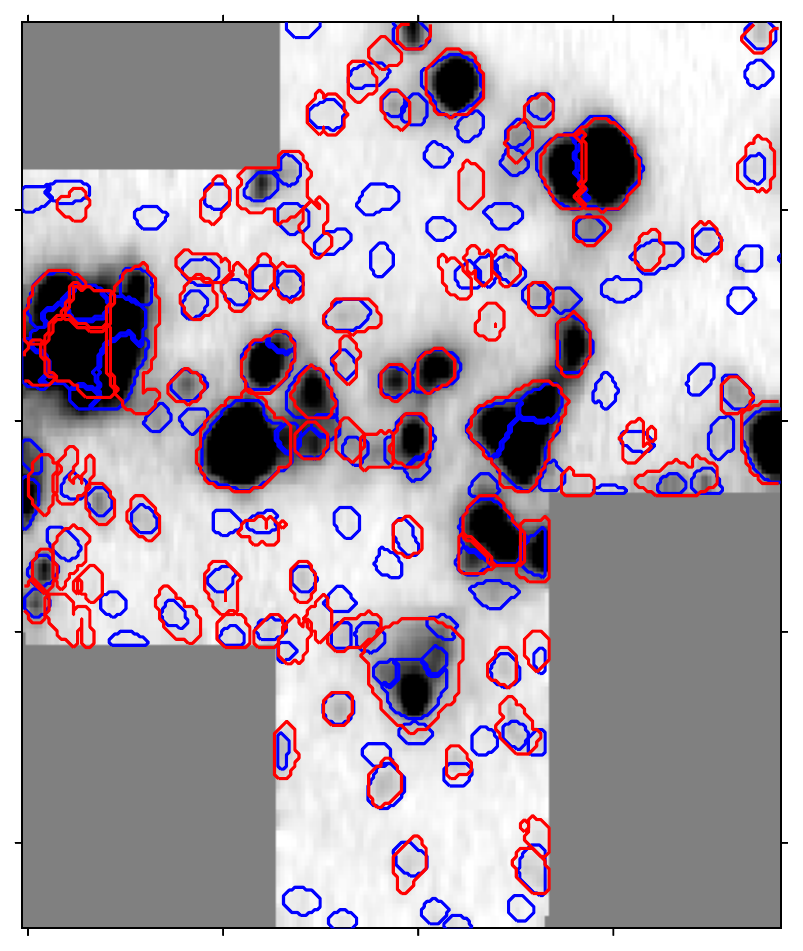}
    \caption{Comparison of \HII\ region boundaries in NGC\,2835, following Figure \ref{fig:hii_masks_1087}.
    \label{fig:hii_masks_2835}}
\end{figure*}

\begin{figure*}[h]
    \centering
    \includegraphics[scale=0.7]{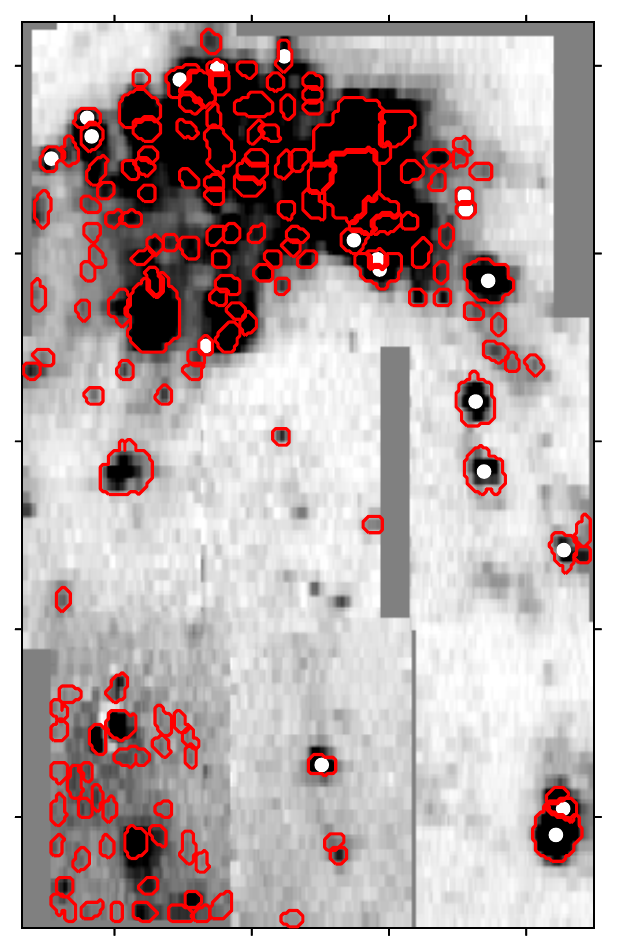}
    \includegraphics[scale=0.7]{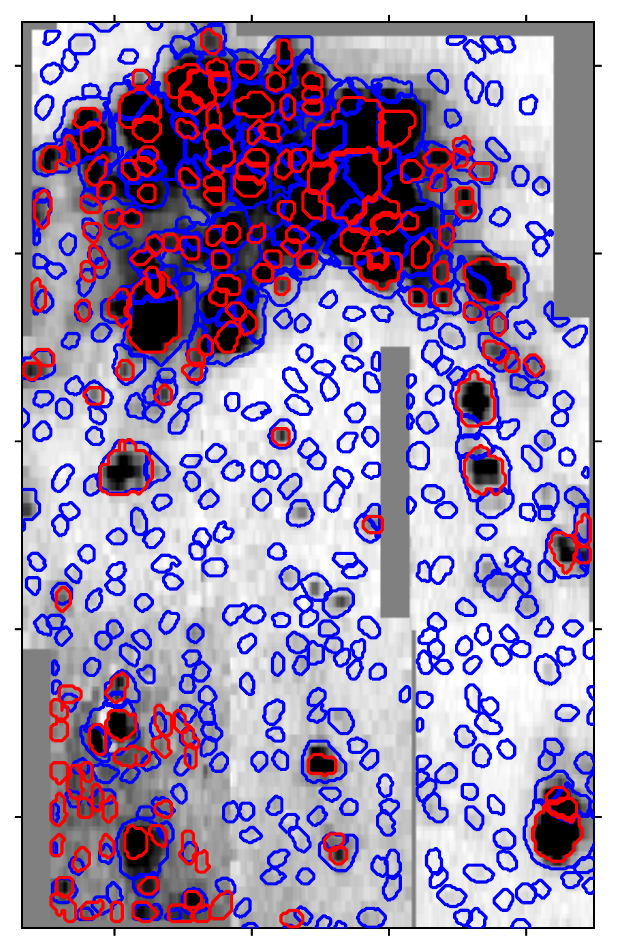}
    \caption{Comparison of \HII\ region boundaries in NGC\,3627, following Figure \ref{fig:hii_masks_1087}.}
    \label{fig:hii_masks_3627}
\end{figure*}

\begin{figure*}[h]
    \centering
    \includegraphics[scale=0.6]{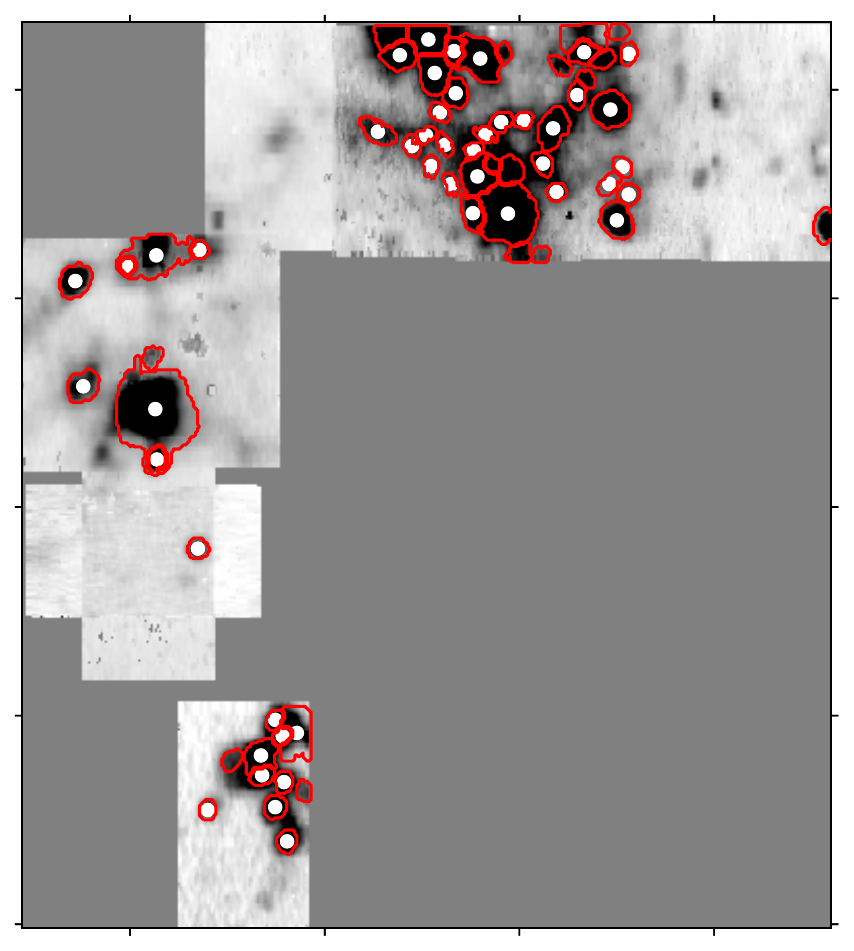}
    \includegraphics[scale=0.6]{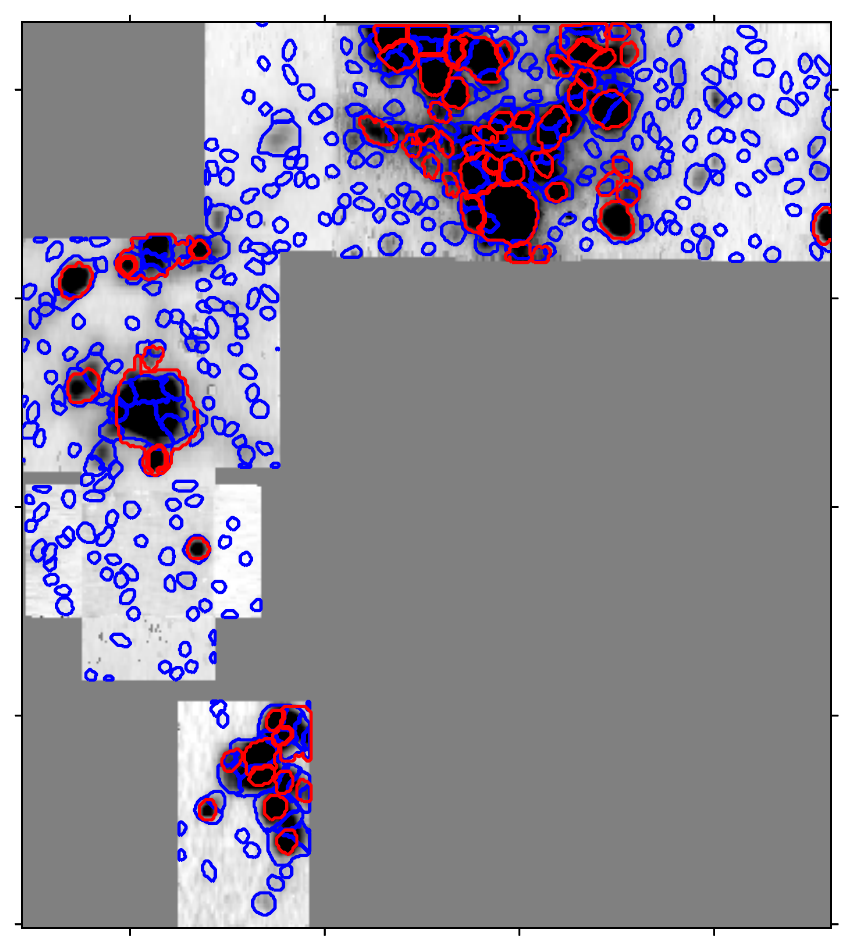}
    \caption{Comparison of \HII\ region boundaries in NGC\,5068, following Figure \ref{fig:hii_masks_1087}.}
    \label{fig:hii_masks_5068}
\end{figure*}

\begin{figure*}[h]
    \centering
    \includegraphics[scale=0.6]{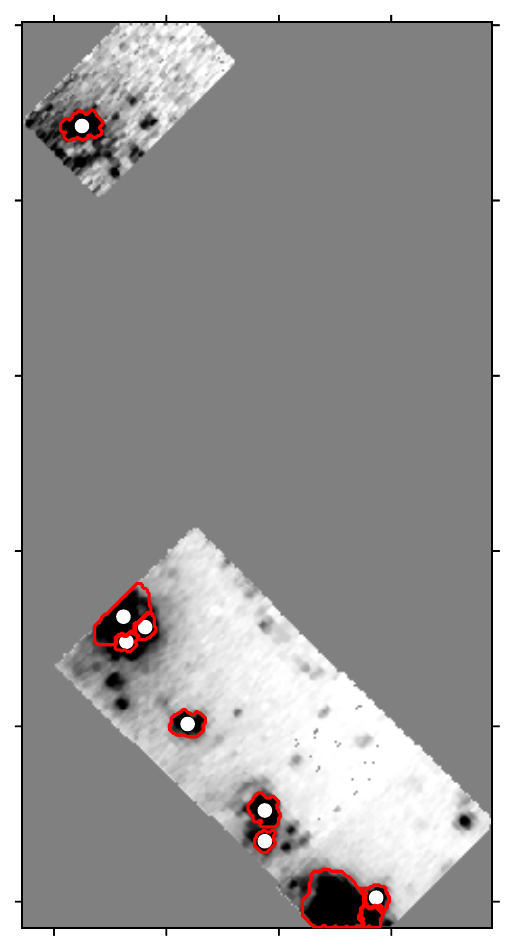}
    \includegraphics[scale=0.6]{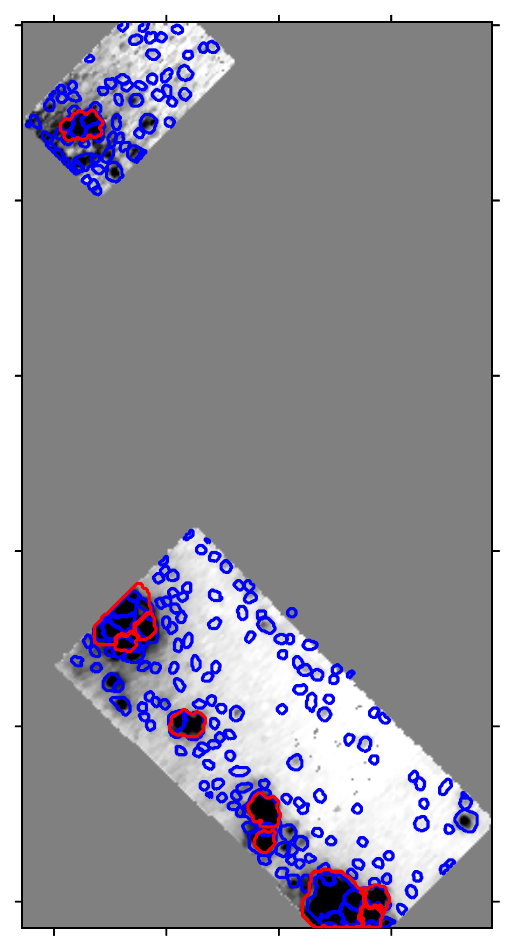}
    \caption{Comparison of \HII\ region boundaries in NGC\,628 following Figure \ref{fig:hii_masks_1087}.}
    \label{fig:hii_masks_0628}
\end{figure*}

We also observe in Figure \ref{fig:lum} that regions with radii less than the KCWI angular resolution are missed KCWI--H$\beta$ region sample. This can be seen clearly in Figures \ref{fig:hii_masks_1087}--\ref{fig:hii_masks_0628}, where we compare the boundaries of the MUSE--H$\alpha$ and KCWI--H$\beta$ regions. In these figures, we see that many of the missed regions are small, and unresolved in the KCWI H$\beta$ map. Also owing to the larger number of detections is that many of the larger KCWI--H$\beta$ regions are resolved into smaller structures in the MUSE--H$\alpha$ regions. 
\begin{deluxetable}{cccc}[h]
\label{tab:total_regions}
\centering
\caption{Total number of regions identified by \texttt{HIIphot} as potential \HII\ regions per galaxy using both KCWI--H$\beta$ and MUSE--H$\alpha$, as well as number of KCWI--H$\beta$ regions with significant with significant auroral lines detections (in 2 or more auroral lines).}
\tablehead{\colhead{Name} & \colhead{$N_{\rm{KCWI}}$} & \colhead{$N_{\rm{MUSE}}$} & \colhead{$N_{ \rm{A}}$}}
\startdata
NGC\,628\tablenotemark{a} & 10  &  230 & 8\\
NGC\,1087 & 173 &  364 & 73 \\
NGC\,1300 & 60  &  191 & 28 \\
NGC\,1385 & 133 &  417 & 58 \\
NGC\,2835 & 87  &  135 & 26 \\
NGC\,3627 & 163 &  451 & 19 \\
NGC\,5068 & 62  &  392 & 48
\enddata
\tablenotetext{a}{ NGC\,628 was imaged in the least ideal observing conditions.}
\end{deluxetable}

%\newpage

%\newpage

%\newpage

%\newpage

\clearpage

\section{Example Gaussian Fits to Auroral Lines}
\label{appn:auroral_fits}
\begin{figure*}[h]
\centering
\gridline{\leftfig{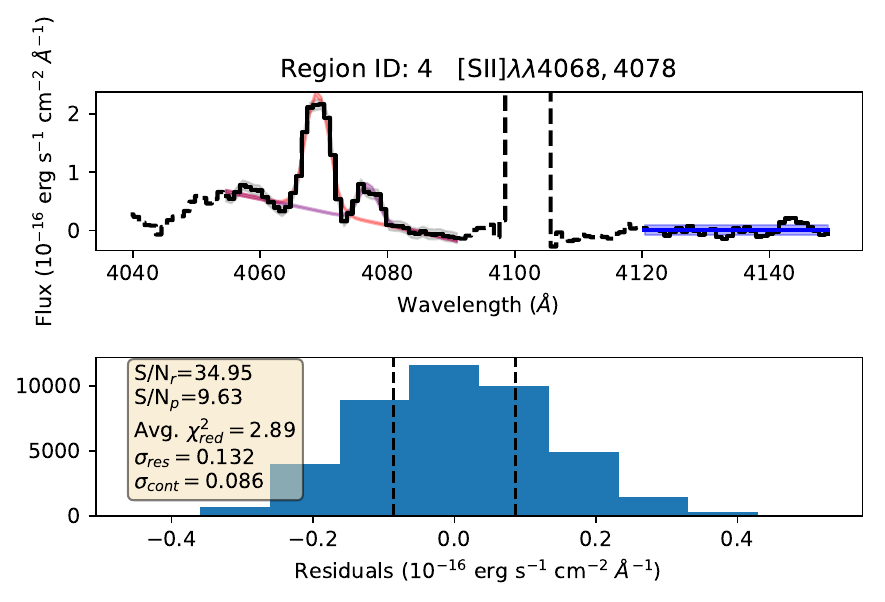}{0.5\textwidth}{}
          \rightfig{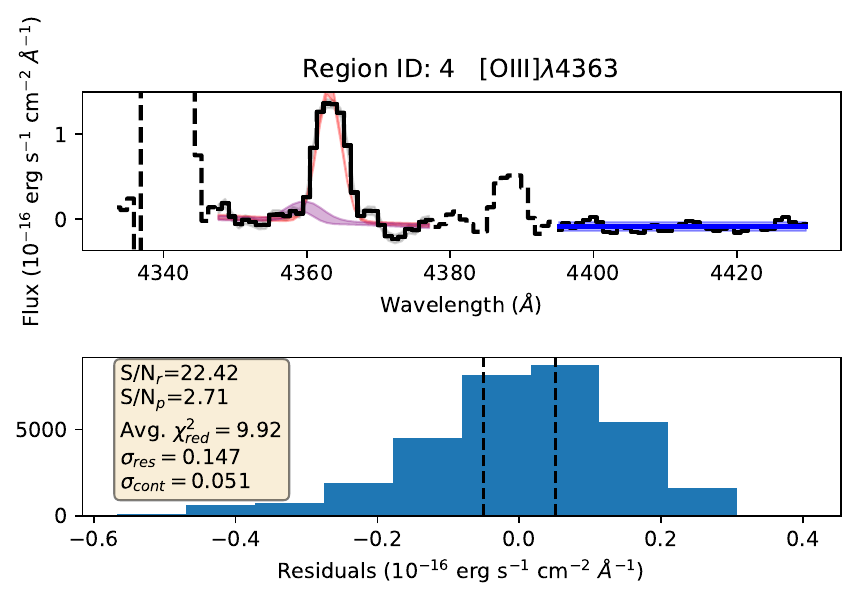}{0.5\textwidth}{}}
\gridline{\leftfig{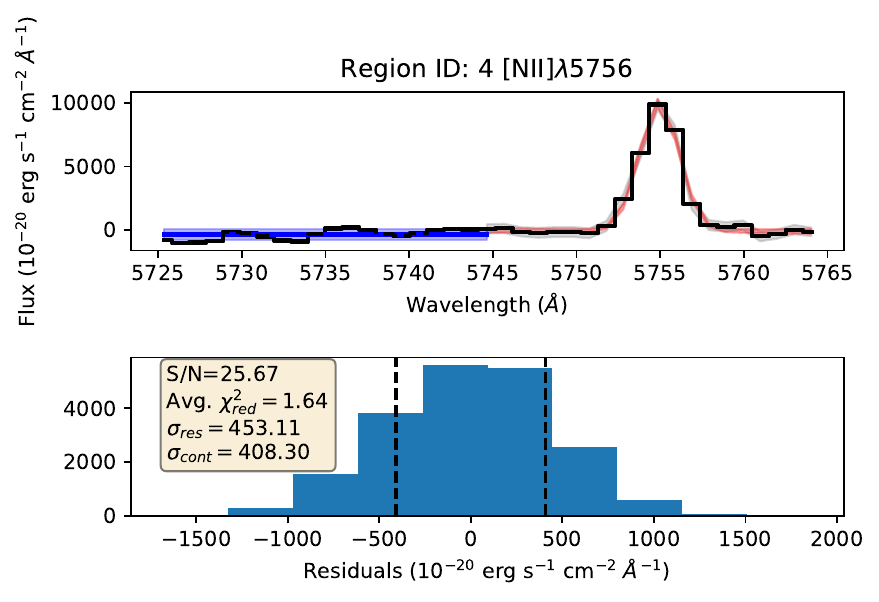}{0.5\textwidth}{}
          \rightfig{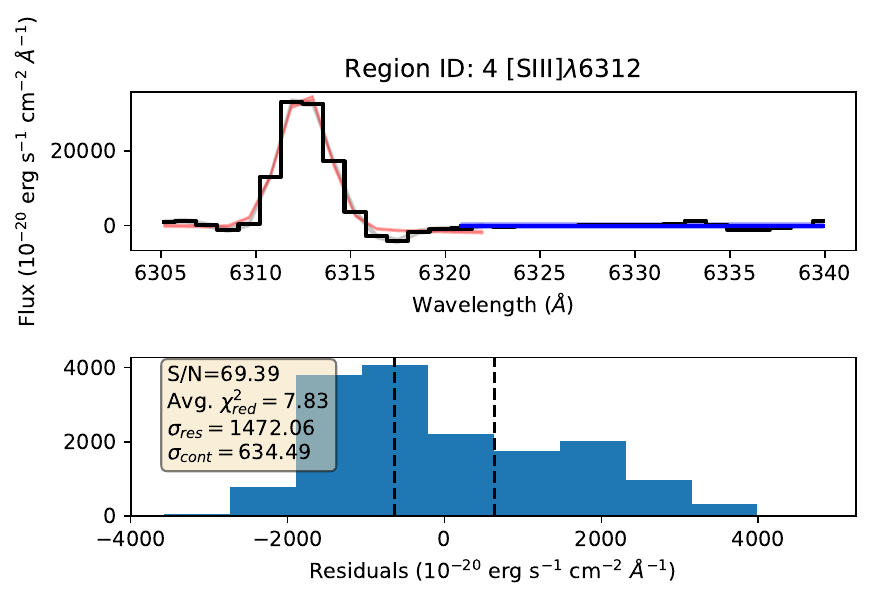}{0.5\textwidth}{}}
          \fig{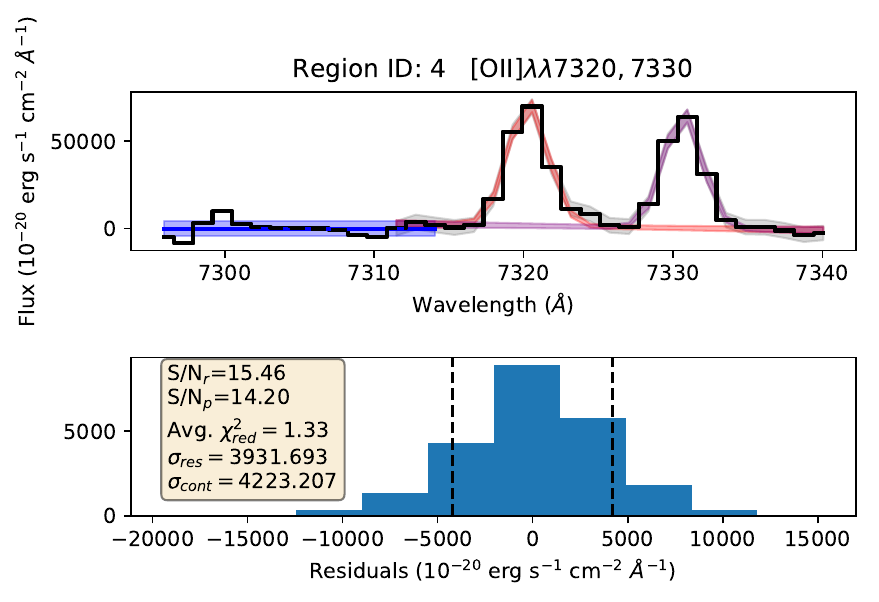}{0.5\textwidth}{}
\caption{Auroral Line fits for an \HII\ region in NGC\,5068. We show in each panel a summary of auroral line fits for a single \HII\ region in NGC\,5068. The top frame in each panel shows the data, (\textit{black-solid}), for the fitting and continuum wavelength ranges. In the wavelength range where $\sigma_{cont}$ is measured, we overlay $\pm \sigma_{cont}$ region (\textit{blue-shaded}) around the line indicating the average value of the continuum (\textit{blue-solid}). The (\textit{red-shaded}) and \textit{purple-shaded} for the double Gaussian fits, show the $1\sigma$ ranges of the fitted models. The bottom frame in each panels shows a histogram of the residuals. We also print text summarizing the S/N and average reduced $\chi^2$ of the fits as well as the $1\sigma$ of the residuals and value of $\sigma_{cont}$}.
\label{fig:auroral_line_fits}
\end{figure*}

\begin{figure*}[h]
\centering
\includegraphics[scale=0.8]{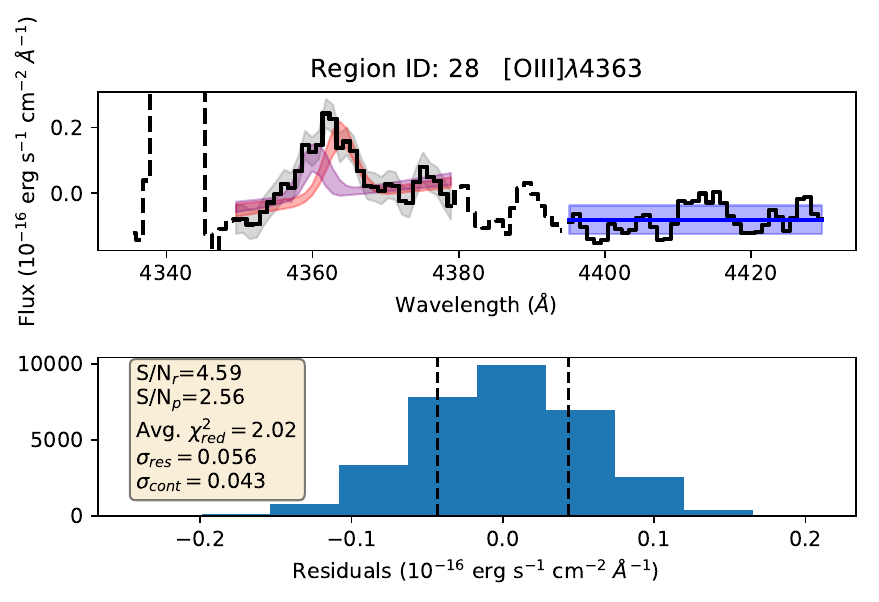}
\caption{Example of auroral line fit measuring [\ion{O}{3}]$\lambda4363$ \AA\ flux with non-negligible [\ion{Fe}{2}]$\lambda4360$ contribution for an \HII\ region in NGC\,1087. Annotations follow those in Figure \ref{fig:auroral_line_fits}.}
\label{fig:auroral_line_fits_lowsn}
\end{figure*}

In this section we show example auroral line fits. In Figure \ref{fig:auroral_line_fits} we show fits to high S/N auroral lines from an \HII\ region in NGC\,5068. We also include in Figure \ref{fig:auroral_line_fits} annotations describing the the standard deviation of the fit residuals, $\sigma_{res}$, the S/N for a single emission line (or in the instance of simultaneous double line fits the S/N of the \textit{red} and \textit{purple} Gaussian fit: S/N$_r$ and S/N$_p$), the continuum noise, $\sigma_{cont}$ and reduced $\chi^2$. In this particular example, the auroral line in these fits are isolated from contaminating sky lines or nearby emission, especially in the case of [\ion{O}{3}]$\lambda4363$ \AA\, where the contribution from [\ion{Fe}{2}]$\lambda4360$ \AA\ is negligible. To demonstrate fitting [\ion{Fe}{2}]$\lambda4360$ alongside [\ion{O}{3}]$\lambda4363$ we show a low S/N detection with non-negligible contamination from [\ion{Fe}{2}]$\lambda4360$ \AA\ in Figure \ref{fig:auroral_line_fits_lowsn}.

\clearpage

\section{Figures of the $\Delta T_e$ and \HII\ Region Property Comparisons}
\label{appn:uncorrelated_properties}
In the section we present the comparisons between  \HII\ region properties that exhibit no significant correlations with $\Delta T_e$.

\begin{figure*}[!h]
    \centering
    \includegraphics[scale=0.8]{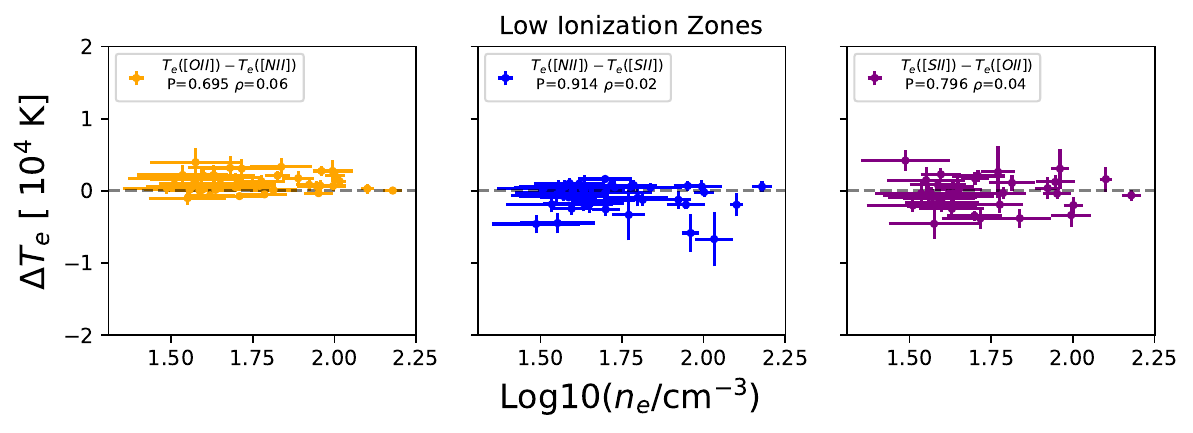}
    \includegraphics[scale=0.8]{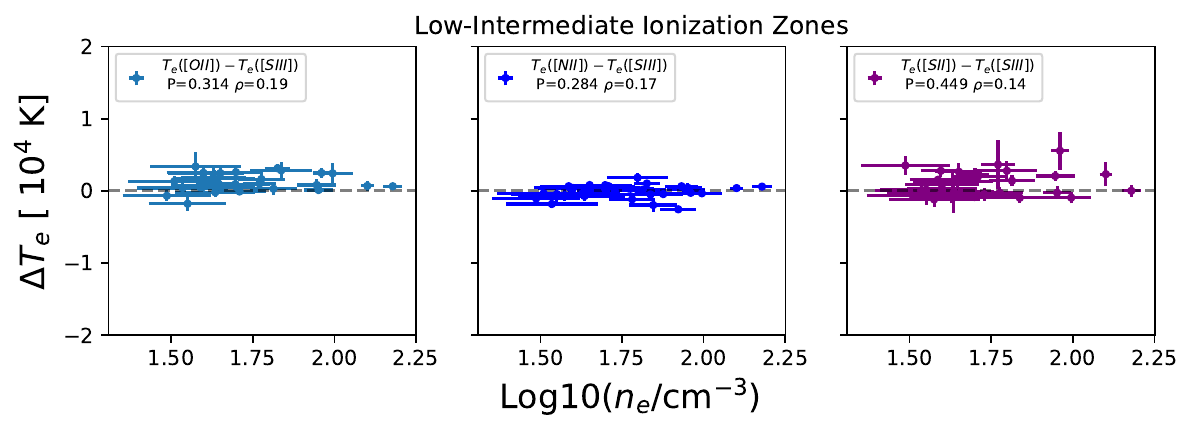}
    \includegraphics[scale=0.8]{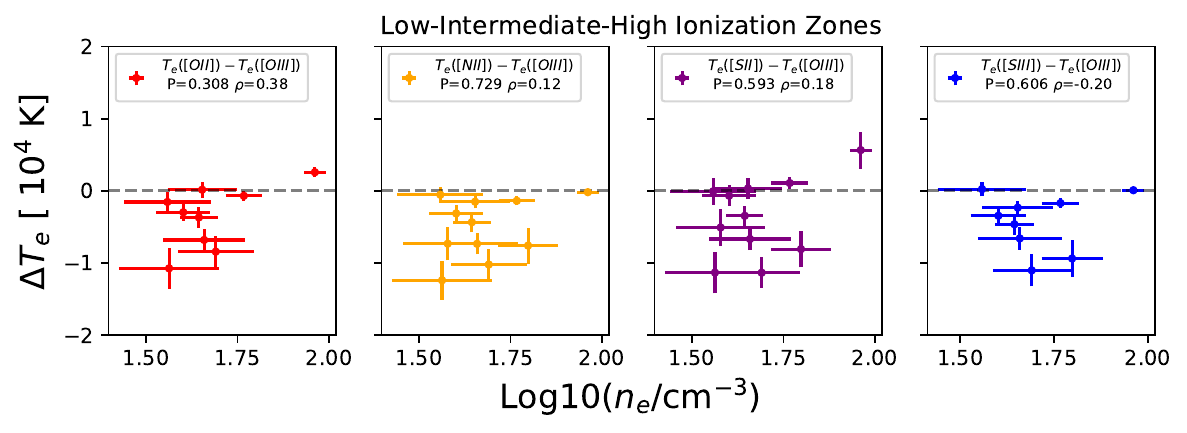}
    \caption{ Electron temperature differences compared to the \HII\ region electron density, $n_{\rm{e}}$. Top: The $\Delta T_e$'s between the low ionization zone temperatures. Middle: The $\Delta T_e$'s between the low and intermediate ionization zone temperatures. Bottom: The $\Delta T_e$'s between the low, intermediate and high ionization zone temperatures.}
    \label{fig:ne_comparisons}
\end{figure*}

\begin{figure*}[!h]
    \centering
    \includegraphics[scale=0.8]{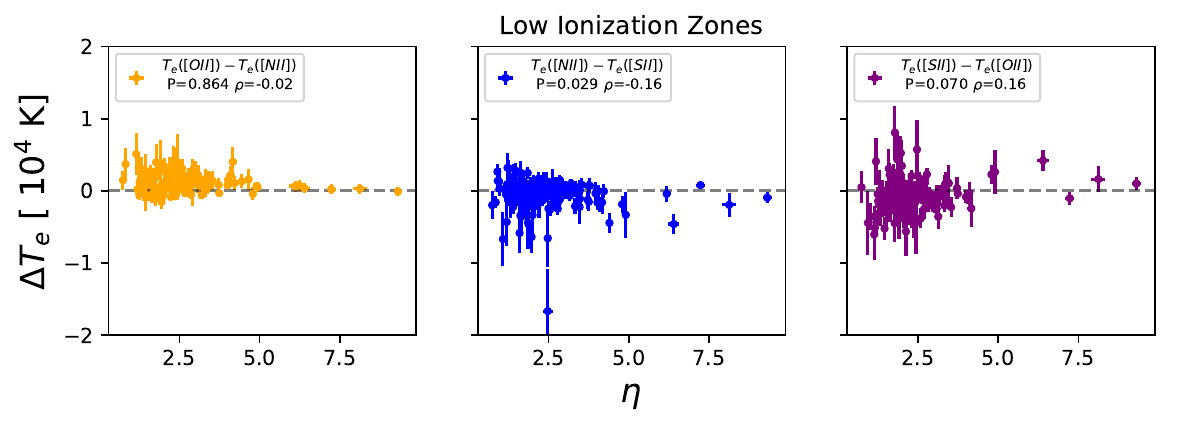}
    \includegraphics[scale=0.8]{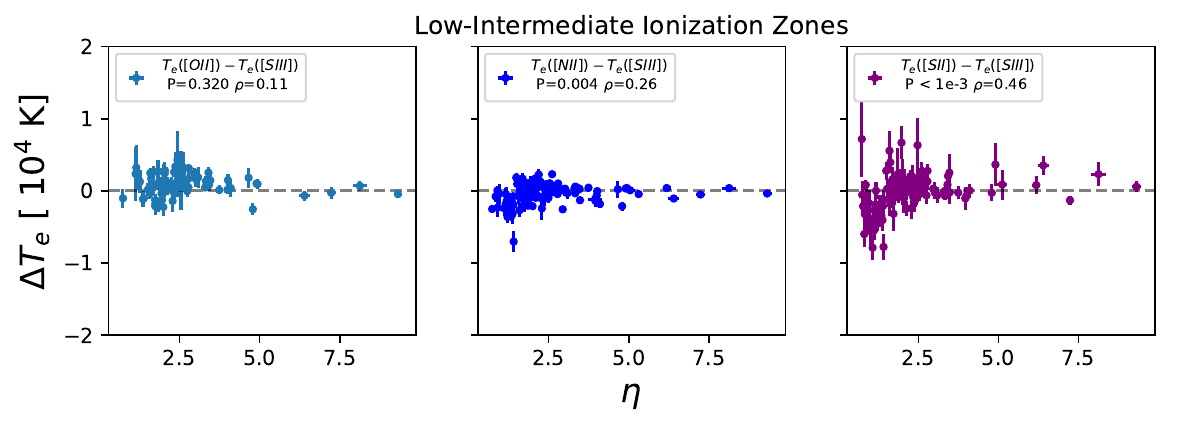}
    \includegraphics[scale=0.8]{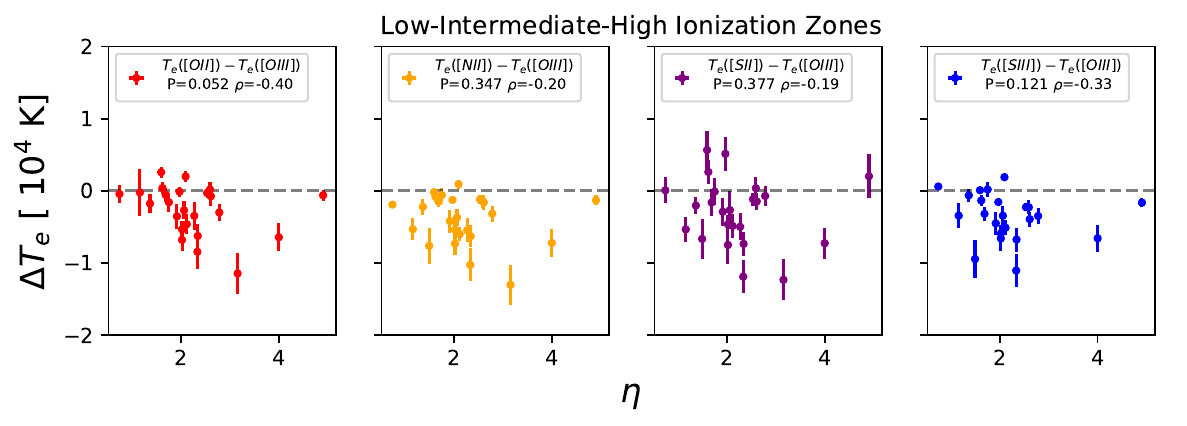}
    \caption{ Electron temperature differences compared to the radiation softness parameter. The order of the panels follow those in Figure \ref{fig:ne_comparisons}.}
    \label{fig:eta_comparisons}
\end{figure*}

\begin{figure*}[h]
    \centering
    \includegraphics[scale=0.8]{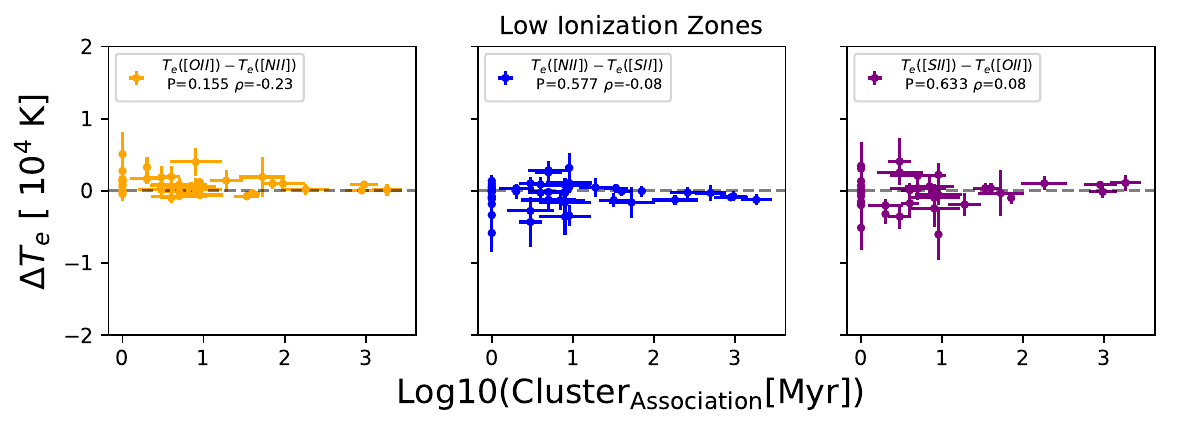}
    \includegraphics[scale=0.8]{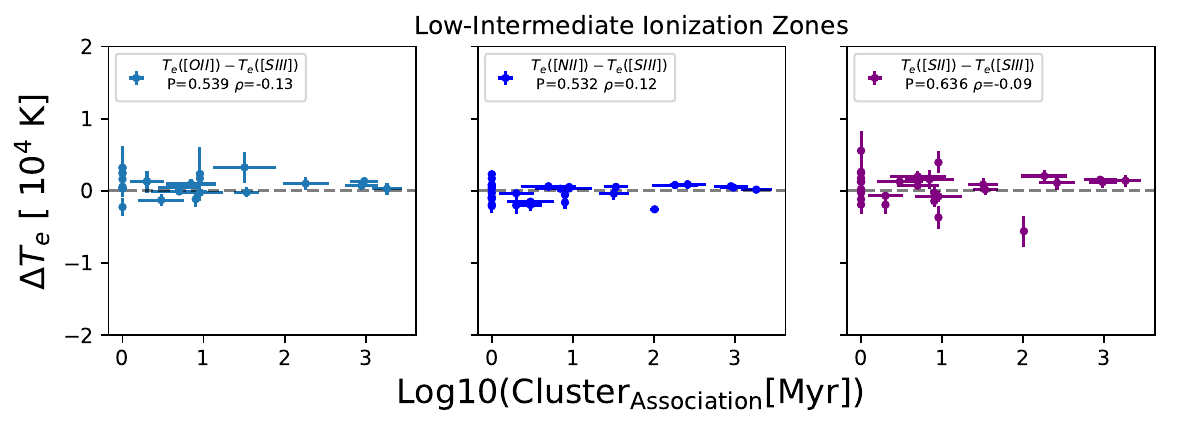}
    \includegraphics[scale=0.8]{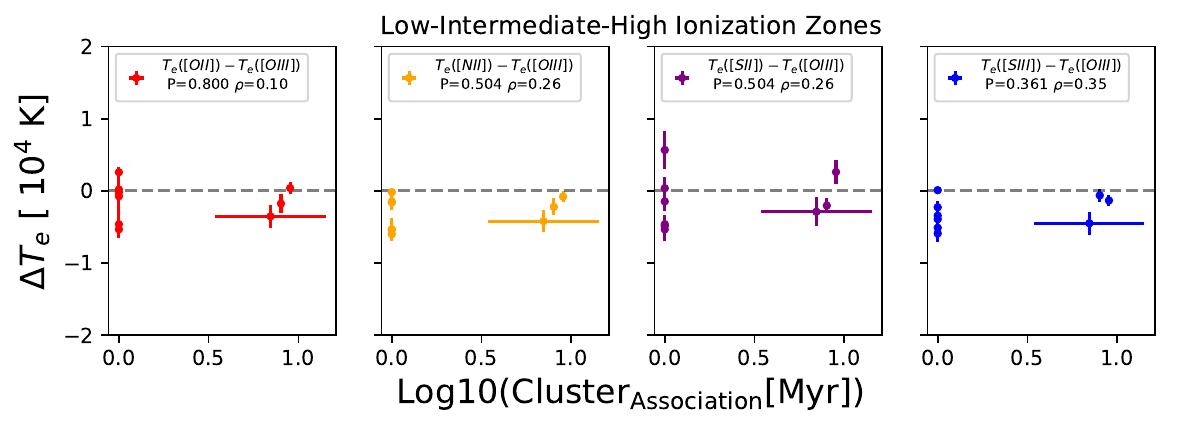}
    \caption{Electron temperature differences compared to the stellar cluster age. The order of the panels follow those in Figure \ref{fig:ne_comparisons}.}
   
    \label{fig:age_cluster_comparisons}
\end{figure*}

\begin{figure*}[h]
    \centering
    \includegraphics[scale=0.8]{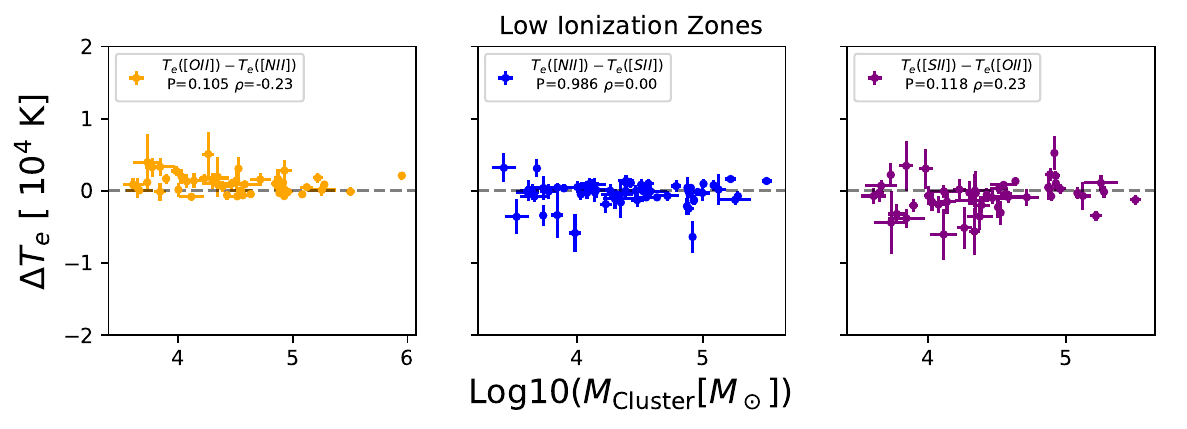}
    \includegraphics[scale=0.8]{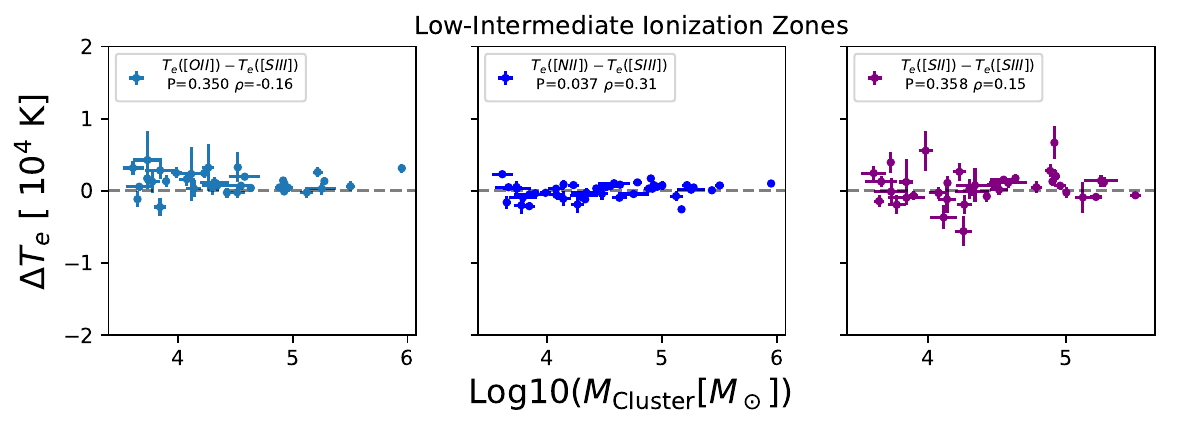}
    \includegraphics[scale=0.8]{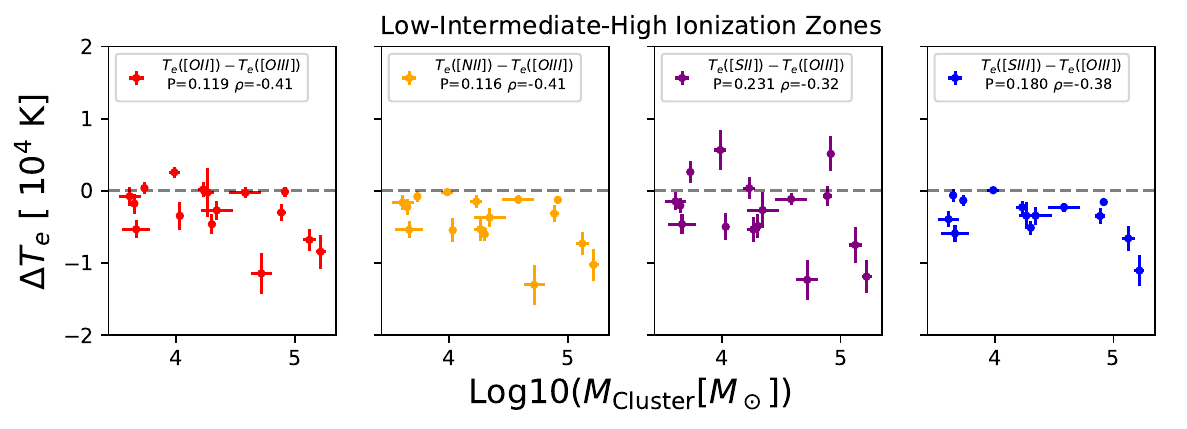}
    \caption{Electron temperature differences compared to the stellar cluster mass. The order of the panels follow those in Figure \ref{fig:ne_comparisons}.}
    \label{fig:mass_cluster_comparisons}
\end{figure*}

\begin{figure*}[!h]
    \centering
    \includegraphics[scale=0.8]{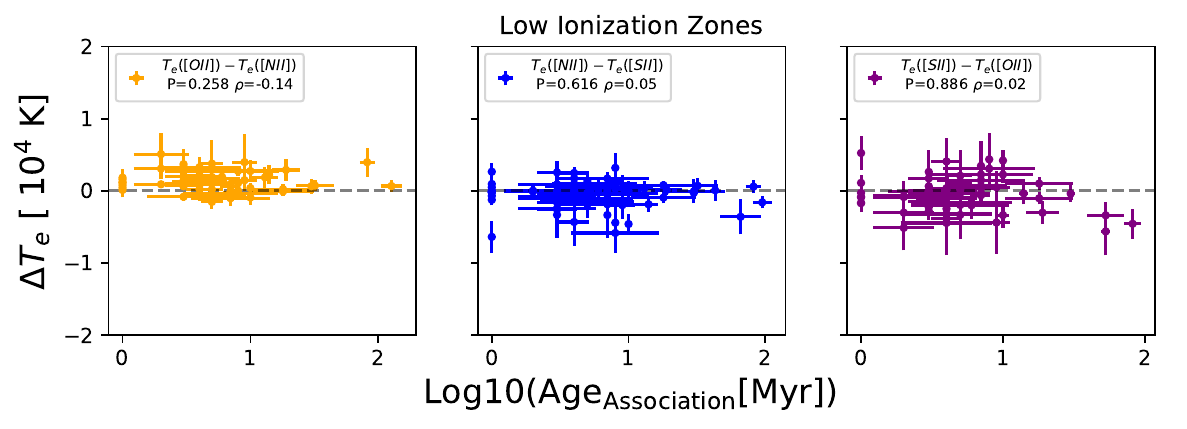}
    \includegraphics[scale=0.8]{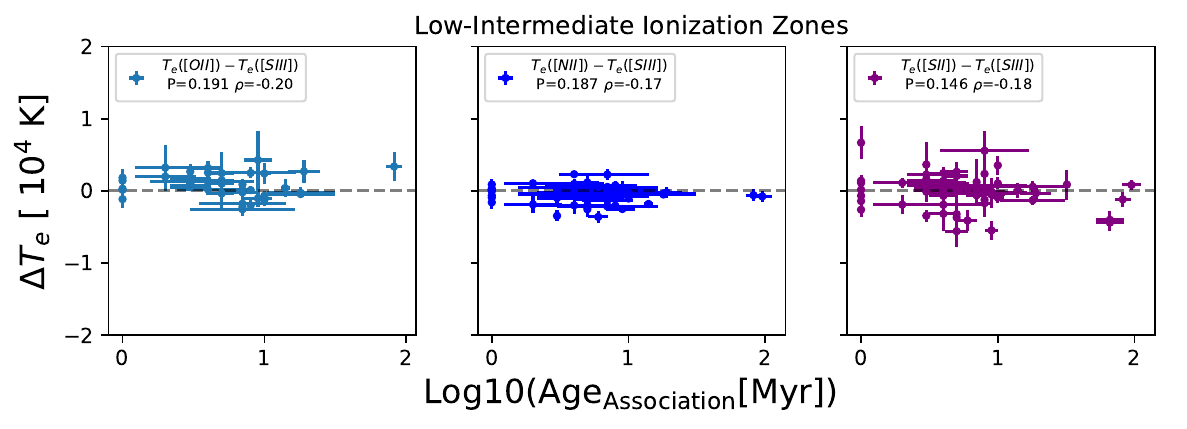}
    \includegraphics[scale=0.8]{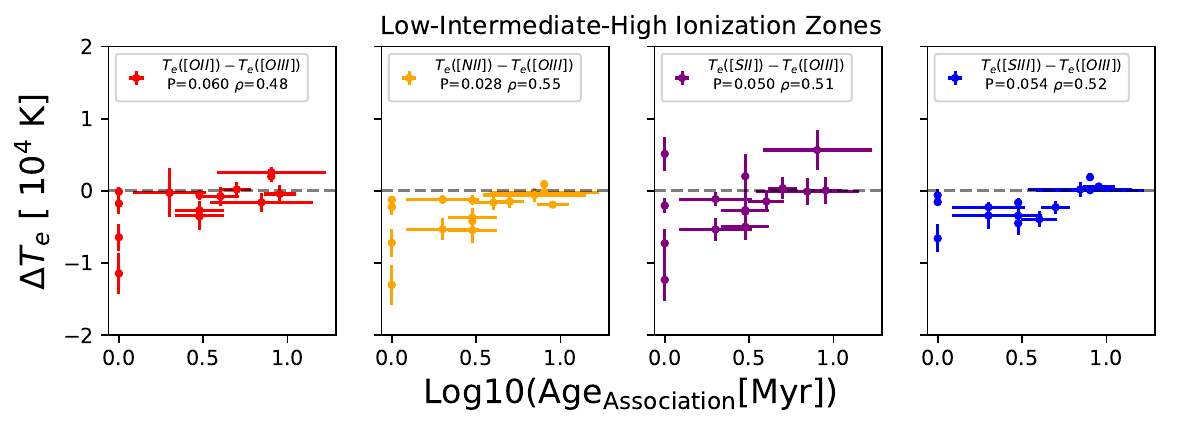}
    \caption{Electron temperature differences compared to the stellar association age. The order of the panels follow those in Figure \ref{fig:ne_comparisons}.}
    \label{fig:age_associations_comparisons}
\end{figure*}

\begin{figure*}[h]
    \centering
    \includegraphics[scale=0.8]{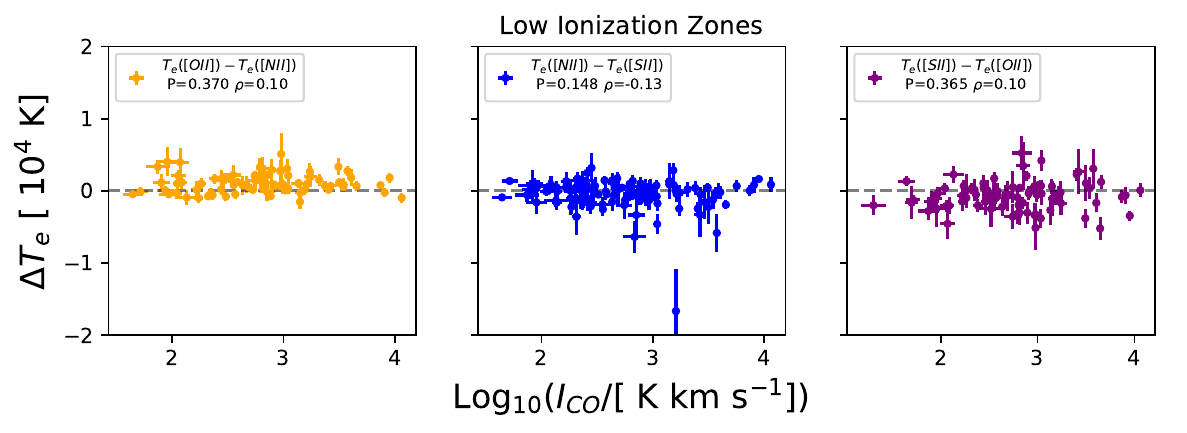}
    \includegraphics[scale=0.8]{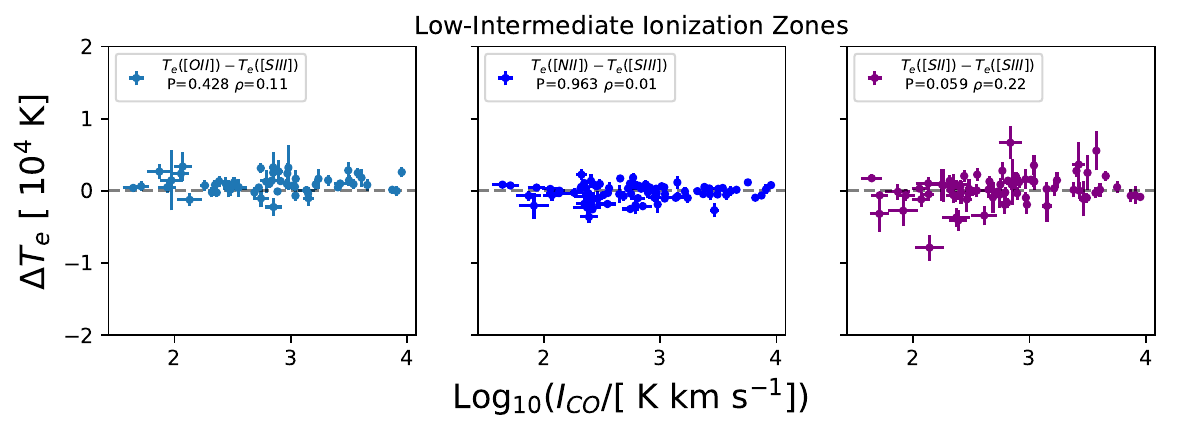}
    \includegraphics[scale=0.8]{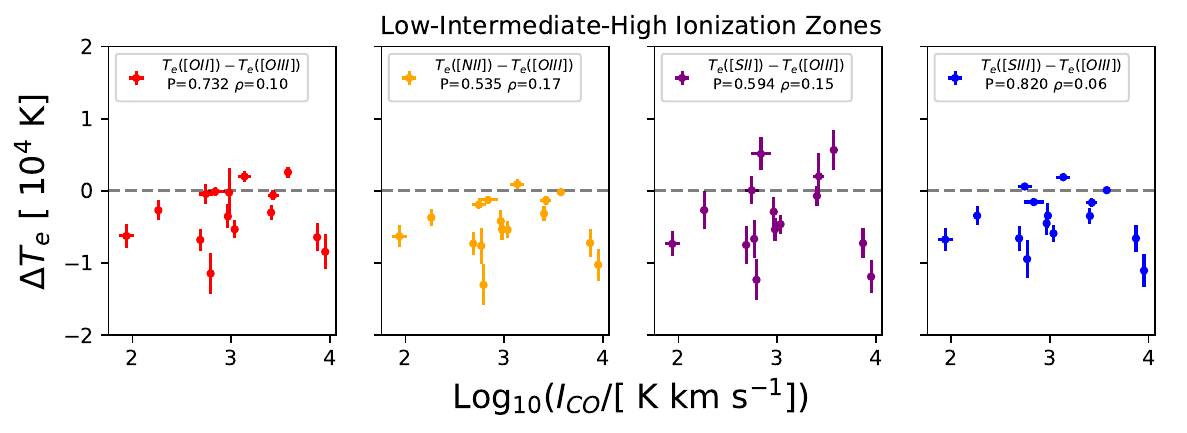}
    \caption{Electron temperature differences compared to the intensity of CO emission, $I_{\rm{CO}}$. The order of the panels follow those in Figure \ref{fig:ne_comparisons}.}
    \label{fig:ico_comparisons}
\end{figure*}

\begin{figure*}[h]
    \centering
    \includegraphics[scale=0.8]{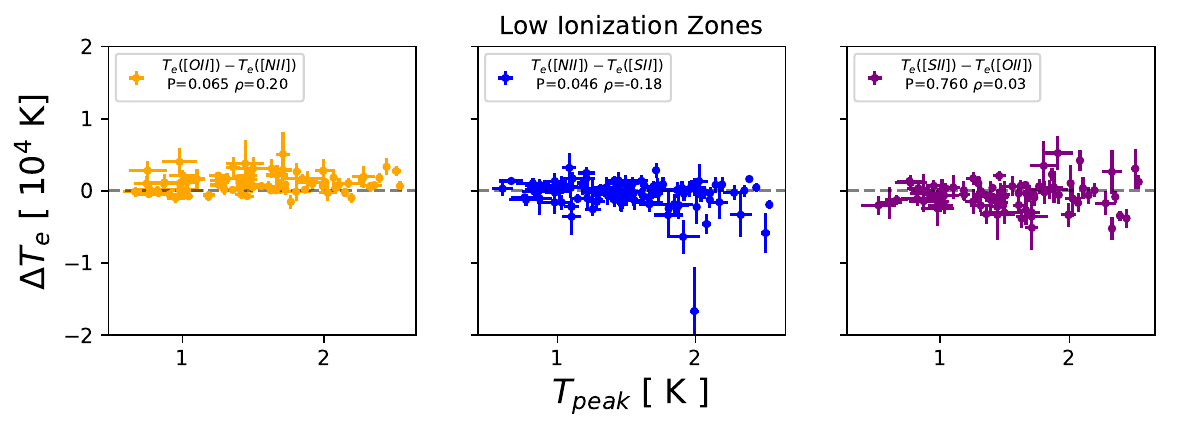}
    \includegraphics[scale=0.8]{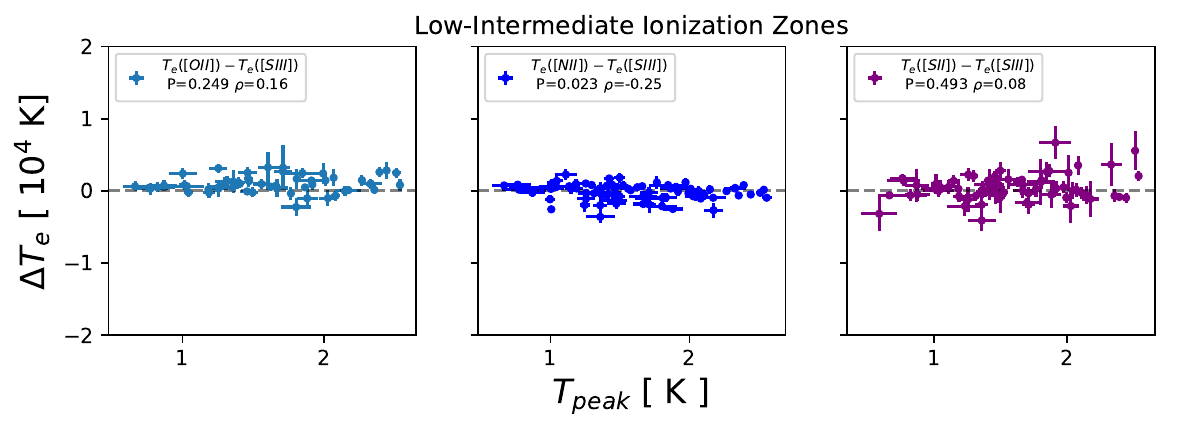}
    \includegraphics[scale=0.8]{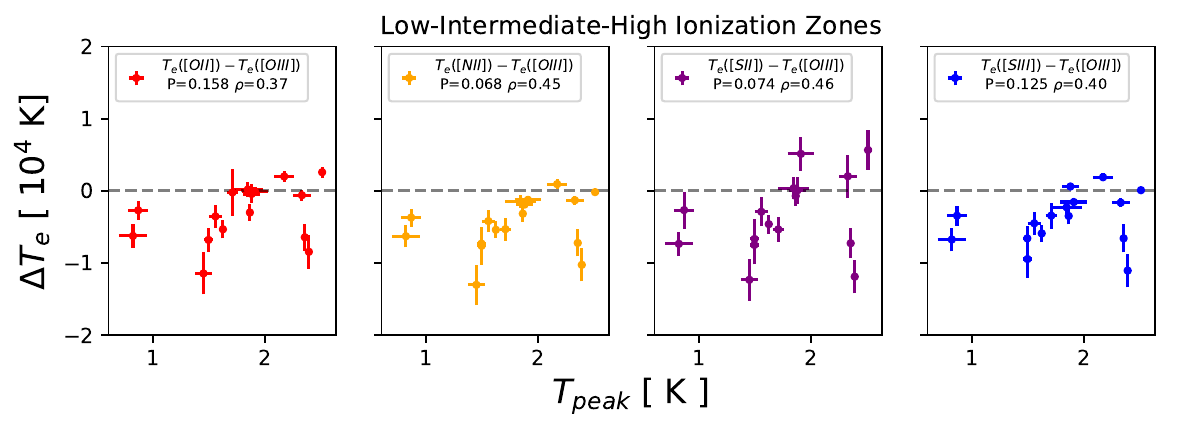}
    \caption{Electron temperature differences compared to the CO peak temperature, $T_{\rm{peak}}$. The order of the panels follow those in Figure \ref{fig:ne_comparisons}.}
    
    \label{fig:ptemp_comparisons}
\end{figure*}

\section{H~II Region Measurements}

\begin{longrotatetable}
\begin{deluxetable*}{ccccccccc}
\tablecaption{Measured and Derived Properties for \HII\ Regions with two more auroral lines.}
\label{tab:auroral_nebular_line_fluxes}
\tablehead{\colhead{ID} &  \colhead{NGC5068\_1} &  \colhead{NGC5068\_2} &  \colhead{NGC5068\_3} & \colhead{NGC5068\_4} &  \colhead{NGC5068\_5} & \colhead{NGC5068\_6} &  \colhead{NGC5068\_7} & \colhead{NGC5068\_8} }
\startdata
$\rm{R.A.}(^{\circ})$ &      199.714 &       199.719 &       199.702 &           199.7 &       199.696 &      199.714 &       199.71 &       199.703 \\
$\rm{Dec.}(^{\circ})$ &      -21.027 &       -21.016 &       -21.012 &         -21.013 &       -21.014 &      -21.015 &      -21.038 &       -21.008 \\
$E(B-V)$(Mag)              &         0.24 &          0.19 &          0.25 &            0.19 &          0.41 &         0.34 &         0.09 &          0.18 \\ \hline
H$\beta$\tablenotemark{a}              &  54.2$\pm$0.1 &   70.7$\pm$0.4 &  140.3$\pm$0.6 &   1385.6$\pm$7.3 &  211.4$\pm$1.4 &  37.6$\pm$0.2 &  27.2$\pm$0.2 &   59.8$\pm$0.3  \\
$\rm{[O~III]}$$\lambda$4363 &   - &    - &    - &     12.8$\pm$0.7 &    - &   - &   - &    -  \\
$\rm{[O~III]}$$\lambda$5007  &  53.6$\pm$0.1 &  125.3$\pm$0.5 &  133.5$\pm$0.6 &  3239.9$\pm$15.0 &  415.7$\pm$1.8 &  35.6$\pm$0.1 &  16.5$\pm$0.2 &   78.2$\pm$0.3 \\
$\rm{[O~II]}$$\lambda$7320   &   0.8$\pm$0.1 &    - &    2.4$\pm$0.8 &     35.7$\pm$2.2 &    3.9$\pm$0.8 &   - &   0.4$\pm$0.1 &    0.9$\pm$0.3 \\
$\rm{[O~II]}$$\lambda$7331   &   0.8$\pm$0.1 &    - &    2.6$\pm$0.8 &     31.1$\pm$2.2 &    3.9$\pm$0.8 &   - &   0.4$\pm$0.1 &    1.1$\pm$0.3  \\
$\rm{[O~II]}$$\lambda$3727   &  79.0$\pm$0.5 &  128.0$\pm$1.4 &  515.4$\pm$2.6 &  3128.6$\pm$25.3 &  639.0$\pm$5.8 &  89.3$\pm$0.6 &  62.1$\pm$0.6 &  194.3$\pm$1.2  \\
$\rm{[S~II]}$$\lambda$4068   &   0.6$\pm$0.1 &    1.5$\pm$0.1 &    3.1$\pm$0.9 &     46.4$\pm$7.4 &    4.0$\pm$1.2 &   0.8$\pm$0.1 &   1.3$\pm$0.4 &    1.4$\pm$0.2  \\
$\rm{[S~II]}$$\lambda$4076   &   0.2$\pm$0.1 &    0.5$\pm$0.1 &    3.1$\pm$1.8 &      3.8$\pm$3.6 &    0.0$\pm$2.4 &   0.3$\pm$0.1 &   0.1$\pm$0.1 &    0.5$\pm$0.2  \\
$\rm{[S~III]}$$\lambda$6313  &   0.5$\pm$0.0 &    0.7$\pm$0.1 &    1.5$\pm$0.1 &     18.9$\pm$0.3 &    2.7$\pm$0.1 &   0.4$\pm$0.0 &   0.2$\pm$0.0 &    0.7$\pm$0.0  \\
$\rm{[S~II]}$$\lambda$6716   &  12.7$\pm$0.1 &   22.6$\pm$0.2 &   60.4$\pm$0.3 &    352.1$\pm$0.9 &   71.0$\pm$0.3 &  14.4$\pm$0.1 &  10.6$\pm$0.1 &   21.5$\pm$0.1  \\
$\rm{[S~II]}$$\lambda$6731  &   9.4$\pm$0.1 &   15.8$\pm$0.1 &   42.1$\pm$0.3 &    259.1$\pm$0.8 &   49.8$\pm$0.3 &  10.1$\pm$0.1 &   7.3$\pm$0.1 &   14.8$\pm$0.1  \\
$\rm{[S~III]}$$\lambda$9069  &  11.9$\pm$0.1 &   12.4$\pm$0.2 &   20.5$\pm$0.2 &    389.1$\pm$1.2 &   42.3$\pm$0.3 &   7.1$\pm$0.1 &   4.5$\pm$0.1 &    9.9$\pm$0.1  \\
$\rm{[N~II]}$$\lambda$5756   &   0.3$\pm$0.0 &    0.3$\pm$0.0 &    0.5$\pm$0.1 &      4.3$\pm$0.1 &    0.8$\pm$0.1 &   0.1$\pm$0.0 &   0.1$\pm$0.0 &    0.2$\pm$0.1  \\
$\rm{[N~II]}$$\lambda$6548   &  10.2$\pm$0.0 &    9.1$\pm$0.1 &   20.3$\pm$0.1 &    152.5$\pm$0.3 &   25.2$\pm$0.1 &   6.2$\pm$0.0 &   4.7$\pm$0.0 &    6.8$\pm$0.0  \\
$\rm{[N~II]}$$\lambda$6584   &  30.1$\pm$0.1 &   26.9$\pm$0.2 &   59.7$\pm$0.3 &    449.8$\pm$1.0 &   74.1$\pm$0.3 &  18.3$\pm$0.1 &  13.9$\pm$0.1 &   20.0$\pm$0.1 \\ \hline
$n_{\rm{e}}$($\rm{cm}^{-3}$)                 &                 99 &                21 &                 13 &                 91 &                20 &                22 &                 12 &                 10  \\
\OIIIOII\             &        0.68$\pm$0.0 &      0.98$\pm$0.01 &        0.26$\pm$0.0 &       1.04$\pm$0.01 &      0.65$\pm$0.01 &        0.4$\pm$0.0 &        0.27$\pm$0.0 &         0.4$\pm$0.0  \\
\SIIISII\             &       1.87$\pm$0.02 &      1.13$\pm$0.02 &        0.7$\pm$0.01 &       2.23$\pm$0.01 &      1.23$\pm$0.01 &      1.01$\pm$0.01 &       0.88$\pm$0.01 &       0.95$\pm$0.01  \\
\temp{N}{2}~(K)              &    8179.0$\pm$291.0 &   8769.0$\pm$563.0 &    8122.0$\pm$489.0 &     8486.0$\pm$90.0 &   9056.0$\pm$458.0 &   7461.0$\pm$657.0 &    8208.0$\pm$778.0 &    8232.0$\pm$858.0  \\
\temp{S}{2}~(K)               &    7580.0$\pm$768.0 &  10364.0$\pm$594.0 &  11467.0$\pm$3028.0 &  14318.0$\pm$2652.0 &  7704.0$\pm$2444.0 &  8661.0$\pm$1044.0 &  14772.0$\pm$4095.0 &  10029.0$\pm$1260.0  \\
\temp{O}{2}~(K)                &  10941.0$\pm$1467.0 &        -&   7999.0$\pm$1134.0 &   11243.0$\pm$700.0 &  8785.0$\pm$1009.0 &        -&   9042.0$\pm$1421.0 &   8120.0$\pm$1143.0  \\
\temp{S}{3}~(K)                &    8513.0$\pm$110.0 &   9528.0$\pm$421.0 &   10235.0$\pm$273.0 &     8755.0$\pm$52.0 &   9807.0$\pm$180.0 &   9249.0$\pm$365.0 &    8438.0$\pm$185.0 &   10250.0$\pm$297.0  \\
\temp{O}{3}~(K)               &         -&        -&         -&    8658.0$\pm$135.0 &        -&        -&         -&         -\\
\enddata
\tablenotetext{a}{Emission line strengths, and uncertainties, are reported in units of $10^{-16}\times \rm{ergs}~\rm{s}^{-1}~\rm{cm}^{-2}$.}
\tablecomments{Table \ref{tab:auroral_nebular_line_fluxes} is published in its entirety in the machine-readable format. A portion is shown here for guidance regarding its form and content.}
\end{deluxetable*}
\end{longrotatetable}
\end{document}